\newcommand{\mum}{\ifmmode{\rm \mu m}\else{$\mu$m}\fi}
\documentclass[iop,numberedappendix,appendixfloats,twocolappendix]{emulateapj}
\usepackage{subfig}
\shorttitle{Dusty SMC OB Stars II: Disks or ISM Hot Spots?}
\shortauthors{Adams et al.}
\slugcomment{{\sc Submitted to \apj:} January 23, 2013}

\begin{document}
\title{Dusty OB stars in the Small Magellanic Cloud - II: Extragalactic Disks or Examples of the Pleiades Phenomenon?\altaffilmark{$\dag$}}
\author{Joshua J. Adams\altaffilmark{1}, Joshua D. Simon\altaffilmark{1}, 
Alberto D. Bolatto\altaffilmark{2}, G~.C. Sloan\altaffilmark{3}, Karin M. Sandstrom\altaffilmark{4,*}, 
Anika Schmiedeke\altaffilmark{4,5}, Jacco Th. van Loon\altaffilmark{6}, Joana M. Oliveira\altaffilmark{6}, 
Luke D. Keller\altaffilmark{7}}
\altaffiltext{1}{Observatories of the Carnegie Institution of Science, 813 Santa Barbara Street, Pasadena, CA 91101, USA; jjadams@obs.carnegiescience.edu}
\altaffiltext{2}{Department of Astronomy, University of Maryland, College Park, MD 20742, USA}
\altaffiltext{3}{Department of Astronomy, Cornell University, 222 Space Sciences Bldg., Ithaca, NY 14853-6801, USA}
\altaffiltext{4}{Max-Planck Institut f\"{u}r Astronomie, K\"{o}nigstuhl 17, D-69117, Heidelberg, Germany}
\altaffiltext{5}{Universit\"{a}t zu K\"{o}ln, Z\"{u}lpicher Strasse 77, 50937 K\"{o}ln, Germany}
\altaffiltext{6}{School of Physical and Geographical Sciences, Lennard-Jones 
Laboratories, Keele University, Staffordshire, ST5 5BG, United Kingdom}
\altaffiltext{7}{Department of Physics, Ithaca College, Ithaca, NY 14850, USA}
\altaffiltext{*}{Marie Curie Fellow}
\altaffiltext{$\dag$}{This work made use of data from \emph{Herschel}, which is an ESA space observatory with science instruments provided by 
European-led Principal Investigator consortia and with important participation from NASA.}
\begin{abstract}
We use mid-infrared \emph{Spitzer} spectroscopy and far-infrared \emph{Herschel} photometry for a sample of 
twenty main sequence O9--B2 stars in the Small Magellanic Cloud (SMC) with strong 
24~\mum\ excesses to investigate 
the origin of the mid-IR emission. Either debris disks around the stars or illuminated patches of dense 
interstellar medium (ISM) can cause such mid-IR emission. In a companion paper, Paper I, we use optical spectroscopy to 
show that it is unlikely for any of these sources to be classical Be stars or Herbig Ae/Be stars. 
We focus our analysis on debris disks and cirrus hot spots. 
The local, prototype objects for these models are the debris disk around Vega and the heated dust cloud surrounding the 
stars in the Pleiades, also known as a cirrus hot spot. These two cases predict different dust masses, radii, origins, and 
structures, but the cleanest classification tools are lost by the poor physical resolution at the distance of the SMC. We also consider 
transition disks, which would have observable properties similar to 
debris disks. 
We begin classification by measuring angular extent in the highest resolution mid-IR images available. We find three out of 
twenty stars to be significantly extended, establishing them as cirrus hot spots. We then fit the IR spectral energy distributions 
to determine dust temperatures and masses. Analysis yields minimum grain sizes, thermal equilibrium distances, 
and the resultant dust mass estimates. We find the dust masses in the SMC stars to be larger than for any known debris disks. 
The difference in inferred properties is 
driven by the SMC stars being hotter and more luminous than known debris disk hosts and 
not in any directly observed dust properties, so this evidence against the debris disk hypothesis is circumstantial. 
Finally, we created a local comparison sample of bright mid-IR OB stars in the Milky Way (MW) by cross-matching the 
\emph{WISE} and \emph{Hipparcos} catalogs. 
We find that of the thousands of 
nearby ($\leq 1$ kpc) hot stars in the MW that show a mid-IR excess, only a small fraction (few percent) 
match the high mid-IR luminosities of the SMC stars. All such local stars in the appropriate luminosity range that can be 
unambiguously classified are young stars 
with optical emission lines or are spatially resolved by \emph{WISE} with sizes too large to be 
plausible debris disk candidates. 
We conclude that the very strong mid-IR flux excesses are most likely explained as cirrus hot spots, although we cannot 
rigorously rule out that a small fraction of the sample is made up of debris disks or transition disks. We present 
suggestive evidence that bow-shock heating around runaway stars may be a contributing mechanism to the 
interstellar emission. These sources, interpreted as cirrus hot spots, offer a new localised probe of 
diffuse interstellar dust in a low metallicity environment. 
\end{abstract}


\section{Introduction}\label{sec:intro}
\setcounter{footnote}{0}
The study of dust around main-sequence stars has made great progress with space-based mid-IR observations, 
primarily through studies with the \emph{Infrared Astronomical Satellite} 
(\emph{IRAS}) all-sky survey 
\citep[e.g.][]{Auman84,Backm93,Gaust93,Sylve96a,Manni98,Rhee07}, the \emph{Spitzer Space Telescope} 
\citep[e.g.][]{Rieke05,Beich06,Su06,Chen06,Carpe06,Hille08,Gaspa08,Rebul08,Trill08,Carpe09,Dahm09,Moral09,Moor11}, and the 
\emph{Herschel Space Observatory} \citep[e.g.][]{Eiroa10,Eiroa11,Matth10,Sibth10,Vande10,Acke12,Wyatt12,Ertel12,Gaspa13,Booth13,Broek13,Eiroa13}. 
Several types of dusty 
objects around stars can produce comparable observational signatures, but the improved spatial resolution 
of \emph{Spitzer} compared to \emph{IRAS} and a number of high-resolution tools (e.g. \emph{Hubble Space Telescope} 
imaging of 
reflection nebulae, coronographic imaging, and interferometry in the optical through sub-mm) 
\citep[e.g.][]{Kalas02,Kalas05,Kalas07,Stark09} have built up large classes of physically 
distinct sources. Four prominent types of objects are found in surveys of main-sequence and younger 
stars: (1) debris disks \citep[e.g.][]{Wyatt08}, (2) patches of overdense ISM heated 
by nearby stars \citep[][hereafter called cirrus hot spots]{vanBu88}, (3) protoplanetary and protostellar disks around pre-main 
sequence objects and young stars \citep[e.g.][]{Willi11}, and (4) classical Be 
stars, which have an excreted gaseous disk 
powered by free-free emission \citep[e.g.][]{Gehrz74,Colli87,Cote87,Water87,Porte03}. 
The cirrus hot spots are of two types: 1) nearly static arrangements of overdense gas \citep[coined as the 
Pleiades phenomenon by][]{Kalas02} or 2) high velocity interactions where the emission is 
enhanced in a bow shock \citep{Kalas07,Hines07,Gaspa08,Debes09,Marti09}. 

The primary discriminating 
observations for nearby samples are spatial extent for the first two categories, signs of 
accretion such as H$\alpha$ emission and age estimates since protoplanetary 
disks have lifetimes of a few Myrs for the third, and the unique spectral energy 
distribution (SED) shape from free-free radiation and H$\alpha$ emission for the fourth. 
Many claims of \emph{IRAS}-discovered debris disks have been overturned with 
higher resolution imaging as instead cirrus hot spots \citep{Kalas02,Gaspa08,Marti09}, but the ambiguity for 
nearby ($<$100 pc) stars with the latest data is minimal. Finally, there will be a small number 
of objects outside of our four classes, such as proplyds, where the irradiated matter 
does not surround the star but is seen in projection. $\sigma$ 
Ori, discussed in Section \ref{sec:WISE_res}, is one such example. However, these 
objects should be rare and often identifiable by H$\alpha$ emission. 

This work seeks to better understand the nature of an intriguing set of dusty objects 
in the Small Magellanic 
Cloud (SMC) identified by the 
\emph{Spitzer} Survey of the SMC \citep[S$^3$MC;][]{Bolat07}. 
\cite{Bolat07} found 193 OB stars in the SMC that show strong mid-IR excesses from the full population of 3800 O9--B3 stars. 
A subsample of 125 from the 193 dusty OB stars were observed by \cite[][hereafter Paper I]{Sheet13} with optical spectroscopy, and 
87 (70\%) were 
established as main-sequence stars. We have obtained both deep far-IR photometry and mid-IR spectra for 11 of these 
87 stars, and mid-IR spectra alone for an additional 9. The advantage of this sample is that we can study dusty clouds 
and disks in a much more metal-poor 
environment than locally, but the challenge is that some of the best classification tools are 
impossible at the SMC distance. The observations in Paper I have shown that the 
majority of this sample and all the sources studied in this work are truly main-sequence stars and 
ruled out classifications as classical 
Be \citep{Miros00,Porte03} or Herbig Ae/Be stars \citep{Hille92,Water98} by the 
absence of Balmer emission. We are 
left with debris disks, transition disks, and cirrus hot spots as the most viable explanations. 

The SMC dusty OB stars share many observational characteristics of debris disks. 
The first debris disk detection was in $\alpha$ Lyrae (Vega) with \emph{IRAS} \citep{Auman84}. 
Since then, $\alpha$ Piscis \citep[Fomalhaut,][]{Holla98}, 
$\beta$ Pictoris \citep{Backm96}, and $\epsilon$ Eridani \citep{Greav05} have joined Vega as debris disk prototypes.
The definitive observation necessary to secure a candidate is to resolve the disk in imaging and, with 
temperature and age information, establish that the reradiating particles are not from the primordial disk but have been  
replenished through collisions. The 
first such confirmation was by \cite{Smith84} for $\beta$ Pictoris, and spatially resolved observations of Vega 
followed \citep{Harv96}. Unfortunately, spatially resolved observations at the distance of the SMC are not feasible 
with current instruments, so we must turn to other, 
less direct methods to infer the nature of the SMC sample. 
Even observations with the forthcoming Mid-Infrared Instrument on the James 
Webb Space Telescope 
would only provide 0\farcs7 (21,350 AU radius) resolution at 20~\mum\ 
and be unable to resolve SMC debris disks. 
Comprehensive \emph{Spitzer} studies of debris disks have focused on A stars \citep{Rieke05,Su05} and
FGK stars \citep{Chen05,Chen06,Hille08,Moor11} or both \citep{Carpe06,Carpe09}. One recent study does 
examine more massive stars \citep[3 M$_{\odot}<$ M$_{*}<$ 10 M$_{\odot}$;][]{Chen12} with B8 as the earliest type found hosting a debris disk. 
The frequency of debris disk detections drops at ages beyond 50 Myr, but appears nearly flat for stellar
types A--K at younger ages \citep{Wyatt08}. 

Transition disks are another source class that may represent some or all of the SMC stars. Transition disks 
were first discovered by \cite{Skrut90} with \emph{IRAS} data. Transition disks are thought be
a phase between protoplanetary disks and debris disks where the star has photoevaporated the gas throughout the
central tens of AU in the disk. The classical T Tauri star TW Hya 
can be considered a prototype object for transition disks \citep{Calve02}. For many of the properties relevant to 
our study, transition disks will have 
similar properties to debris disks. They may have vast sizes around massive stars. The key feature for transition 
disks is understanding how rapidly they are destroyed and what remnants they leave behind. We will discuss the 
results from numerical simulations in Section \ref{sec:proto}, which inform that understanding. 

Alternatively, it is also possible that these SMC stars are surrounded by cirrus hot spots. The 
reflection nebula in the Pleiades star cluster is the prototypical cirrus hot spot \citep{Arny77,Kalas02,Gibso03a,Gibso03b}. 
We note that debris disks have also been found around some stars in the Pleiades cluster \citep{Gorlo06,Sierc10}, so the two 
phenomena are not mutually exclusive. 
The observable properties of cirrus hot spots have 
considerable overlap with debris disks \citep{Backm93}. The study by \cite{Gaust93} compiled stars with mid-IR excess  
in \emph{IRAS} data and by analyzing 1753 hot stars, found that 302 show extended emission characteristic of cirrus hot spots. Common 
properties for the cirrus hot spots are halos that extend from 1,000 to 100,000 AU from the star, 
dust masses of order 1 to 100 M$_{\rm \earth}$, temperatures of order 100 to 150 K, and 
luminosities relative to the
star of $10^{-5}<{\rm L}_{\rm disk}/{\rm L}_{\rm *}<10^{-2}$ 
\citep[properties based on one K dwarf, two G dwarfs, and three 
late-B dwarfs;][]{Kalas02}. 
For comparison, debris disks parent bodies commonly extend from 10 to 
several hundred AU while some cases contain a halo of small grains out to 
1,000 AU, dust masses that
range from 10$^{-3}$ to a few Earth masses, cold component temperatures from 50 to 100 K, and luminosities relative to the
star of $10^{-5}<{\rm L}_{\rm disk}/{\rm L}_{\rm *}<10^{-2}$ \citep{Backm93,Chen06,Carpe09,Krivo10}. Some debris disks 
also show a warm component with a temperature of $\approx$200K \citep{Moral11}. 
The larger mass estimates generally come from submillimeter data, and analysis of the mid-IR range generally 
implies $\sim$10$^{-1}$ M$_{\earth}$ for debris disks. The population of dusty OB stars in the SMC is rather rare. 
\cite{Bolat07} find that there are 3800 O9--B3 stars in the SMC, and only 193 show strong mid-IR excesses. 
70\% of these are normal, main-sequence stars (Paper I), so any explanation must only explain 
an occurrence fraction of 2--4\% in massive stars. 
The incidence of cirrus hot spots in the MW around massive stars 
\citep[17\%;][]{Gaust93} is more than 
enough to explain this occurrence rate. How the SMC may be different is still uncertain. 
Cirrus hot spots in the SMC might be expected to be fainter than the MW ones based on 
the SMC's lower dust-to-gas ratio \citep[10$\times$ lower;][]{Leroy07}. However, 
the Local Bubble is also underdense and still contains many cirrus hot spots \citep{Kalas02}. 
The relative number of debris disks and cirrus hot spots, 
even in the MW, is not well determined for the range of hot stars (O9--B3) present in 
our S$^3$MC subsample. 

In this work, we seek to better characterize the properties of these dusty OB stars and 
discriminate between the debris disk and cirrus hot spot models with new data and a large, 
newly collected local sample including information on spatial extent. We start by describing the 
photometric and spectroscopic data in Section \ref{sec:data}. The angular extent of the dust is measured in three out of twenty of the 
stars. This permits their classification as cirrus hot spots, but the classification of the remaining seventeen is uncertain. 
Next, we fit simple models with dust 
emission to the spectral energy distributions in Section \ref{sec:sed}. 
We collect literature and catalog data for local sources that may serve as analogs in Section \ref{sec:MW}. 
In Section \ref{sec:discuss}, we discuss whether the evidence favors 
either of the proposed models. 
The SMC stars stand out in mid-IR color from any published lists of debris disks or cirrus 
hot spots, but a sample of nearby early-type stars from the \emph{Hipparcos} and \emph{Wide-field Infrared Survey Explorer} 
(\emph{WISE}) catalogs does match 
the SMC stars. Finally, we present our conclusions in Section \ref{sec:conc}. 
All magnitudes and colors listed are in the Vega system, and we have assumed an SMC 
distance of 61 kpc throughout \citep{Weldr04}. All images are displayed with equatorial North up and East to the left. 

\section{Observational Data}\label{sec:data}

We present new data and review old data on a subset of stars that Paper I classify as main sequence O9--B2 stars. 
We selected the sample for mid- and far-IR followup 
observations by requiring a flux cut of $f_{\nu}(24\mum) > 700 \mu$Jy, the sources to be unresolved in the 24\mum\ imaging,
no H$\alpha$ detections in narrow-band imaging to be present, and minimal 
excess at 8\mum\ in the IRAC bands. Figure 16 of Paper I shows the 
distribution of relative fluxes for all the dusty SMC sources, and the sources we study 
here fall in the two central bins. Our subsample ought to represent the 
most compact and coolest dusty OB stars in the SMC. The basic properties 
of our targets are listed in Table \ref{tab:sedprop1}, with more details on the columns given in Section \ref{sec:ancil}. 
\subsection{Spitzer Photometry}

Photometry is measured with data taken by the \emph{Spitzer Space Telescope} \citep{Werne04}. 
\cite{Bolat07} have described the observations from the Infrared Array Camera \citep[IRAC;][]{Fazio04} and the 
Multiband Imaging Photometer for Spitzer \citep[MIPS;][]{Rieke04}. We 
have examined new peak-up images at 16~\mum\ taken with the 
Infrared Spectrograph \citep[IRS][]{Houck04} as part of the GO program 50088 (P.I. J.~Simon). 
These images are analyzed for their spatial extent and discussed in Section \ref{sec:peakup}. 
IRAC and MIPS 24~\mum\ photometry are taken from Paper I, where the combined 
catalog for S$^3$MC \citep{Bolat07} and SAGE-SMC \citep{Gordo11} observations is used 
for the best depth. The catalogs were produced with point spread function (PSF) photometry from MOPEX \citep{Makov05}. None of 
our sources are detected in the MIPS 70~\mum\ band to the catalog's faint flux limit of 35 mJy for unresolved sources.  

\begin{centering}
\begin{deluxetable*}{crrccrr@{$\pm$}lcc}
\tabletypesize{\scriptsize}
\tablecaption{Stellar properties of the dusty OB stars\label{tab:sedprop1}} \tablewidth{0pt}
\tablehead{     \colhead{ID No.}     &
                \colhead{$\alpha$}      &
                \colhead{$\delta$}      &
                \colhead{Spectral}      &
                \colhead{A(V)$_{\rm SMC}$}      &
                \colhead{M$_{\rm V}$}      &
                \multicolumn{2}{c}{a$_{\rm min}$}      &
                \colhead{Herschel}      &
                \colhead{Resolved}      \\
                \colhead{}     &
                \colhead{(J2000)}      &
                \colhead{(J2000)}      &
                \colhead{Type}      &
                \colhead{(mag)}      &
                \colhead{(mag)}      &
                \multicolumn{2}{c}{(mm)}      &
                \colhead{Observation?}      &
                \colhead{at 16~\mum?}      }
\startdata

B004 & 11.40574 & $-$73.21014 & O9 & 0.25 & $-$4.10 & 0.95 & 0.13 & Y & N \\
B009 & 11.73266 & $-$73.08136 & B0$-$B2 & 0.07 & $-$3.87 & 0.43 & 0.06 & Y & N \\
B011 & 11.73879 & $-$73.30242 & \ldots & 0.06 & $-$2.95 & 0.43 & 0.06 & Y & N \\
B014 & 11.82917 & $-$73.11972 & B0$-$B2 & 0.07 & $-$4.52 & 0.43 & 0.06 & Y & N \\
B021 & 11.92833 & $-$73.00271 & B0 & 0.22 & $-$5.00 & 0.52 & 0.07 & N & N \\
B024 & 11.95000 & $-$73.07528 & B0$-$B2 & 0.06 & $-$3.56 & 0.43 & 0.06 & Y & N \\
B026 & 11.96667 & $-$73.35611 & B0$-$B2 & 0.08 & $-$3.93 & 0.43 & 0.06 & Y & N \\
B029 & 12.04036 & $-$73.40363 & B0 & 0.17 & $-$5.04 & 0.52 & 0.07 & Y & Y \\
B034 & 12.12513 & $-$73.30265 & B0 & 0.09 & $-$3.14 & 0.52 & 0.07 & N & N \\
B087 & 13.25087 & $-$72.67440 & B0 & 0.00 & $-$4.63 & 0.52 & 0.07 & Y & Y \\
B096 & 13.50825 & $-$72.70609 & B0 & 0.00 & $-$3.68 & 0.52 & 0.07 & N & N \\
B100 & 13.72685 & $-$72.53598 & B1 & 0.07 & $-$3.36 & 0.43 & 0.06 & N & N \\
B102 & 13.76366 & $-$72.92260 & O9 & 0.09 & $-$4.43 & 0.95 & 0.13 & N & N \\
B137 & 14.74641 & $-$72.74281 & B1 & 0.09 & $-$4.91 & 0.43 & 0.06 & N & N \\
B148 & 15.22989 & $-$72.13463 & B0 & 0.03 & $-$4.47 & 0.52 & 0.07 & N & N \\
B154 & 15.67326 & $-$72.01004 & B1 & 0.14 & $-$3.38 & 0.43 & 0.06 & Y & N \\
B159 & 15.75023 & $-$72.46061 & B0 & 0.00 & $-$4.13 & 0.52 & 0.07 & N & N \\
B182 & 16.22062 & $-$71.91369 & B0 & 0.00 & $-$4.98 & 0.52 & 0.07 & Y & N \\
B188 & 17.05033 & $-$71.97593 & OBe? & 0.01 & $-$3.97 & 0.95 & 0.13 & Y & Y \\
B193 & 19.19872 & $-$73.14434 & B0 & 0.29 & $-$4.53 & 0.52 & 0.07 & N & N 
\enddata
\tablecomments{The spectral types in column 4 are estimated from optical spectroscopy in Paper I and 
determine the stellar parameters. For the O9, B0, and B1 stars we consider, the masses are 20.3, 17.5, and 14.2 M$_{\rm \odot}$, 
the luminosities are 11.1$\times$10$^4$, 5.3$\times$10$^4$, and 3.6$\times$10$^4$ L$_{\rm \odot}$, 
the radii are 8.0, 7.4, and 6.5 R$_{\rm \odot}$, and the temperatures are 3.7$\times$10$^4$, 3.2$\times$10$^4$, 
and 3.1$\times$10$^4$ K, respectively. We have made the small corrections due to 
MW and SMC extinction as described in the text. The V-band absolute magnitudes for O9, B0, and B1 stars in 
\cite{Schmi82} are -4.25, -4.0, and -2.07, and the match to the observed values in column 6 
provide confidence that the stars are all dwarfs. 
In column 7, a$_{\rm min}$ is the minimum grain size that can 
remain stable around the star against radiation pressure. Column 8 lists whether the 
star was observed with \emph{Herschel}. Column 9 states if the mid-IR emission was found to be 
significantly extended in the IRS peak-up image. If so, it is too large to be a debris disk. 
}
\end{deluxetable*}
\end{centering}

The broadband data need color corrections since our SED fits are made to data 
evaluated at discrete wavelengths. 
We show that the corrections are small and the exact values 
adopted are unimportant to our results. Tabulated color corrections exist for power-law functions and 
single-temperature blackbodies. There is no simple color correction for us to make over all the 
wavelengths our data span, so we make a piecewise correction. We fit the data with high-temperature 
blackbodies for the stellar photospheres and modified and straight blackbody functions 
for the mid-IR excess (Section \ref{sec:sed}), which 
commonly have temperatures of 100 K and relative IR to bolometric flux ratios of 10$^{-4}$. 
We therefore split into three wavelength ranges for the purpose of color correction: 
at $\lambda\leq$15~\mum\ a Rayleigh-Jeans function representing the stellar photosphere is used, 
for $\lambda>$15~\mum\ and $\lambda\leq$30~\mum\ a 100 K blackbody is used, and finally a 
power law function with f$_{\rm \nu} \propto \nu^3$ is used at $\lambda>$30~\mum\ as the Rayleigh-Jeans 
regime for the dust, which radiates inefficiently under an assumed emissivity of $\beta_{\rm em}=1$. 
In Section \ref{sec:sed}, we make fits with both an unmodified blackbody function and with an 
emissivity of $\beta_{\rm em}=2$, so this evaluation for the color correction is a compromise with 
differences too small to make an impact on our results. We have taken color corrections from Table 4.4 of the IRAC Instrument 
Handbook\footnote{http://irsa.ipac.caltech.edu/data/SPITZER/docs/irac/\\
iracinstrumenthandbook/1} and list them in Table \ref{tab:col_cor}. The 24~\mum\ color correction is from 
the MIPS Instrument Handbook\footnote{http://irsa.ipac.caltech.edu/data/SPITZER/docs/mips/\\
mipsinstrumenthandbook/1}. The effect of choosing corrections for 
different spectral shapes on the final results is smaller than the statistical uncertainties.  
\begin{deluxetable}{crrr}
\tabletypesize{\scriptsize}
\tablecaption{Photometry correction terms\label{tab:col_cor}}
\tablewidth{0pt}
\tablehead{
\colhead{Band} & \colhead{Effective} & \colhead{Aperture} & \colhead{Color} \\
\colhead{} & \colhead{$\lambda$ (\mum)} & \colhead{Correction} & \colhead{Correction} }
\startdata
IRAC 1 & 3.550 & \ldots\tablenotemark{*} & 1.0111 \\
IRAC 2 & 4.439 & \ldots\tablenotemark{*} & 1.0121 \\
IRAC 3 & 5.732 & \ldots\tablenotemark{*} & 1.0115 \\
IRAC 4 & 7.872 & \ldots\tablenotemark{*} & 1.0337 \\
MIPS 1 & 23.68 & \ldots\tablenotemark{*} & 0.947 \\
PACS 1 & 70.00 & 0.701 & 1.04 \\
PACS 3 & 160.0 & 0.759 & 1.14
\enddata
\tablenotetext{*}{These data were fit with PSF photometry instead of aperture 
photometry. Aperture corrections were made internally in MOPEX, but 
they are not meaningful to list without a full description of the PSF and 
thus not listed here. See \cite{Bolat07} for details.}
\end{deluxetable}

\subsection{16~\mum\ Peak-up Images}
\label{sec:peakup}
The 16~\mum\ peak-up images were taken since the camera has the best resolution available (${\rm FWHM}=3\farcs8$) 
at wavelengths where the dust emission dominates. A 6-position dither pattern was cycled for total exposures of 30 s 
per star. The images are shown in Figures \ref{fig:photim} and \ref{fig:peak}. We have made detections and 
measurements on the images with SExtractor \citep{Berti96}. We use the Kron aperture ($r_{\rm Kron}$)
as the metric to determine angular extent. 
The Kron aperture is a circle with the same first moment as the light profile and a scaling factor. 
SExtractor uses the factor $\times2.5$ to enclose 96\% of an object's flux. 
\cite{Kron80} notes that the first moment of the light 
profile is approximately the size of the half-light radius, so our reported radii will be 2.5 half-light radii. 

We have fit the appropiate point-response-function (PRF) image in the same manner. The 
PRF with the standard pixel scale, dated Feb. 2009, was used\footnote{http://irsa.ipac.caltech.edu/data/SPITZER/docs/\\
irs/calibrationfiles/peakuppsfprf/}. We measure $r_{\rm Kron}=6\farcs17$ for the PRF, which we subtract 
in quadrature to determine the intrinsic stellar values ($r_{\rm Kron,i}$). The determination of 
size in the peak-up images is dominated by the photon-noise with a 
median error of 5\arcsec, which is roughly one pixel when convolved with 
the PRF. We find three 
stars that are significantly extended ($>3\sigma$) after 
correcting for the PRF: B029 with $r_{\rm Kron,i}=9\farcs5\pm1\farcs9$, 
B087 with $r_{\rm Kron,i}=11\farcs6\pm2\farcs7$, and B188 with $r_{\rm Kron,i}=8\farcs6\pm2\farcs8$. These 
sources have dust clouds of size $\sim$ 3 pc, which makes them far larger than debris disks. The 
size constraints on the remaining sources are not small enough to determine their nature. The extended objects 
are flagged by column 9 in Table \ref{tab:sedprop1}.

\begin{figure*}
\centering
   \subfloat{
      \includegraphics [scale=0.45,angle=0]{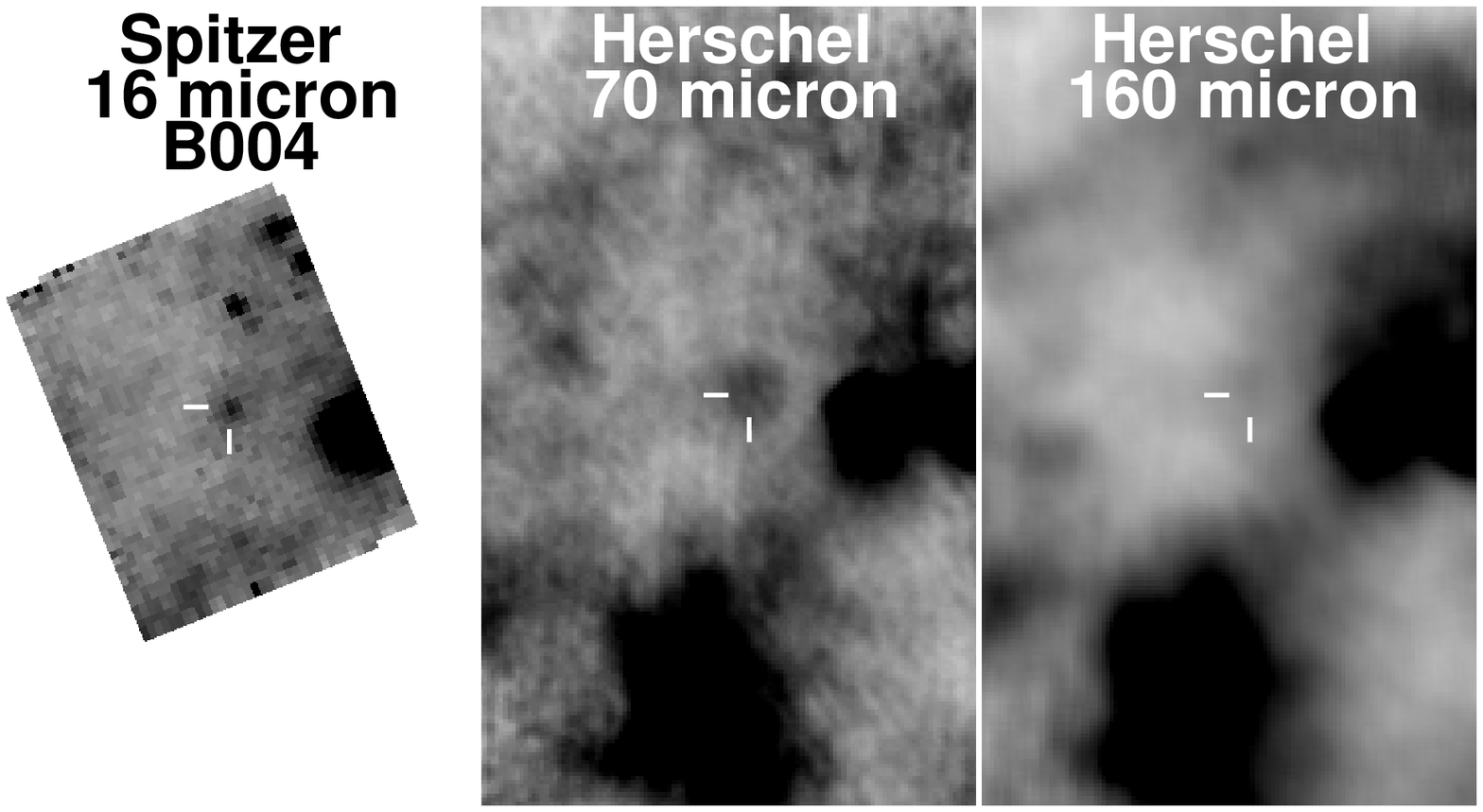}
   }
   \subfloat{
      \includegraphics [scale=0.45,angle=0]{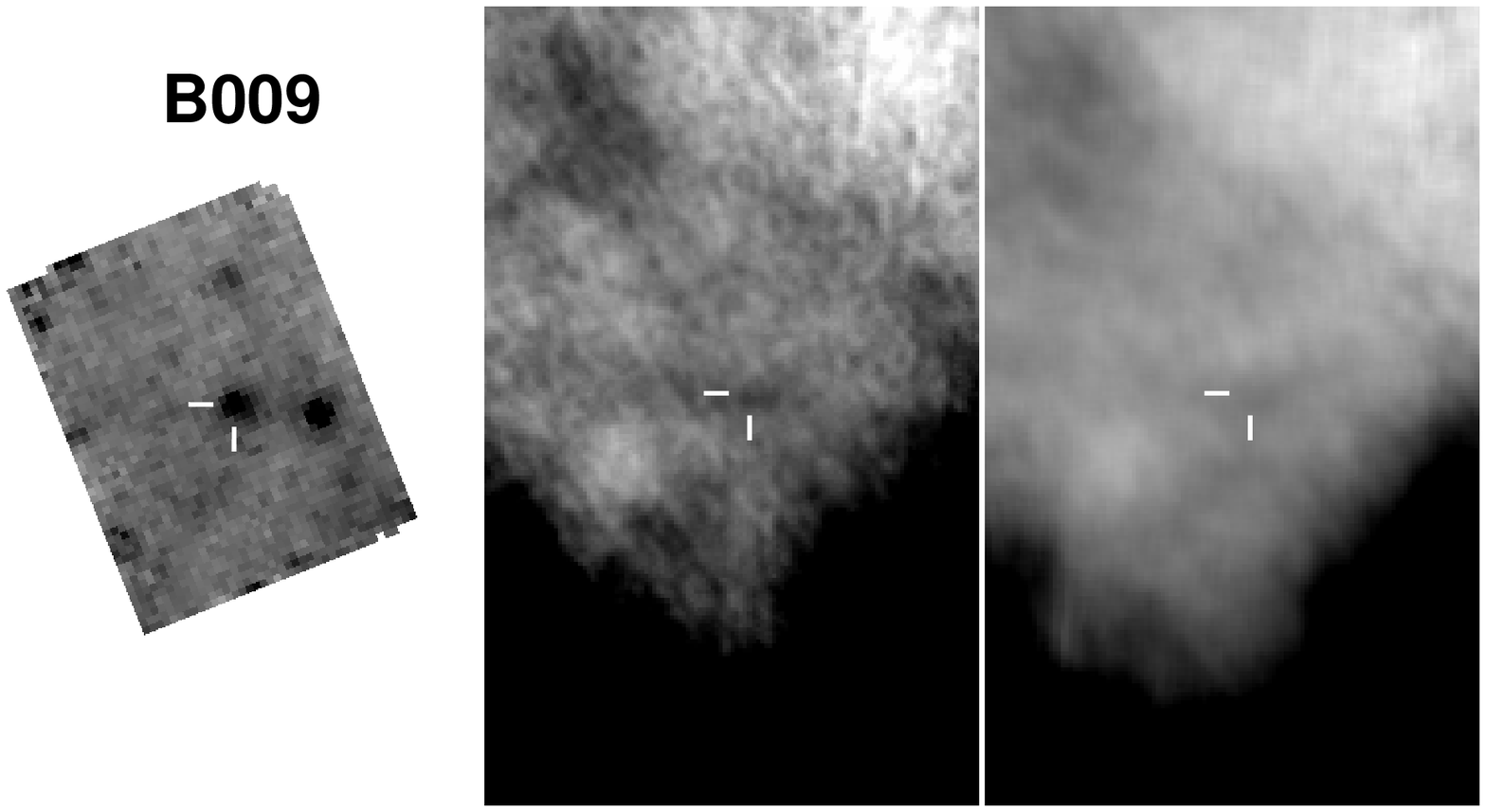}
   }\\
   \subfloat{
      \includegraphics [scale=0.45,angle=0]{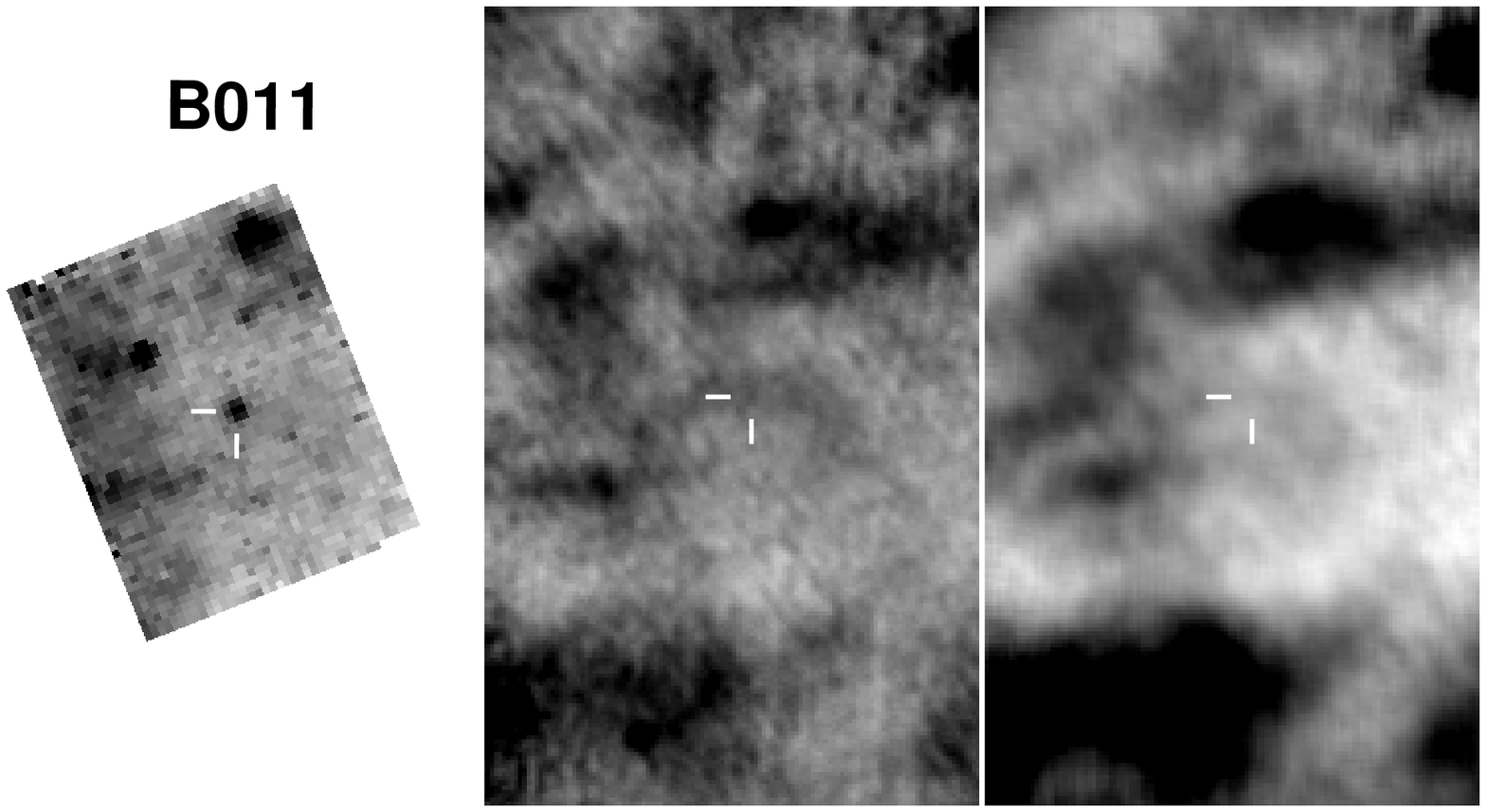}
   }
   \subfloat{
      \includegraphics [scale=0.45,angle=0]{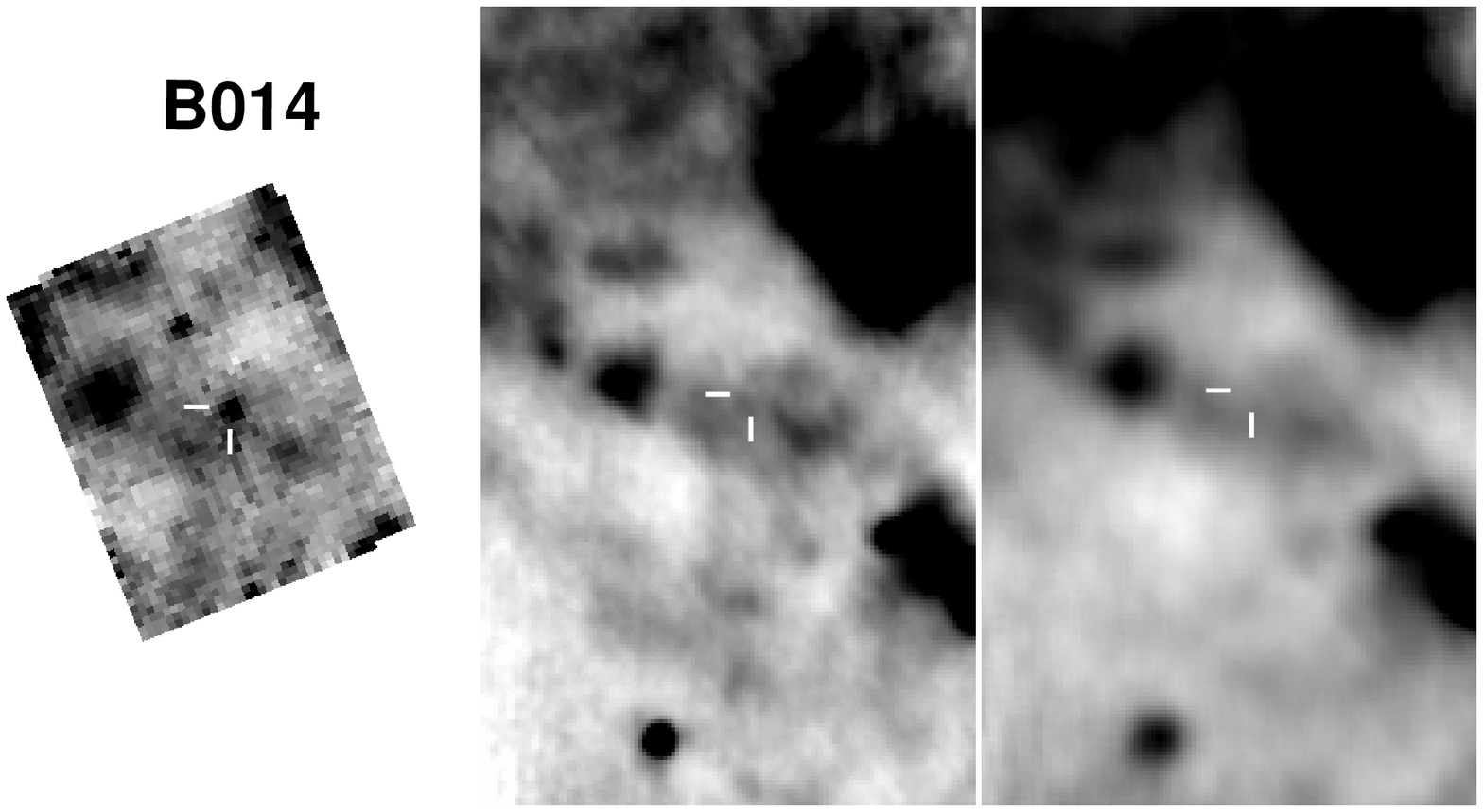}
   }\\
   \subfloat{
      \includegraphics [scale=0.45,angle=0]{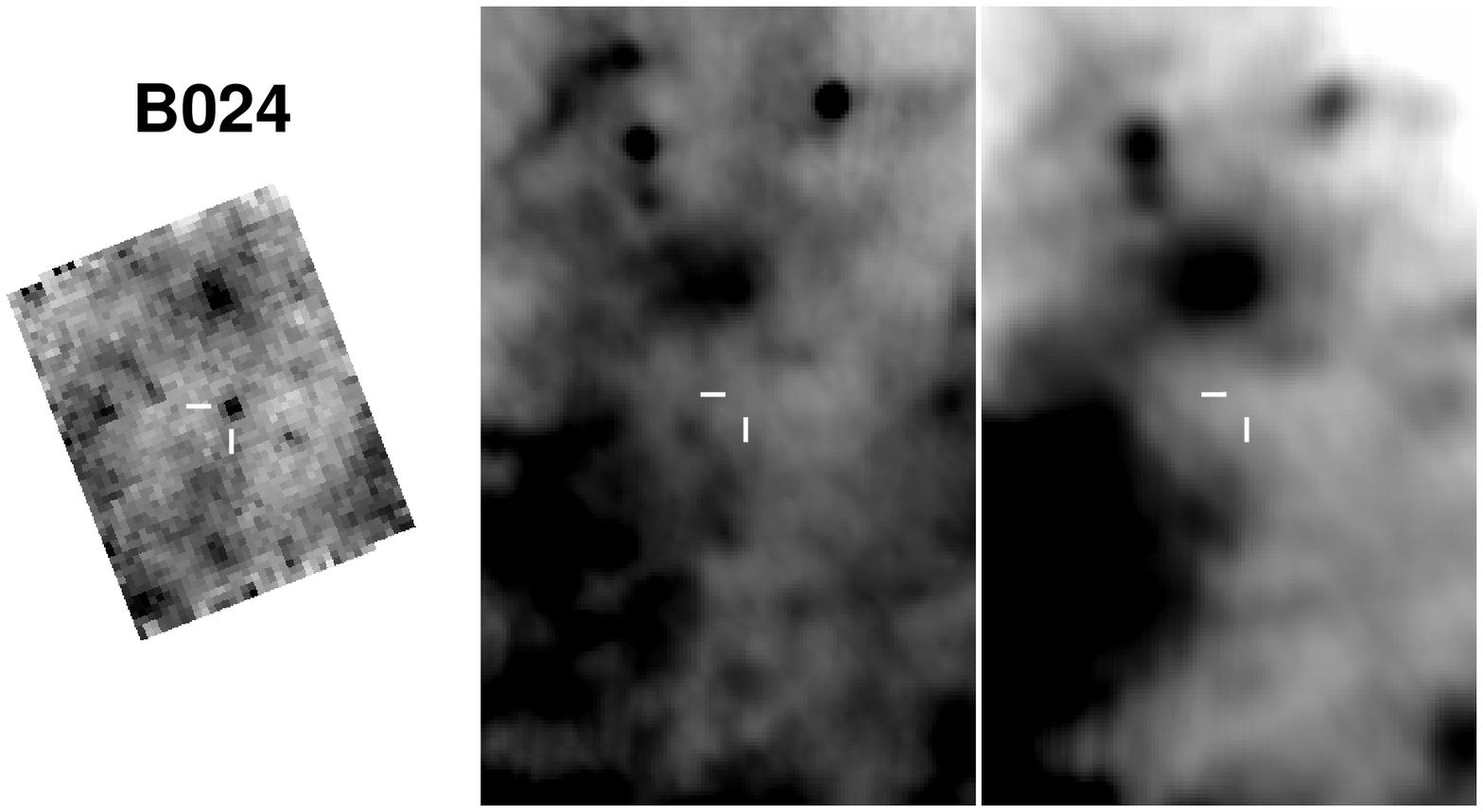}
   }
   \subfloat{
      \includegraphics [scale=0.45,angle=0]{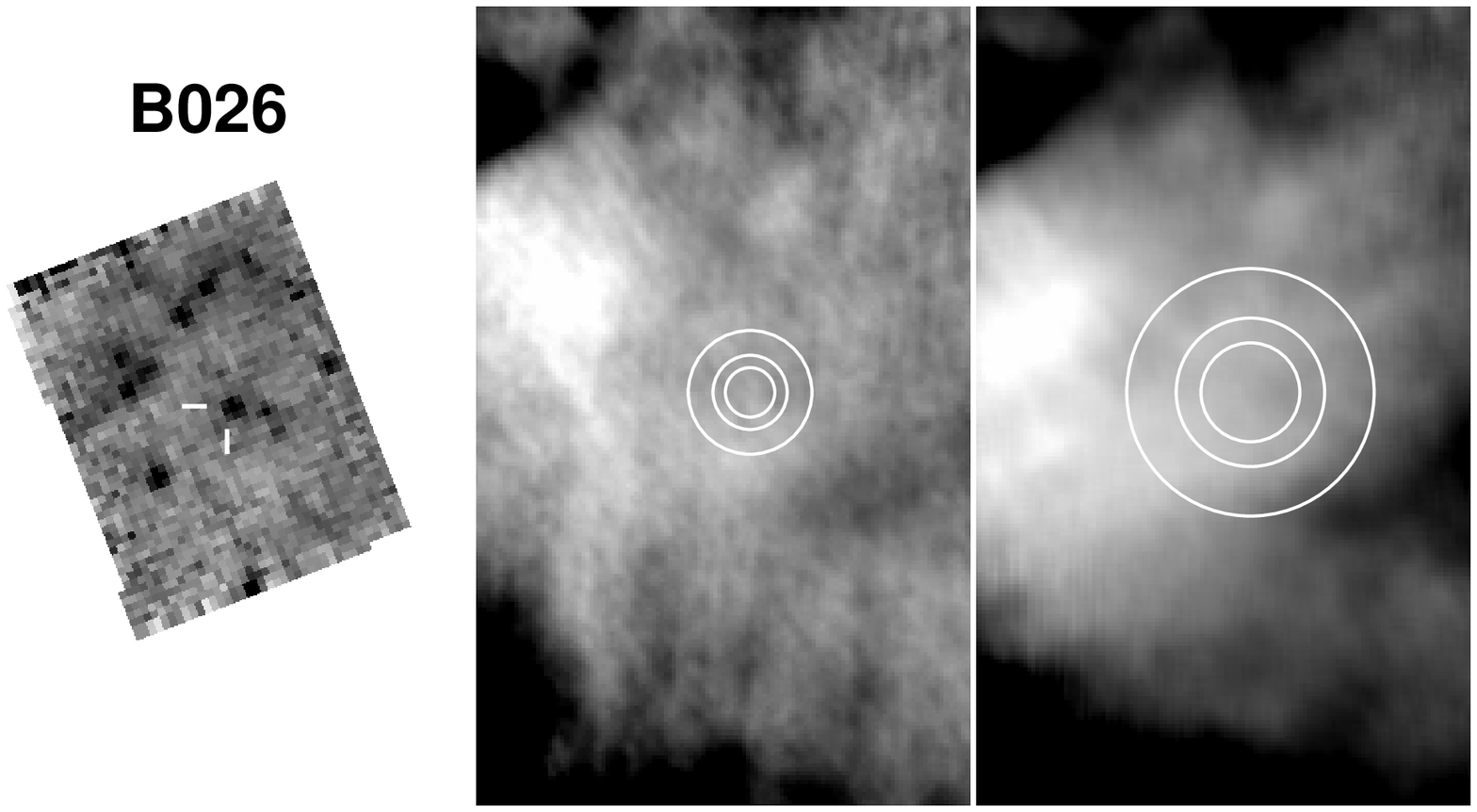}
   }
   \caption{The newly acquired images for the target SMC stars. The stellar 
positions are indicated by white tick marks, and the photometry apertures 
adopted for the \emph{Herschel} data are shown over the B026 images.
A gap in the apertures is used between the
target aperture and the outer, background annulus. 
\textit{Left} The IRS peak-up images at 16~\mum. The two white target lines are 
6\arcsec\ long. The PRF FWHM is roughly 3\arcsec. The 
image size is 1\farcm2$\times$1\farcm5. \textit{Middle} The \emph{Herschel} PACS 70~\mum\ image 
with a target aperture radius of 6\arcsec. 
\textit{Right} The \emph{Herschel} PACS 160~\mum\ image with a target aperture radius of 12\arcsec. 
The portion of the \emph{Herschel} images shown covers 2\farcm2$\times$3\farcm2. 
The only significant \emph{Herschel} detection is in the 70~\mum\ band for B004. 
There is possibly a source in B014 lost to the local background, and 
B009 has a promising, but low significance, 70~\mum\ detection. 
The only stars significantly
extended in the IRS peak-up 16~\mum\ images are B029, B087, and B188.
}
\label{fig:photim}
\end{figure*}

\begin{figure*}
   \centering
   \ContinuedFloat
   \subfloat{
      \includegraphics [scale=0.45,angle=0]{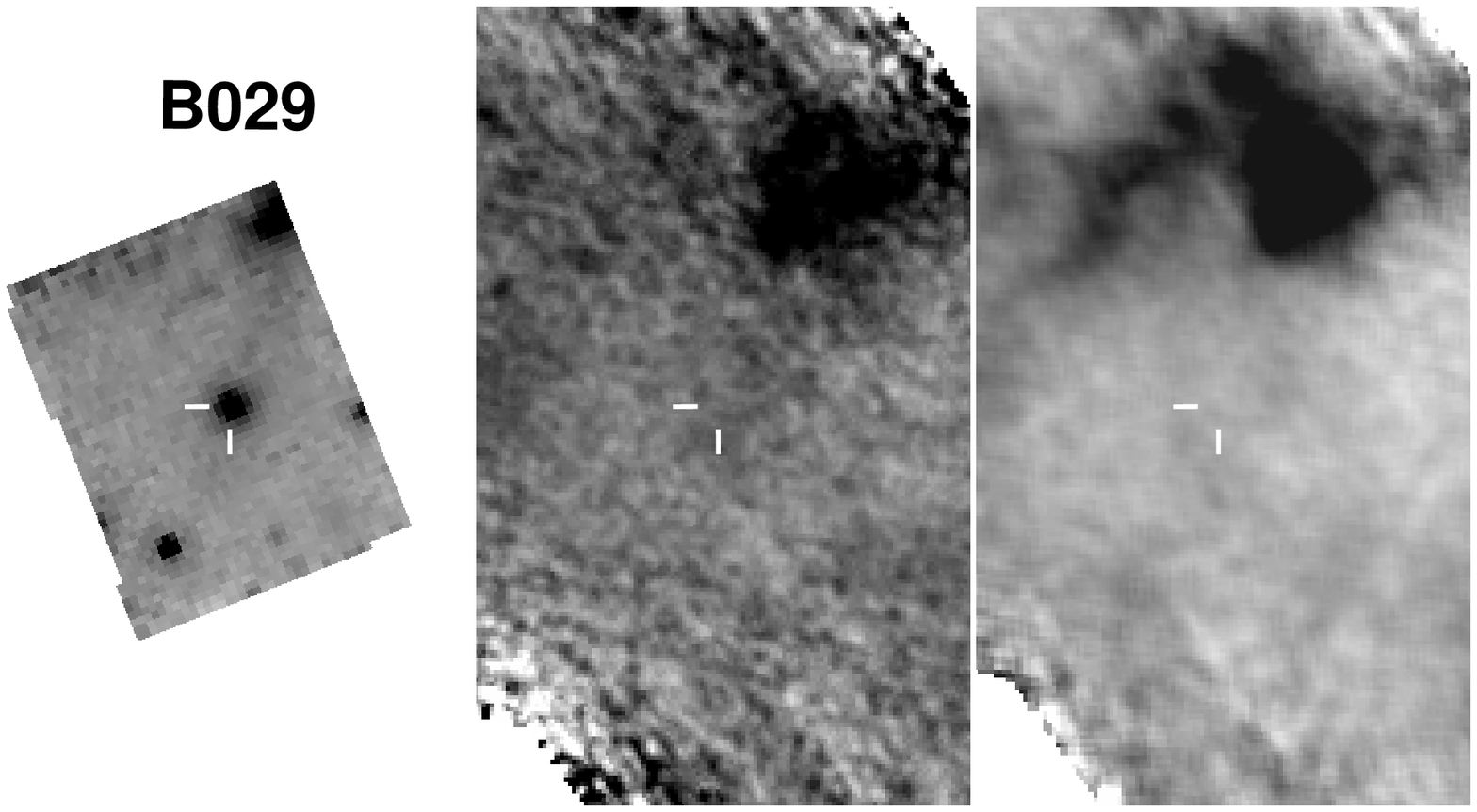}
   }
   \subfloat{
      \includegraphics [scale=0.45,angle=0]{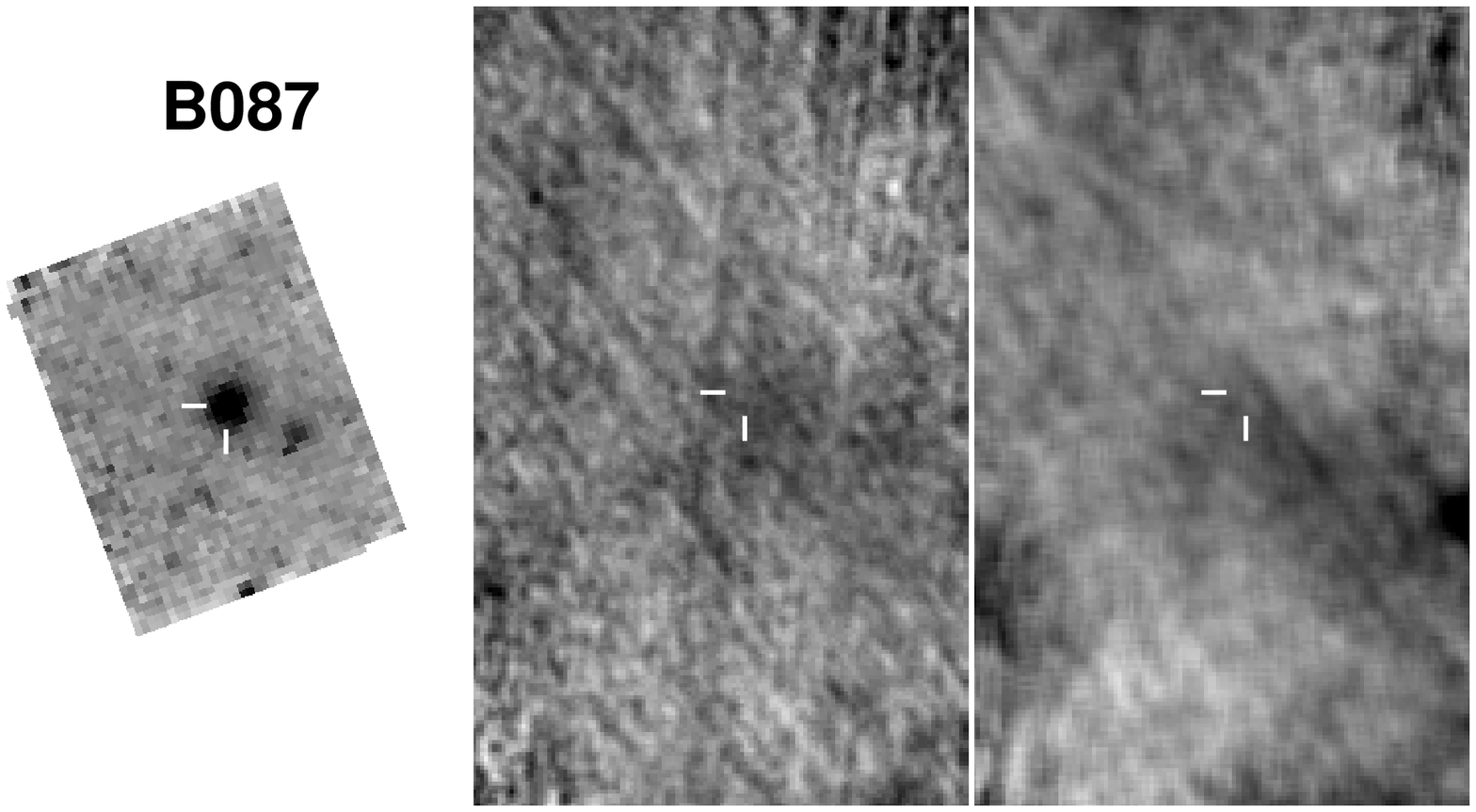}
   }\\
   \subfloat{
      \includegraphics [scale=0.45,angle=0]{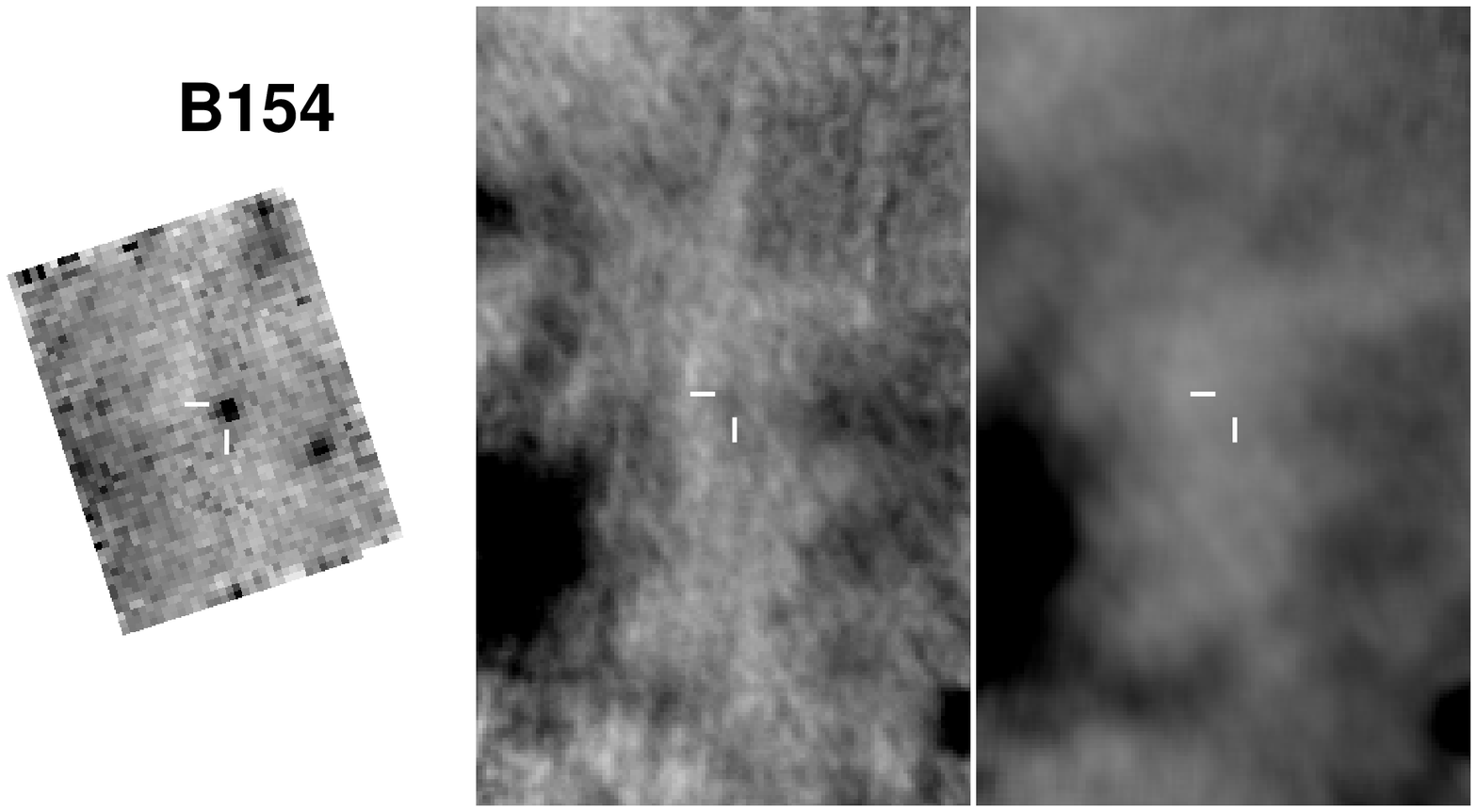}
   }
   \subfloat{
      \includegraphics [scale=0.45,angle=0]{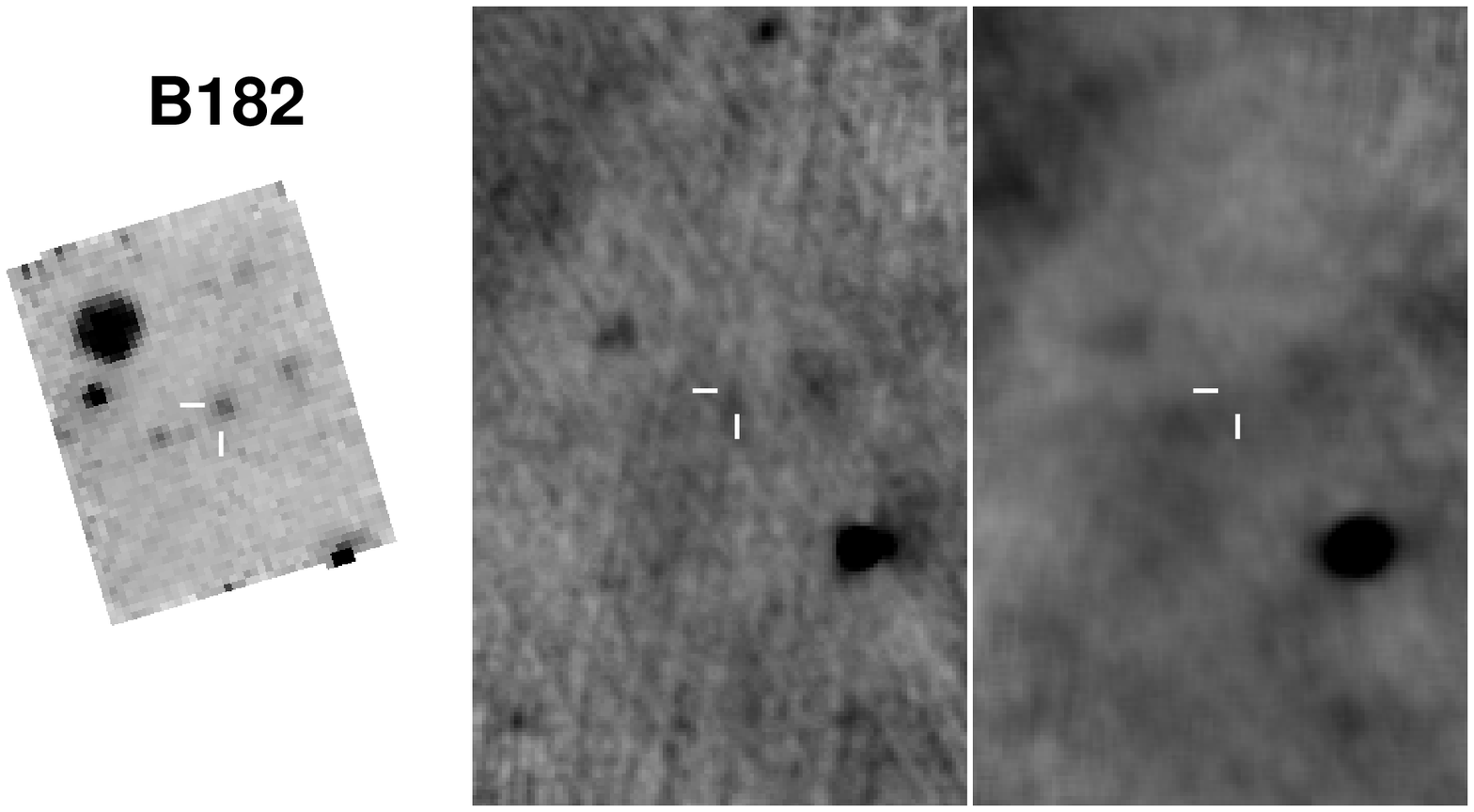}
   }\\
   \subfloat{
      \includegraphics [scale=0.45,angle=0]{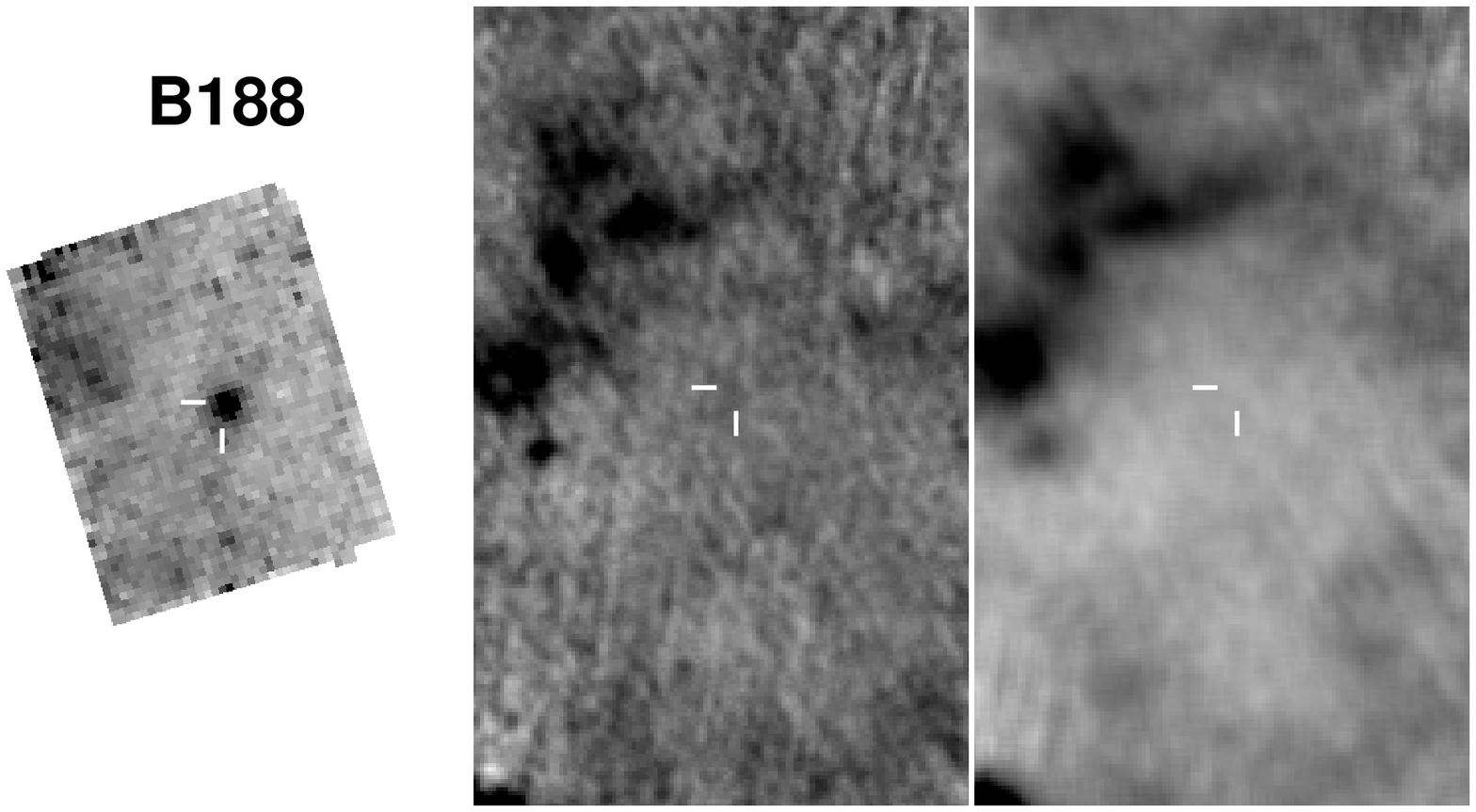}
   }
   \caption{(Continued)} 
\end{figure*}

\begin{figure}
\centering
\includegraphics [scale=0.45,angle=0]{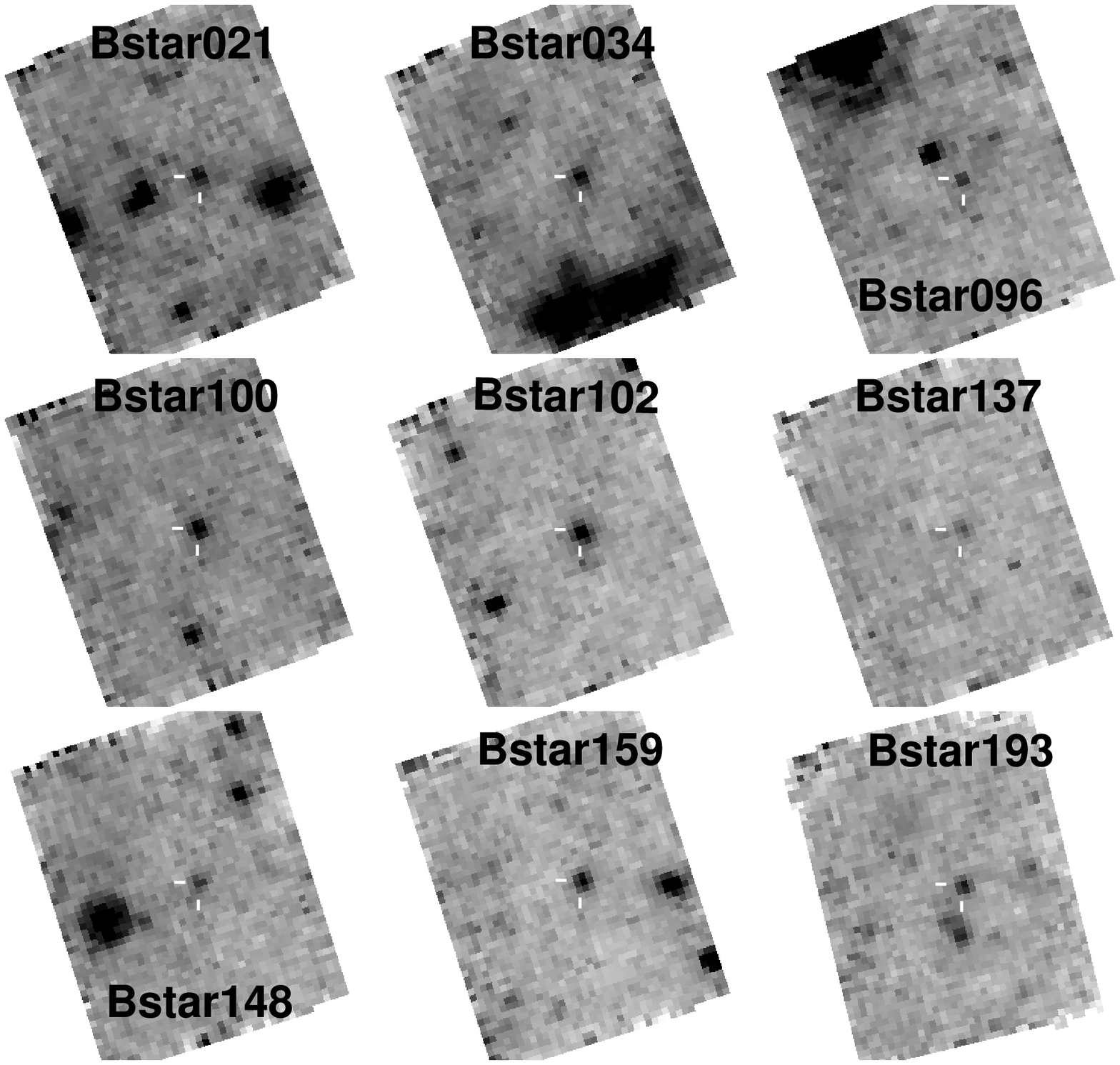}
\caption{The IRS peak-up images at 16~\mum\ are shown for the objects 
lacking \emph{Herschel} observations. The two white target lines have 
6\arcsec\ lengths, and the PRF FWHM is 3\arcsec. None of 
these stars are significantly ($>3\sigma$) more extended than the PRF. 
}
\label{fig:peak}
\end{figure}
 
\subsection{Spitzer Spectroscopy}
\label{sec:spsp}
A subset of 23 stars from the Paper I sample of main sequence OB stars 
were observed in low-resolution mode with the IRS
in guest observer (GO) program 50088. Two stars, B083 and B161, are 
not analyzed here as they are spatially resolved at the IRS resolution. 
B161 also has a complicated background from a bright, nearby source. 
One star, B112, had insufficient
signal-to-noise (S/N), so we present the 20 others. The IRS observed all sources in
staring mode using just the Long-Low (LL) module, which produces
spectra with wavelength coverage from 14 to 35~\mum.  The exposure
time was tailored to achieve S/N $\sim$ 5 based on the measured 8 and
24~\mum\ fluxes of the sources.

We generated the spectra using optimal extraction \citep{Lebou10}, which
reduces the impact of noise in pixels not exposed to the core of the
point-spread function from the source. The starting point was the
S18.5 version of the pipeline from the Spitzer Science Center (SSC).
We observed all sources in both the first- and second-order LL apertures.
For most targets, the image with the source in the other nod position
in the same aperture served as the background, but in some cases,
complex backgrounds forced us to use images with the source in the
other aperture as the background. Images were cleaned of known
rogue pixels and pixels flagged by the pipeline as bad using software
similar to the IRSclean code available from the SSC. 
Spectra were extracted from coadded images in each nod position. 
Where necessary, a polynomial was fit to the background in the spatial
direction, at each wavelength, to better isolate the spectral
emission from the source. When spectra from the
separate nods were combined, sharp features (such as spikes and divots)
in one spectrum but not present in the other were ignored.  The spectra
combined from the two LL orders were calibrated using similarly
observed and processed spectra of the standard stars HR 6348 (K0 III)
and HD 173511 (K5 III). We tested tapered-aperture algorithms in place 
of the adopted optimal-extraction one and found no significant differences. 

The spectra were taken with the goal of better determining spectral slopes for temperature 
measurements and not to obtain a 
S/N sufficient to reveal any detailed spectral structure related to the chemical composition of the 
dust \citep[e.g.][]{Chen06}. \cite{Dahm09} find that B and A stars with debris disks  
show featureless continua at $\lambda\ge8~\mum$ unlike later-type stars with 
prominent silicate absorption at 10~\mum\ and 20~\mum, 
making it unlikely that these SMC stars would possess strong spectral features even with deeper data.  
IRS spectra for cirrus hot spots are not available in large numbers, but we can 
guide our expectations from the spectra of reflection nebulae \citep{Werne04b,Sellg07}. 
\cite{Boers10} attempt to identify the PAH emission features in 
reflection nebulae and other sources 
from 15--20~\mum\ and find that the strongest features are at 16.4, 17.4, and 
18.9~\mum. The features show large variation of strength with position. 
There is not a conclusive chemical identification of the 
carriers for these features, and they are much weaker than the better 
known PAH features from 5--15~\mum. They are weak enough that we should 
not detect them in our individual IRS spectra, but some of the 
stronger regions of emission might barely be detectable in a stack. We searched our data 
with fits to various templates for (PAH) features, emission lines, and specific grain spectra 
with the PAHFIT software \citep{Smith07} and found no 
significant features in any spectra, nor in an optimally weighted stack of all the spectra. 
Table \ref{tab:IRS} presents a brief observing log, and the LL spectra appear as insets in Figure 
\ref{fig:sed1}. 

\begin{deluxetable}{crrrr}
\tabletypesize{\scriptsize}
\tablecaption{\emph{Spitzer} IRS Data Log\label{tab:IRS}}
\tablewidth{0pt}
\tablehead{
\colhead{ID} & \colhead{Start} & \colhead{Exposure} & \colhead{f$_{\rm \nu}$\tablenotemark{*}} & \colhead{S/N\tablenotemark{*}}\\
\colhead{} & \colhead{Date (UTC)} & \colhead{Time (s)} & \colhead{(mJy)} & \colhead{pixel$^{-1}$} }
\startdata
B004 & 2008-12-07 & 609.5 & 2.44 & 5.0 \\
B009 & 2008-12-08 & 243.8 & 3.63 & 4.2 \\
B011 & 2008-12-08 & 365.7 & 2.15 & 4.2 \\
B014 & 2008-12-08 & 487.6 & 4.10 & 6.0 \\
B021 & 2008-12-07 & 1462.8 & 1.16 & 2.9 \\
B024 & 2008-12-07 & 1219.0 & 1.12 & 2.9 \\
B026 & 2008-12-07 & 1097.1 & 1.09 & 3.4 \\
B029 & 2008-12-08 & 121.9 & 7.09 & 6.6 \\
B034 & 2008-12-07 & 731.4 & 2.87 & 2.4 \\
B087 & 2008-12-08 & 243.8 & 4.25 & 5.8 \\
B096 & 2008-12-03 & 2438.0 & 1.04 & 2.3 \\
B100 & 2008-12-07 & 853.3 & 1.59 & 4.1 \\
B102 & 2008-12-07 & 487.6 & 2.18 & 4.8 \\
B137 & 2008-12-02 & 2560.0 & 0.87 & 3.0 \\
B148 & 2008-12-06 & 1584.7 & 1.00 & 2.9 \\
B154 & 2008-12-07 & 609.5 & 1.27 & 1.4 \\
B159 & 2008-12-07 & 731.4 & 1.58 & 5.3 \\
B182 & 2008-12-06 & 1097.1 & 1.67 & 3.3 \\
B188 & 2008-12-08 & 243.8 & 3.75 & 4.3 \\
B193 & 2008-12-06 & 1584.7 & 2.23 & 3.8

\enddata
\tablenotetext{*}{Values are averages over 20~\mum--30~\mum.}
\end{deluxetable}

\begin{figure*}
\centering
   \subfloat{
      \includegraphics [scale=0.4,angle=0]{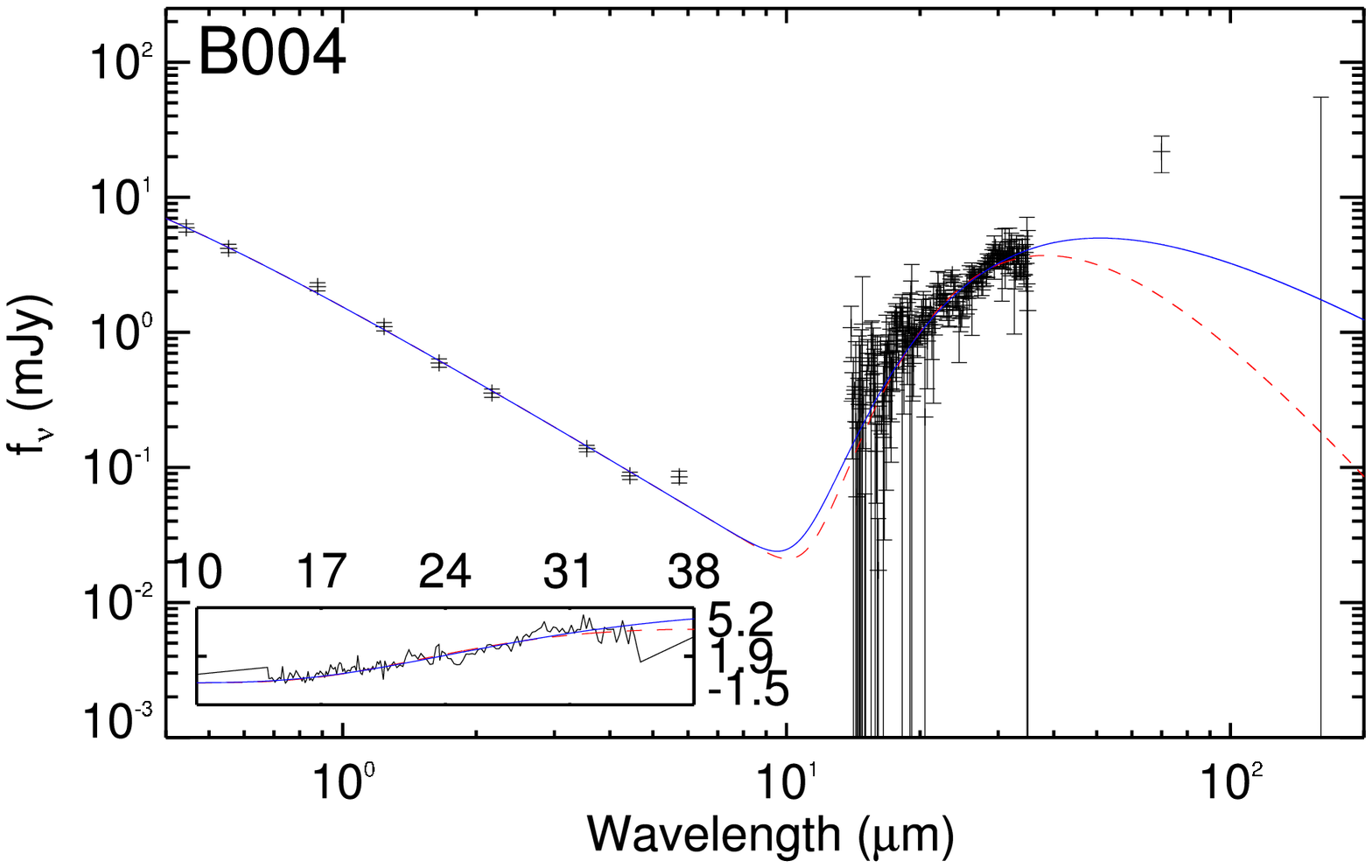}
   }
   \subfloat{
      \includegraphics [scale=0.4,angle=0]{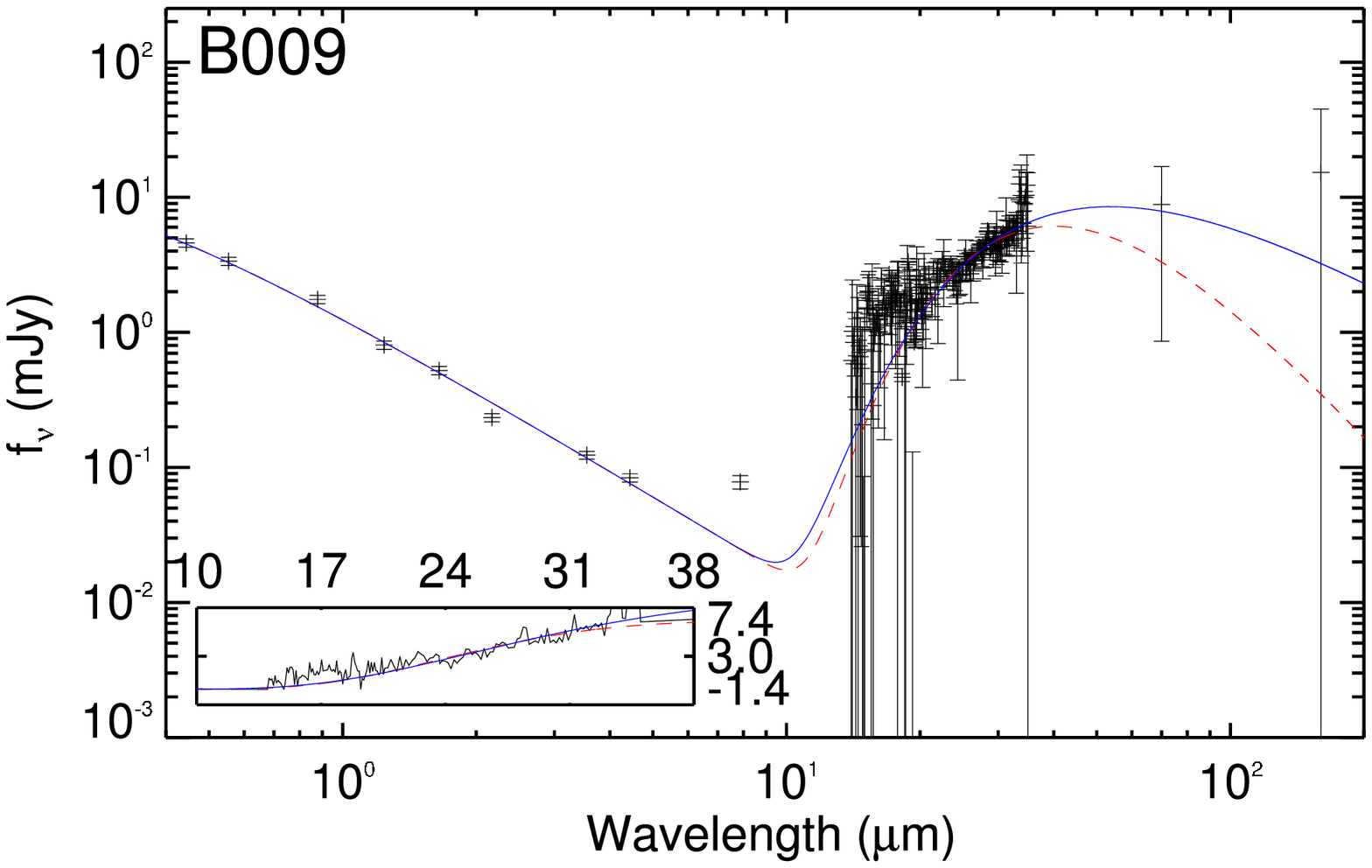}
   }\\
   \subfloat{
      \includegraphics [scale=0.4,angle=0]{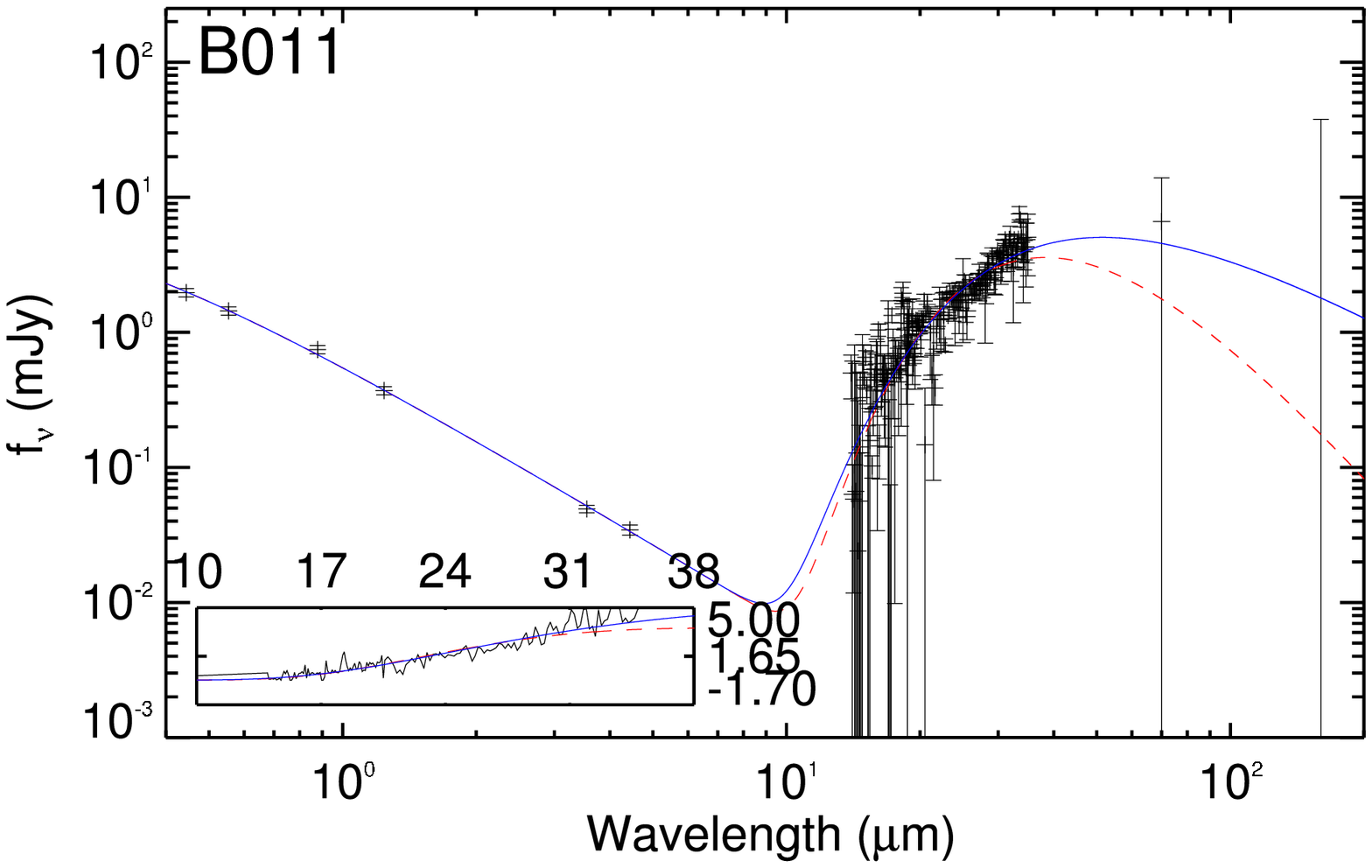}
   }
   \subfloat{
      \includegraphics [scale=0.4,angle=0]{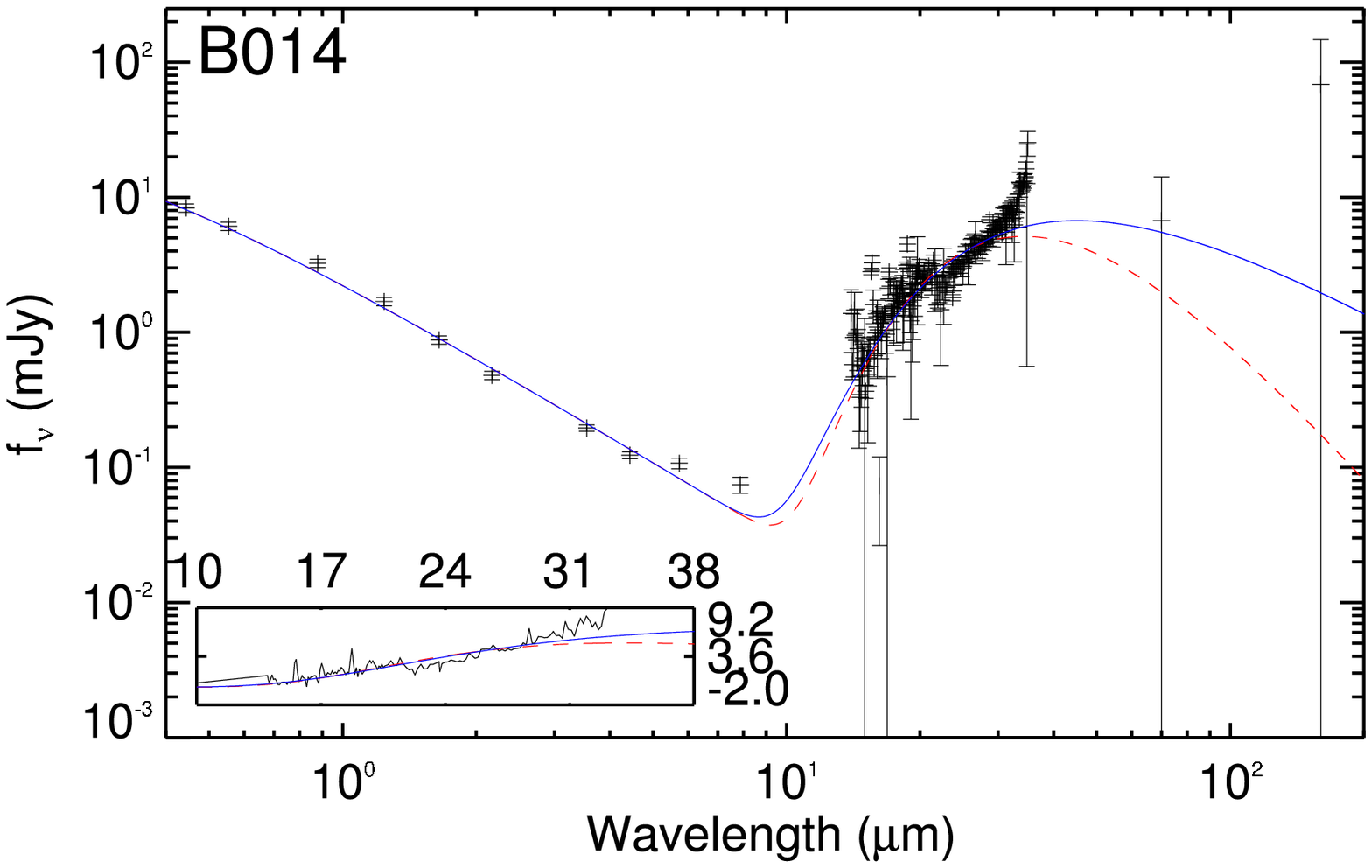}
   }\\
   \subfloat{
      \includegraphics [scale=0.4,angle=0]{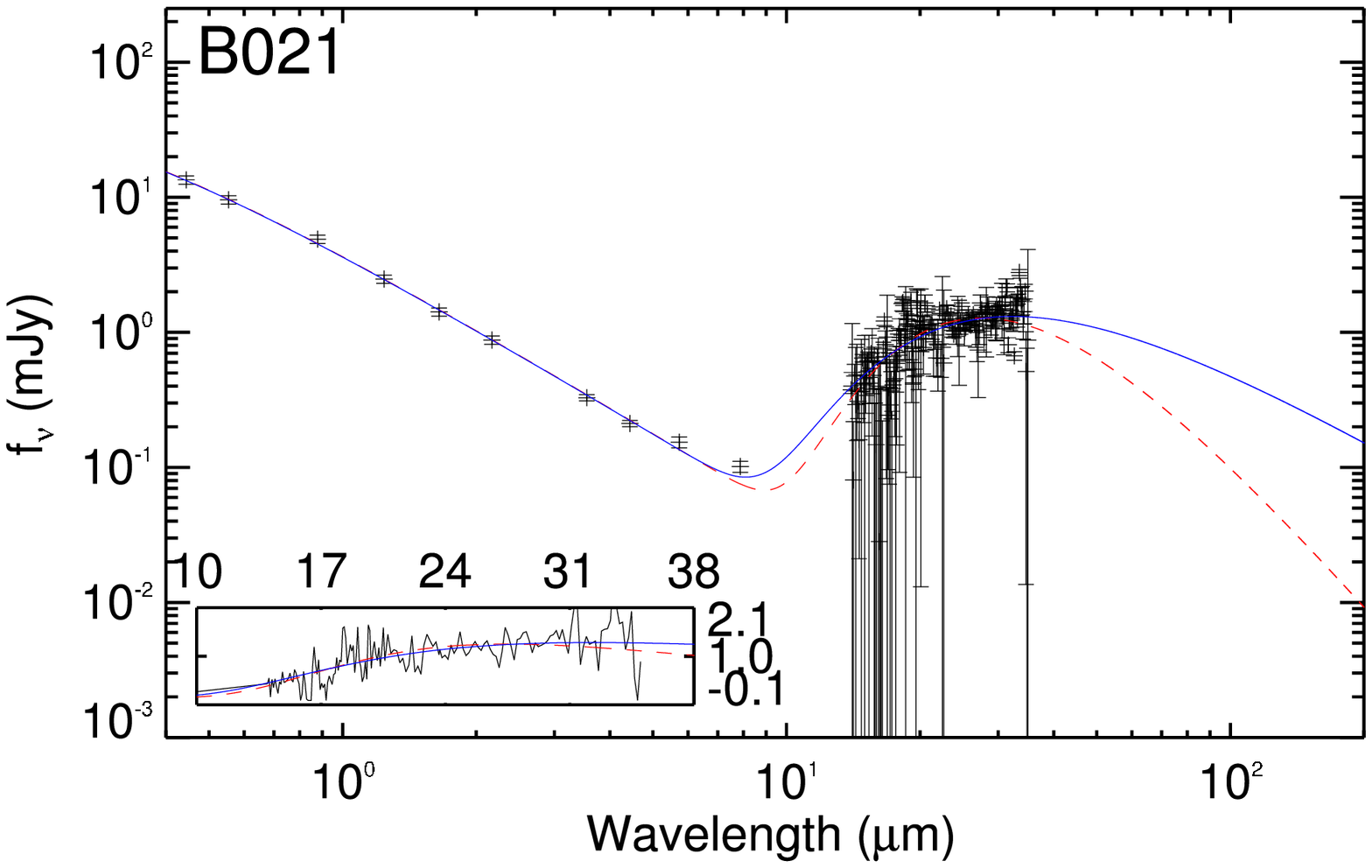}
   }
   \subfloat{
      \includegraphics [scale=0.4,angle=0]{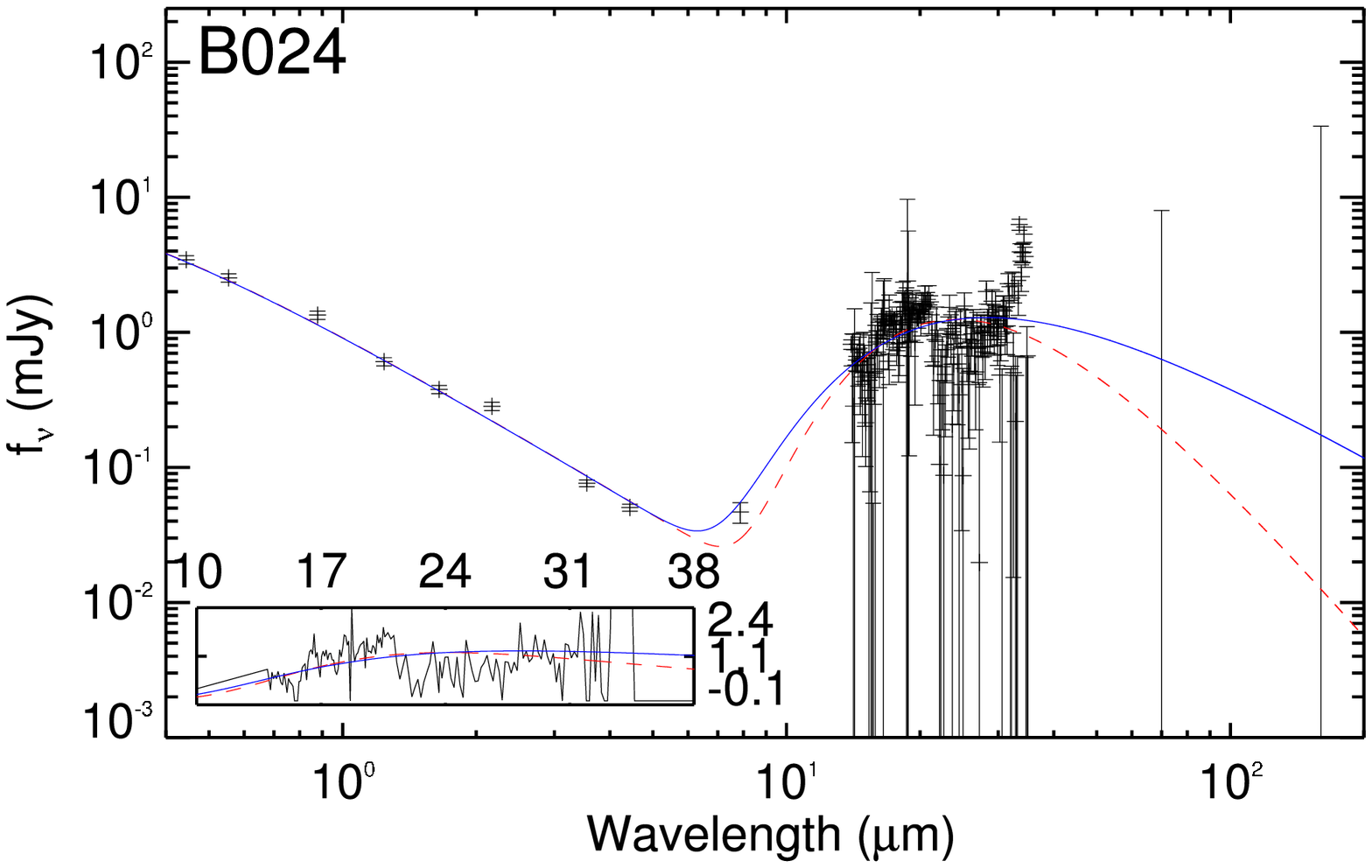}
   }\\
   \subfloat{
      \includegraphics [scale=0.4,angle=0]{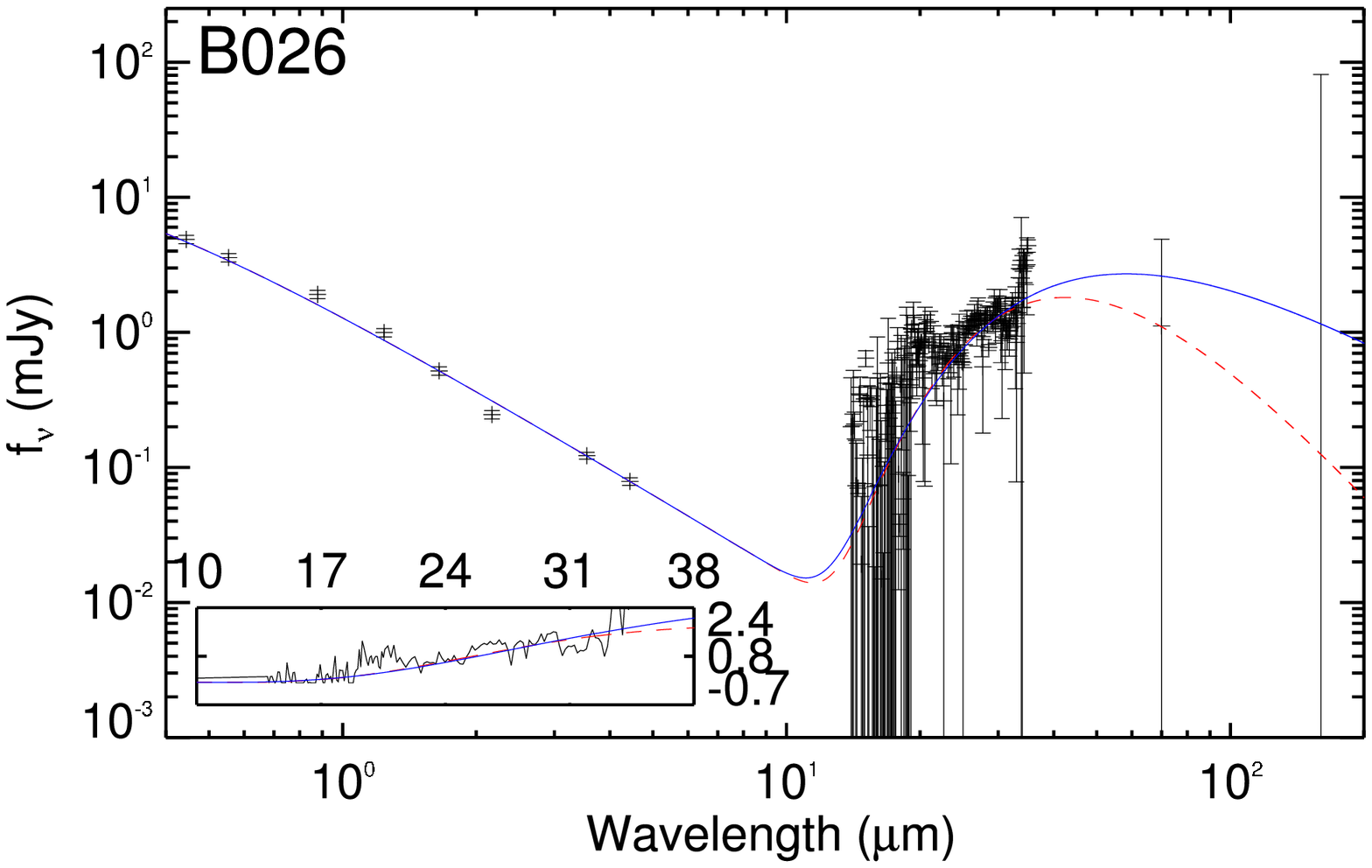}
   }
   \subfloat{
      \includegraphics [scale=0.4,angle=0]{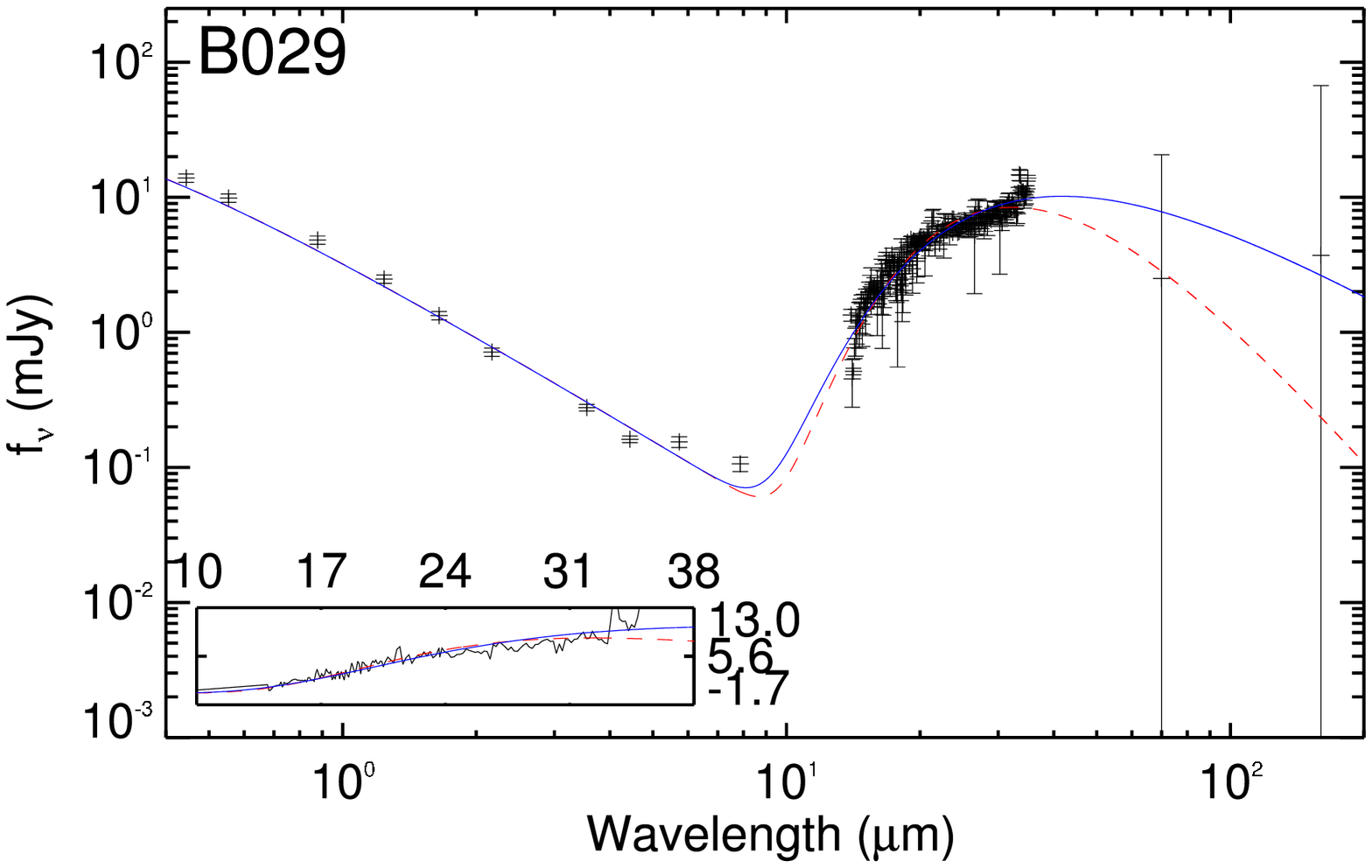}
   }
   \caption{The spectral energy distributions and best-fit dust models 
for the sources observed with IRS using a 
single temperature for the dust and a hot component for the 
stellar atmosphere. 
The insets show the IRS data on a linear scale. 
The solid blue curve is the fit with a blackbody function, and 
the dashed red curve is the fit with a modified blackbody function ($\beta_{\rm em}=2$). The 
data at 5.8 and 8.0~\mum\ are often underestimated by the models 
possibly because we have neglected PAH emission or possibly 
because there is a weaker, warmer dust component. 
Notice that except for 70~\mum\ in B004, all the \emph{Herschel} data are 
upper limits and sufficiently fit by all the models. The B004 70~\mum\ 
data is underfit at 3$\sigma$ significance.}
\label{fig:sed1}
\end{figure*}
\begin{figure*}
   \centering
   \ContinuedFloat
   \subfloat{
      \includegraphics [scale=0.4,angle=0]{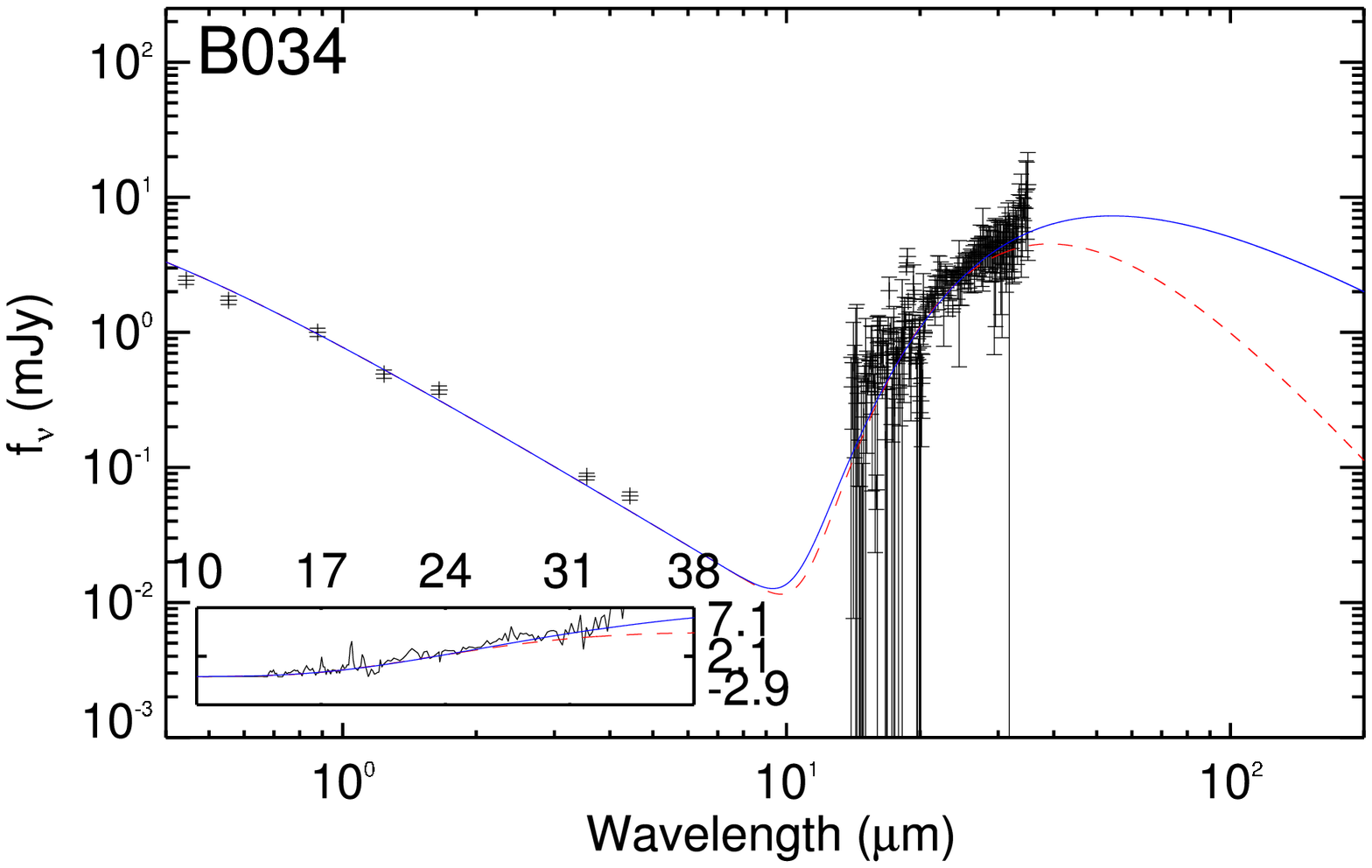}
   }
   \subfloat{
      \includegraphics [scale=0.4,angle=0]{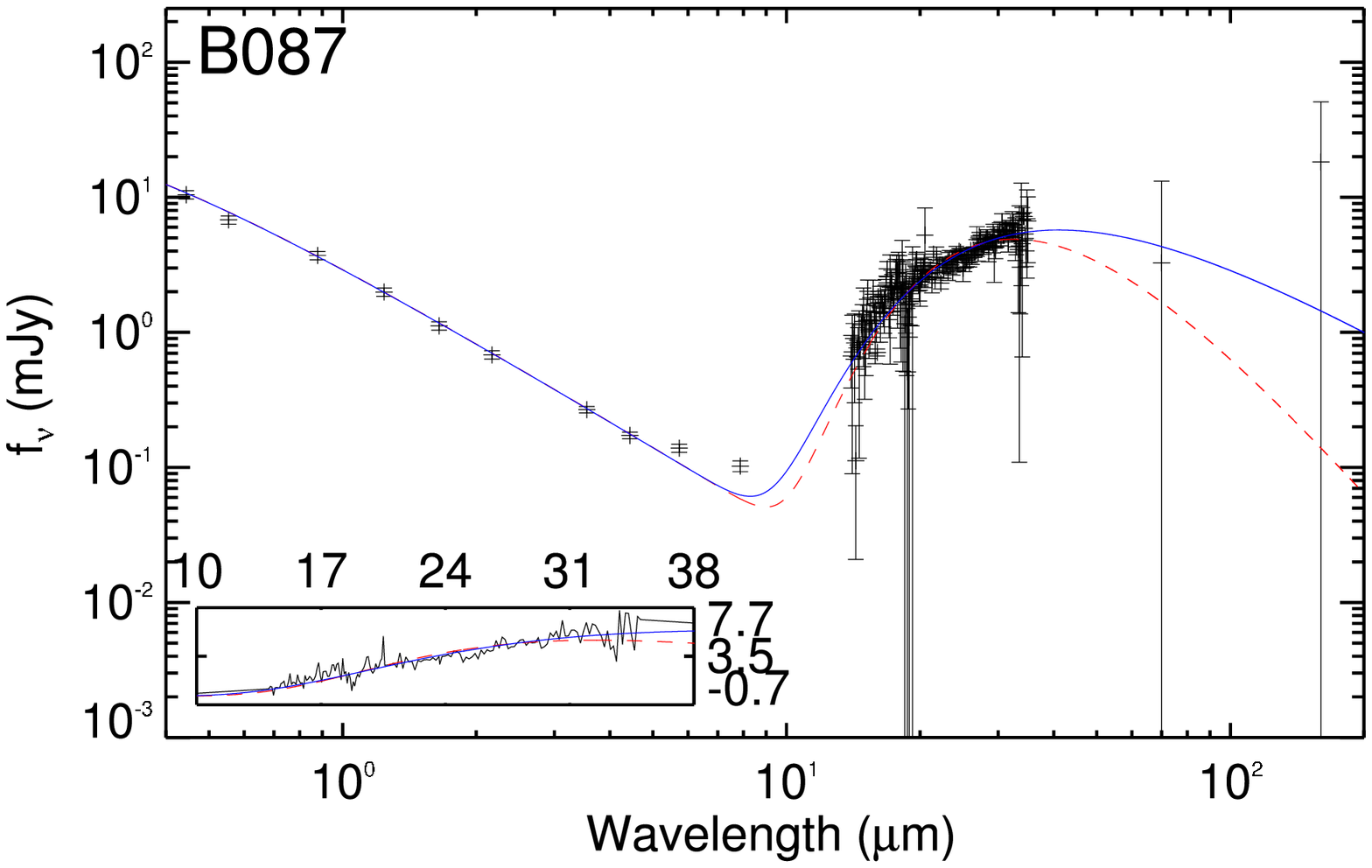}
   }\\
   \subfloat{
      \includegraphics [scale=0.4,angle=0]{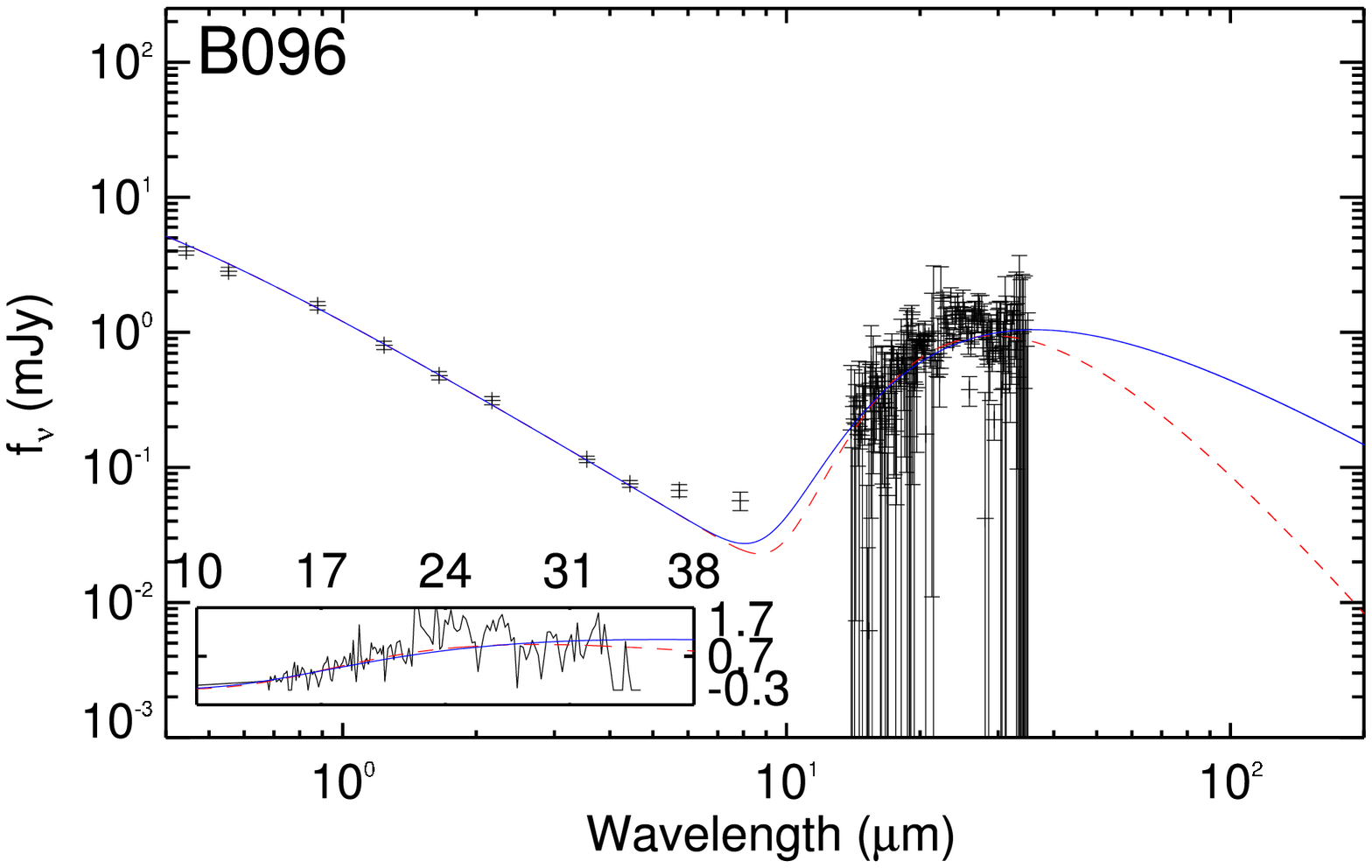}
   }
   \subfloat{
      \includegraphics [scale=0.4,angle=0]{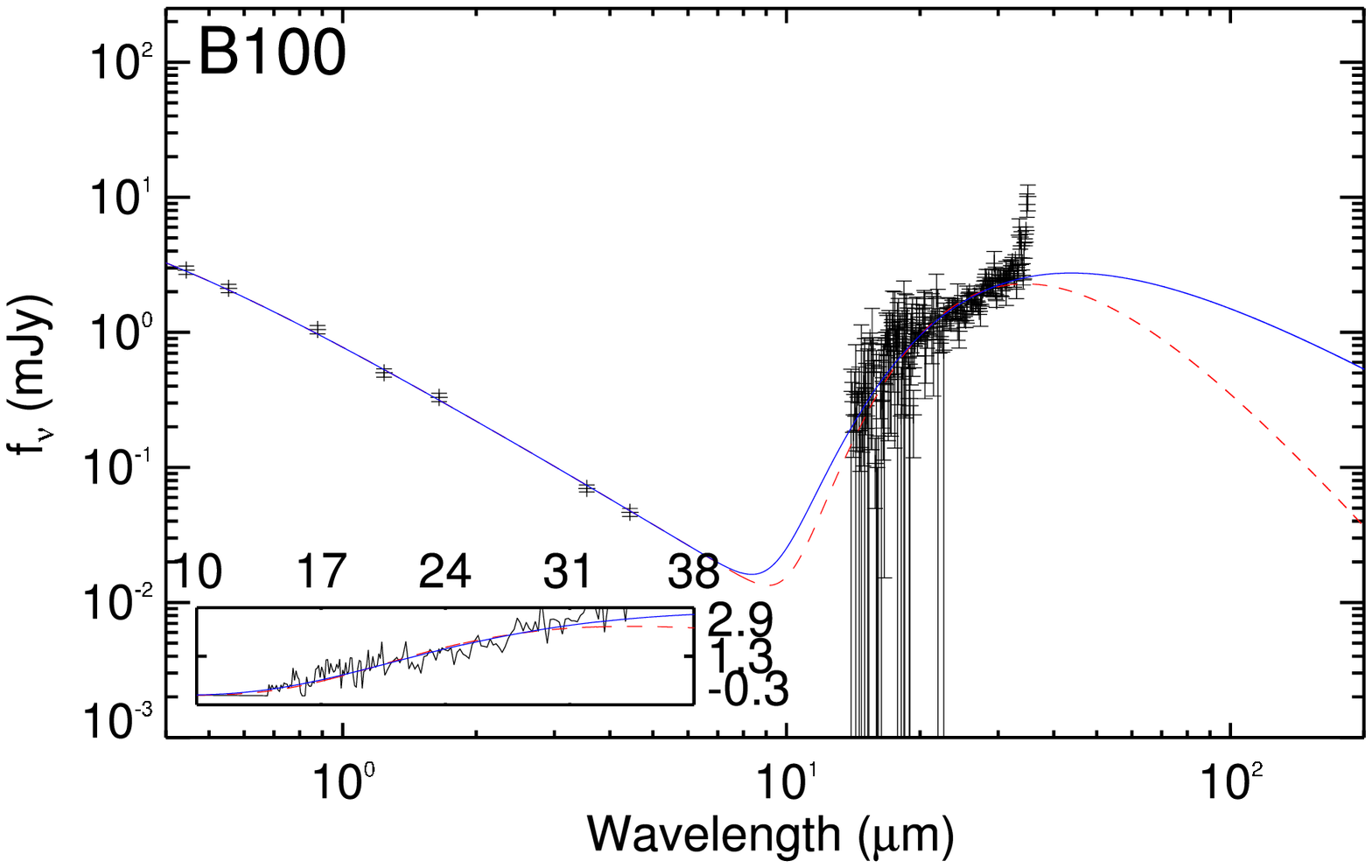}
   }\\
   \subfloat{
      \includegraphics [scale=0.4,angle=0]{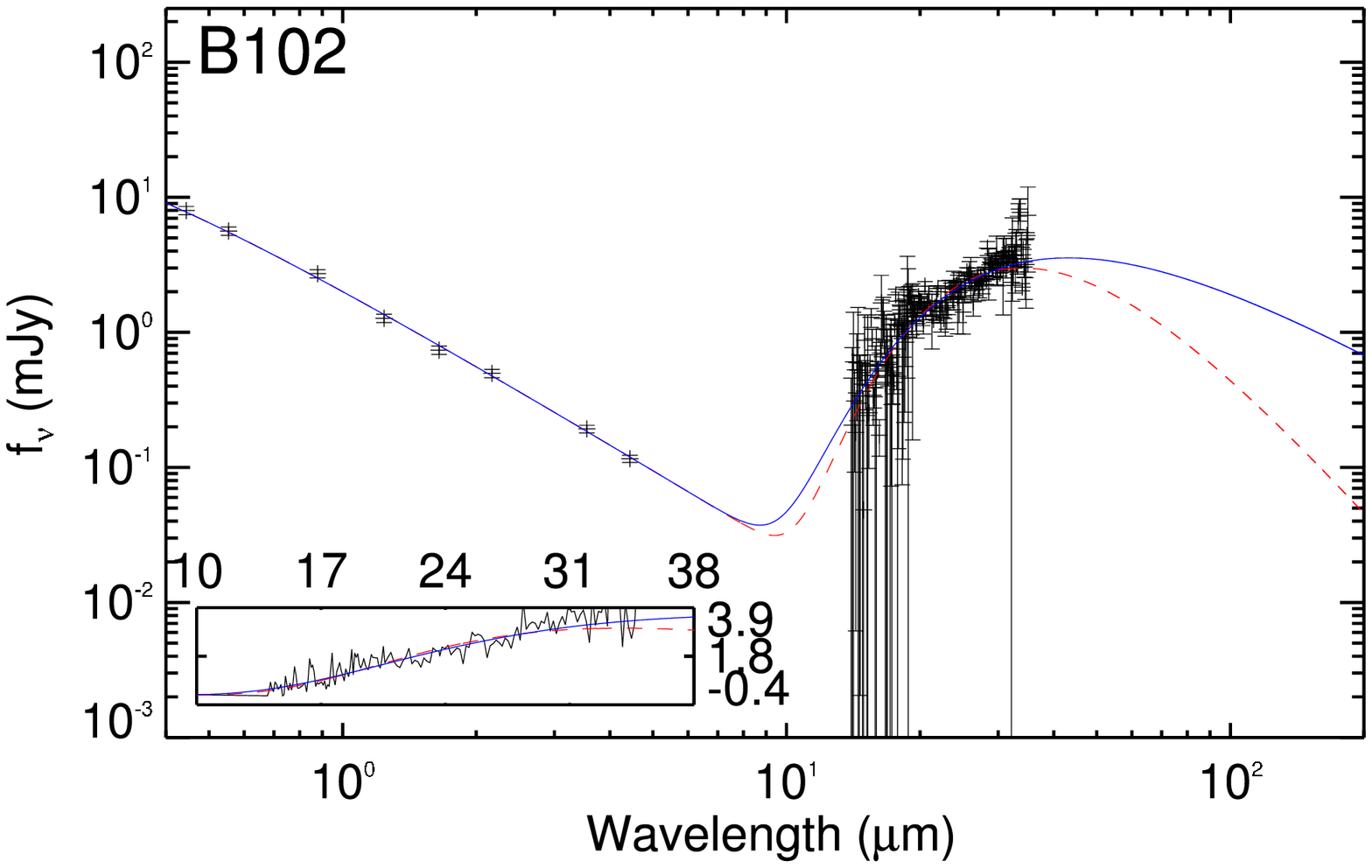}
   }
   \subfloat{
      \includegraphics [scale=0.4,angle=0]{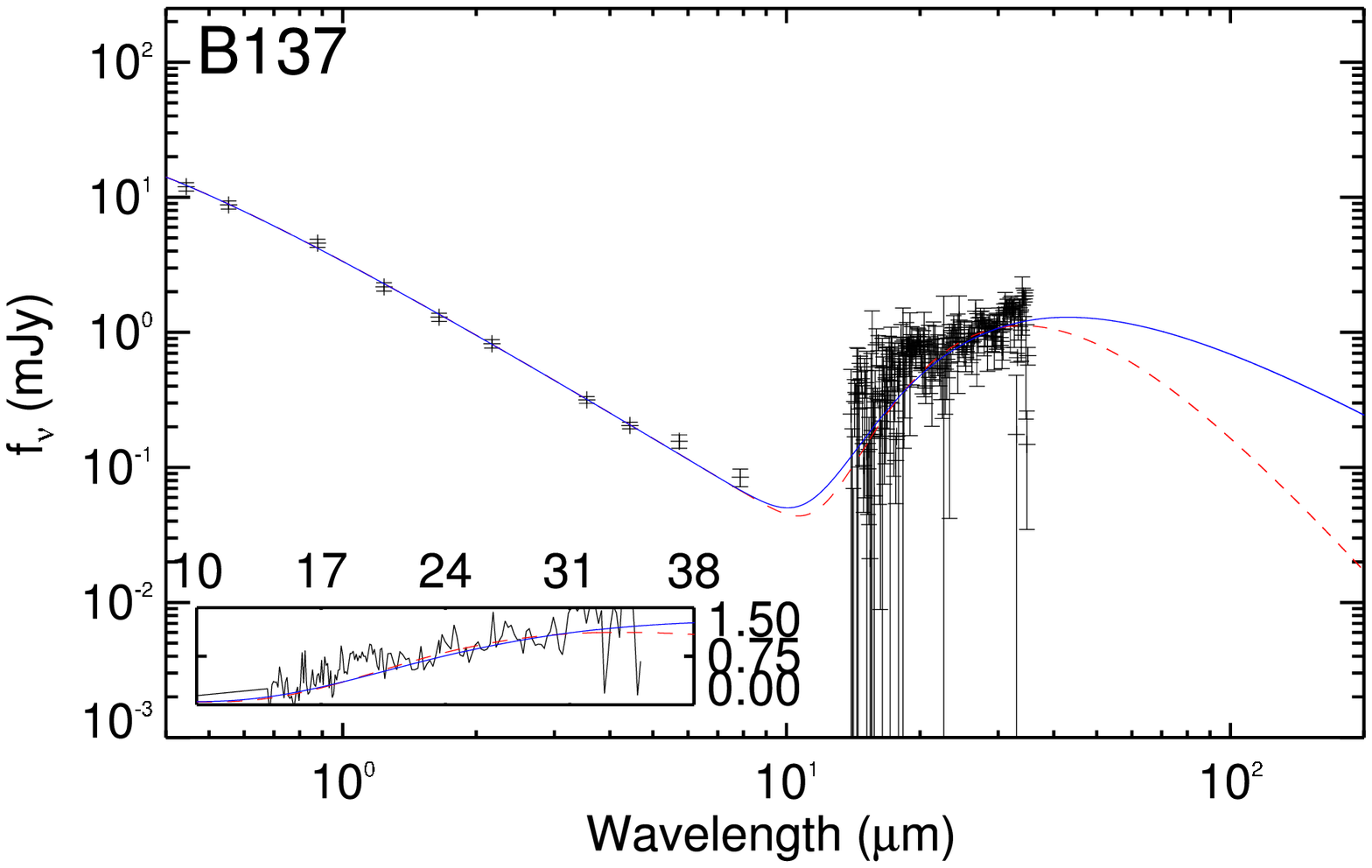}
   }\\
   \subfloat{
      \includegraphics [scale=0.4,angle=0]{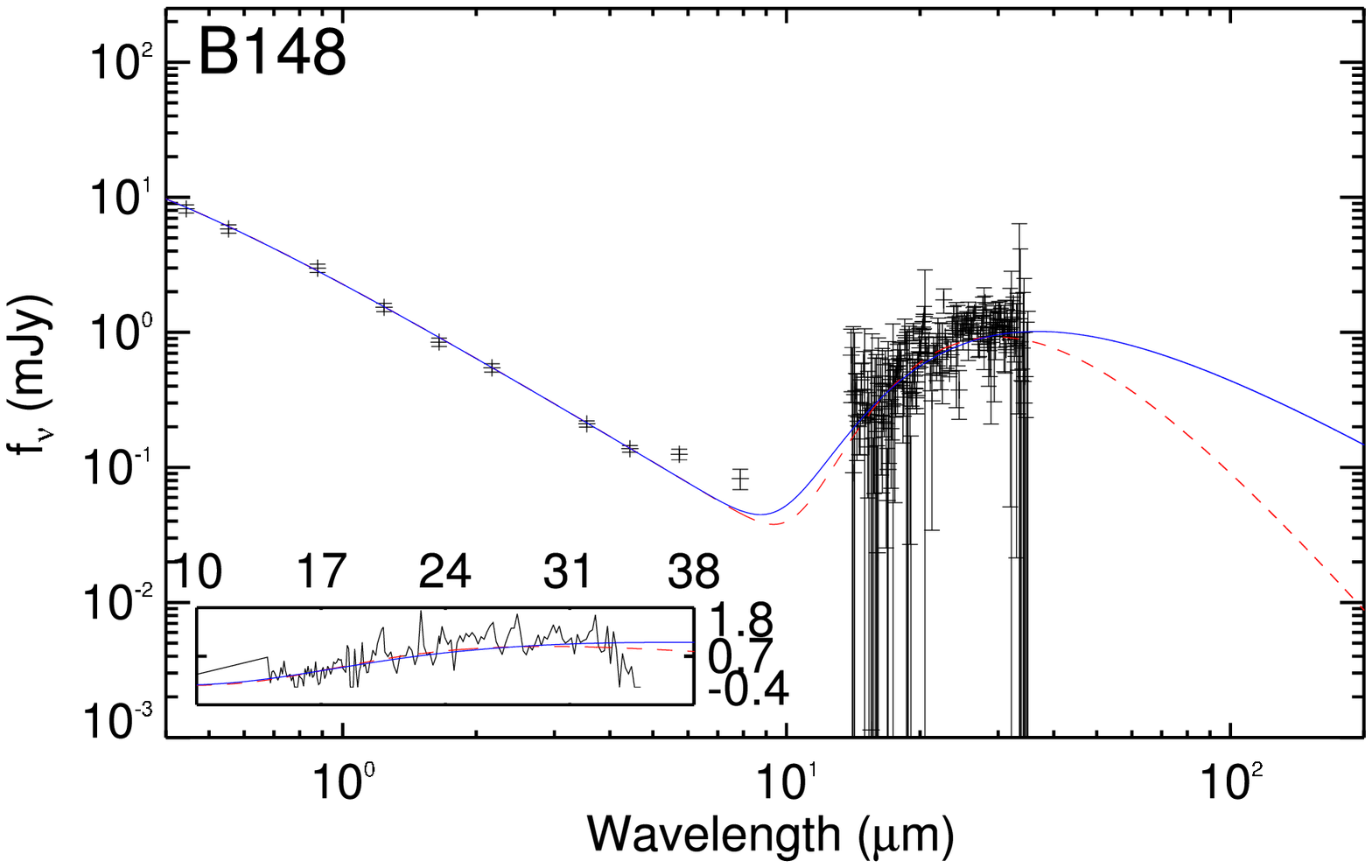}
   }
   \subfloat{
      \includegraphics [scale=0.4,angle=0]{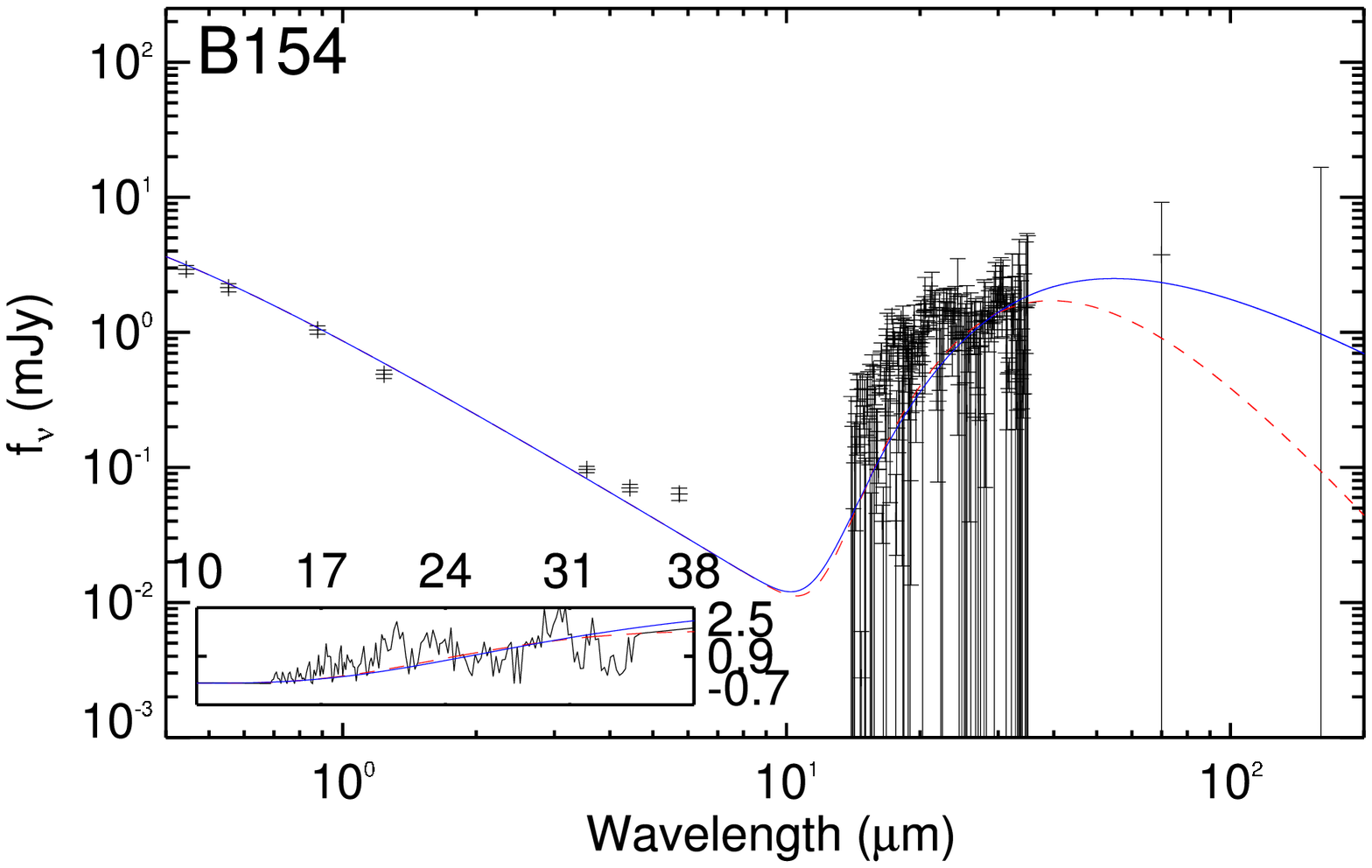}
   }
   \caption{(Continued)} 
\end{figure*}
\begin{figure*}
   \centering
   \ContinuedFloat
   \subfloat{
      \includegraphics [scale=0.4,angle=0]{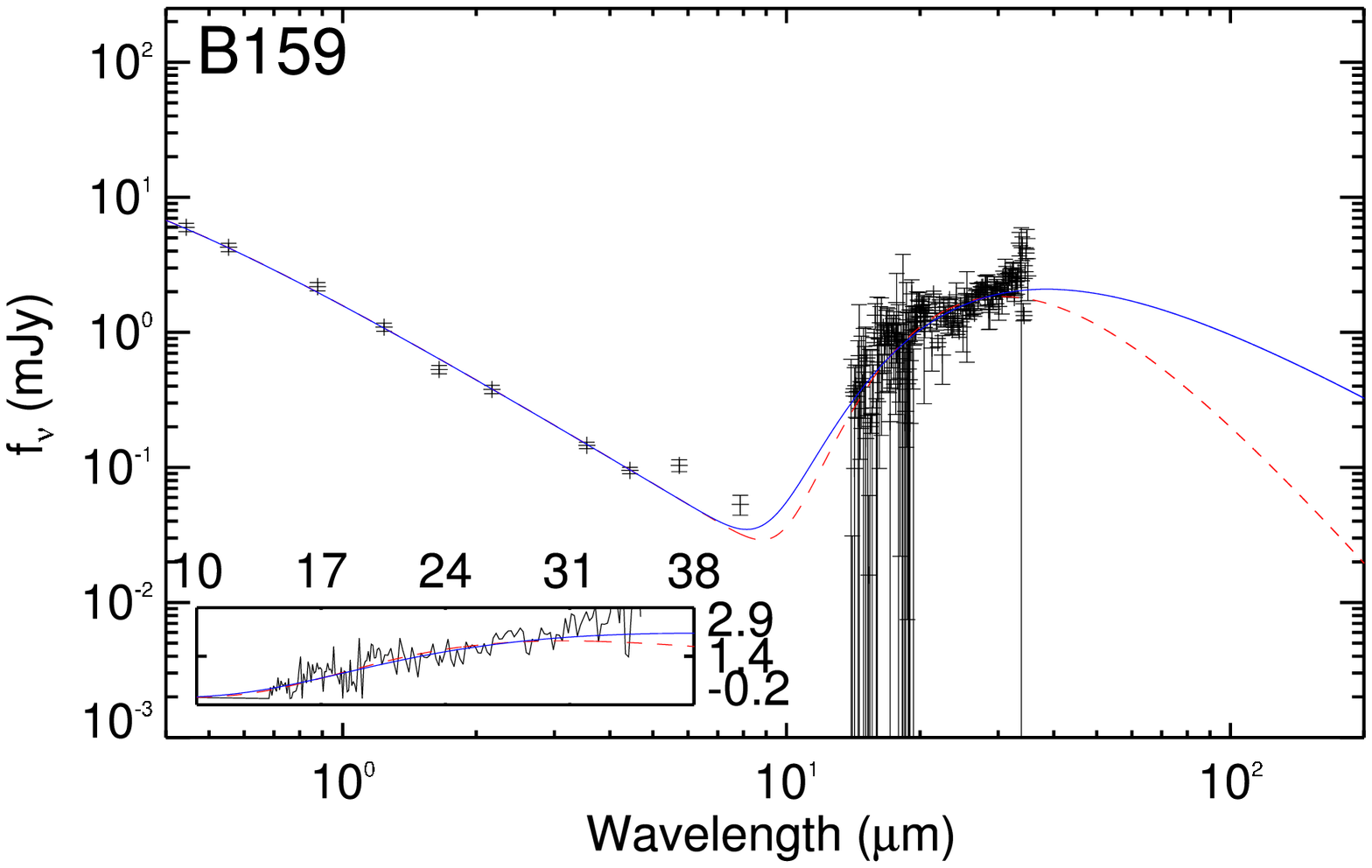}
   }
   \subfloat{
      \includegraphics [scale=0.4,angle=0]{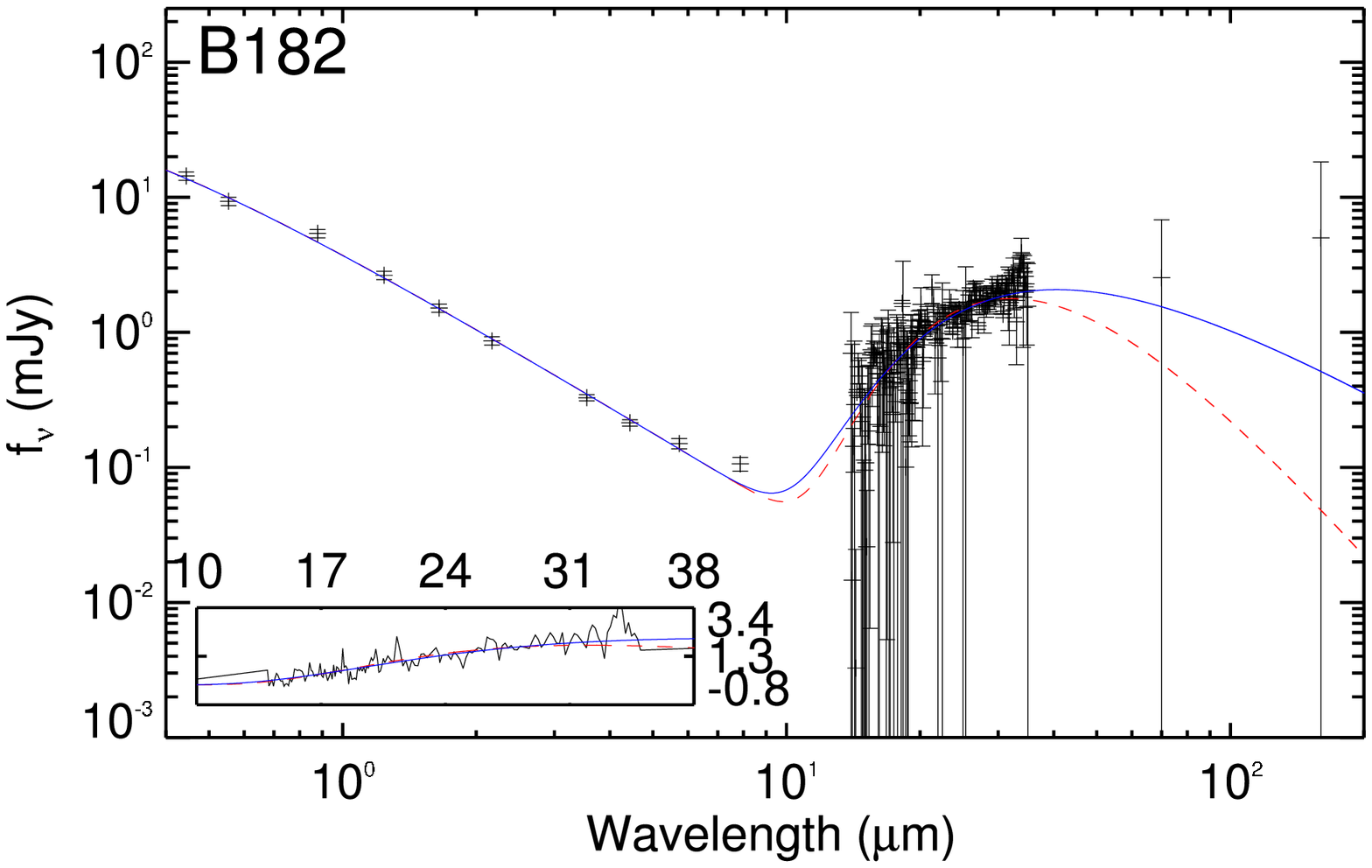}
   }\\
   \subfloat{
      \includegraphics [scale=0.4,angle=0]{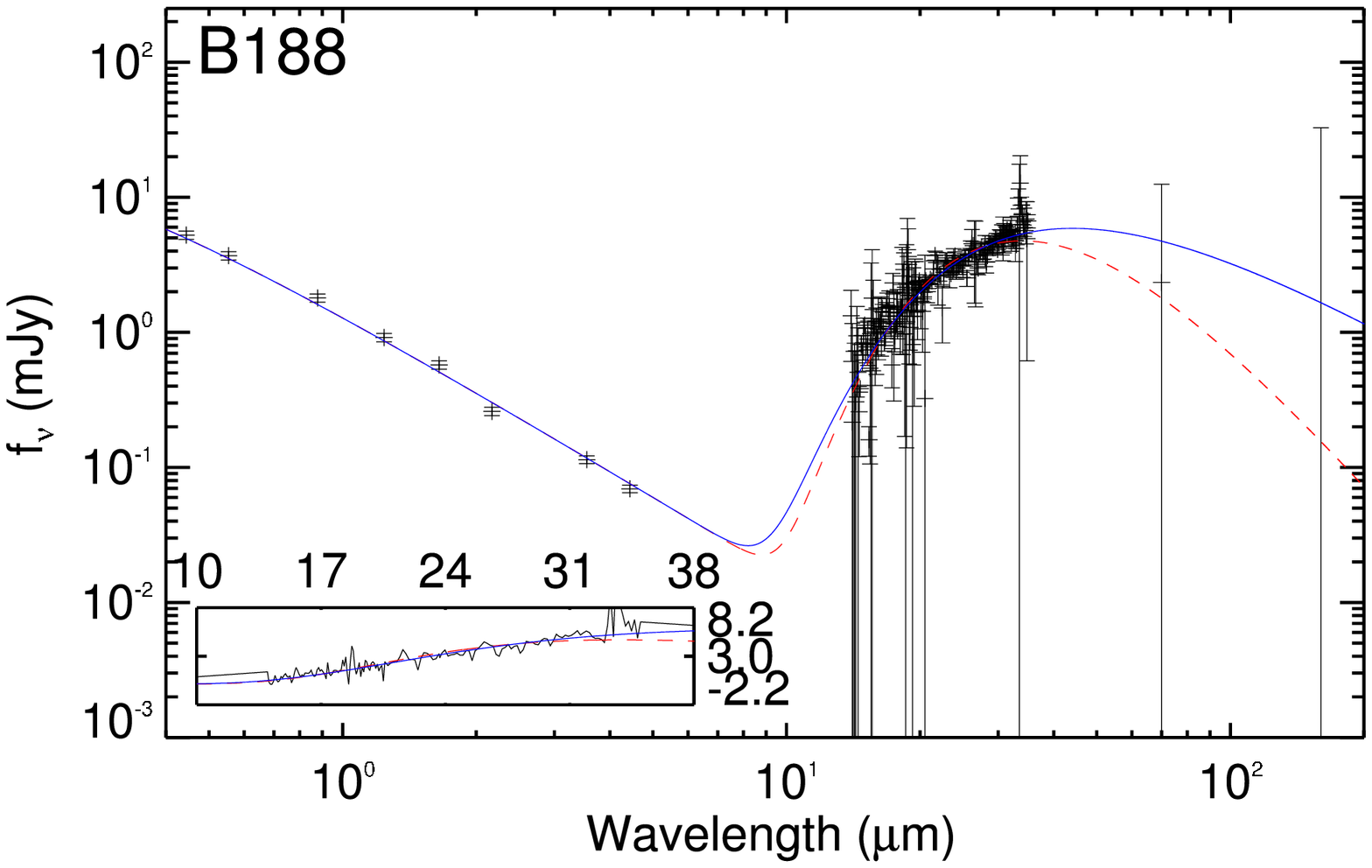}
   }
   \subfloat{
      \includegraphics [scale=0.4,angle=0]{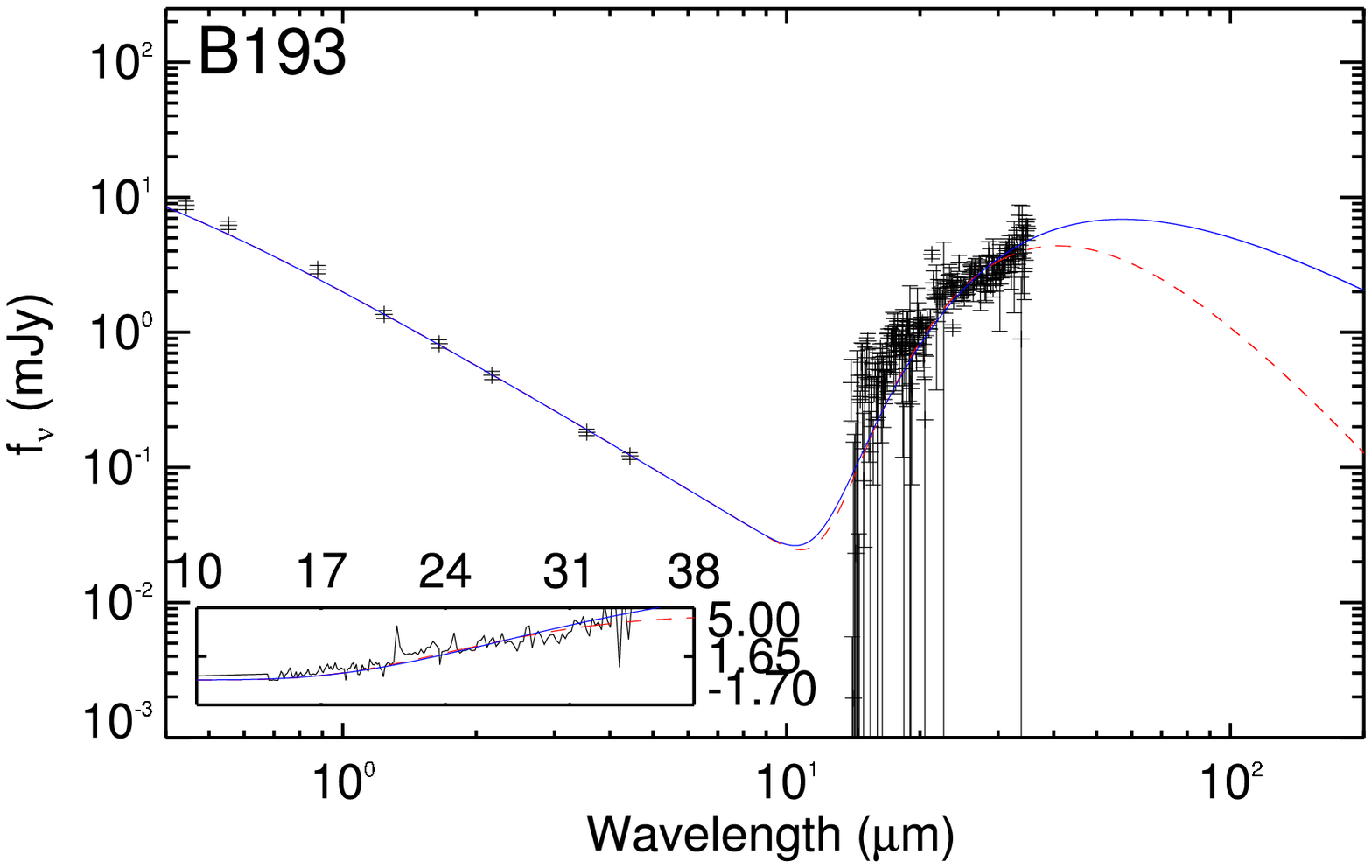}
   }
   \caption{(Continued)} 
\end{figure*}

\subsection{\emph{Herschel} Photometry}

We use observations taken with the European Space Agency's (ESA) \emph{Herschel 
Space Observatory} \citep{Pilbr10} and its Photometer Array Camera 
and Spectrometer (PACS) instrument \citep{Pogli10}. The observations were taken during 
program cycle OT1 (P.I. J.~Simon) with 9 hours on 11 targets in the 70~\mum\ and 
160~\mum\ bands as mini scan maps under median scan speed with four 4\arcmin~scan legs and 2\arcsec~cross-scan 
steps. Integration times were chosen so as to be able to detect the emission at 5$\sigma$ 
significance at 70~\mum\ if half of the mid-IR excess were coming from cold (55K) dust. 

The data were reduced to level 1 with the \emph{Herschel} Interactive Processing 
Environment (HIPE) v.8.0 developer build 3459 \citep{Ott10} using the 
PACS calibration tree 32. Apart from the standard reduction steps, we 
applied `$2^{\text{nd}}$ level deglitching' to remove outliers in the time 
series data (`time-ordered' option) by $\sigma$-clipping (25 $\sigma$ 
threshold) the flux values which contribute to each pixel.
Based on the level 1 data, we have generated final level 2 maps using 
Scanamorphos version 16 \citep{Rouss12} with the `minimap' keyword. 
We used a final map pixel scale of 1\arcsec\ at 70 \mum\ and 
160 \mum. 

The color corrections were taken from \cite{Pogli10}, and the aperture corrections for 
radii of 6\arcsec~and 12\arcsec~in the 70~\mum\ and 160~\mum\ bands are gathered 
from the PACS Observer's Manual\footnote{herschel.esac.esa.int/Docs/PACS/pdf/pacs\_om.pdf} and 
listed in Table \ref{tab:col_cor}. During the reductions, the 
plate scales were resampled to 1\arcsec~pixel$^{-1}$ from the native 
3\farcs2 and 6\farcs4 scales of the blue and red channels. The pipeline calculates the 
photon shot noise and detector noise, but because the data is resampled, the 
noise map is correlated. Flux measurements must consider this as a correlation correction term that 
scales up the total variance over an aperture in the noise map. 
We follow the formalism from the \emph{WISE} 
Supplement\footnote{http://wise2.ipac.caltech.edu/docs/release/allsky/expsup/\\
sec2\_3f.html} to estimate the correlation corrections for this resampled data. 
Under the assumption of a Gaussian point-spread function and 
\emph{Herschel} parameters, this correction evaluates to 69.6 and 278.5 for the 
blue and red bands. We use background subtraction annuli of radii from 9\arcsec~to 15\arcsec~and 
18\arcsec~to 30\arcsec~for the blue and red bands, respectively. We measure source flux in 
apertures of 6\arcsec~and 12\arcsec~for the 70~\mum\ and 160~\mum\ bands. 

Since the fields are not entirely uncrowded, 
we carefully consider the background measured in the annulus. We found all 5$\sigma$ significant 
point sources in the image other than our target after smoothing the images with the point spread functions, and if 
close we masked the overlap of the sky annulus and the contaminating sources. By adjusting the 
precise values of the sky annulus size, we found only small changes consistent with the errors. 
We only detect one source with meaningful significance, but the upper limits for the others still add useful constraints to 
the dust temperature. The only significant detection is 
at 3.4$\sigma$ in the 70~\mum\ band for source B004. 
The 11 SMC sources were chosen to be outside of obvious star clusters and outside the halos from luminous neighbors
based on the MIPS 70~\mum~ data. We make estimates of the confusion noise from each image for 
this reason. 

The confusion noise is found in the images themselves by identifying all point 
sources, laying down randomly placed, non-overlapping apertures that did not intersect the point sources out to 
their background annuli, and finding the 
excess variance above the Poisson uncertainies. This error estimate corrects for faint, unresolved point 
sources in the background and extended dust emission. 
Most of the 160~\mum\ data go deep enough to have 
significant confusion noise, but only one of the 70~\mum\ images with a 
long exposure is affected by confusion. 
We list the corrected photometric measurements in 
Table \ref{tab:hersphot} 
with the error budget broken into two components. The first component 
follows from the values in the uncertainty maps under the apertures and sky annuli, and the second 
is from confusion. The total error follows by combining them in quadrature. 
\begin{deluxetable}{crrr@{$\pm$}c@{$\pm$}lr@{$\pm$}c@{$\pm$}l}
\tabletypesize{\scriptsize}
\tablecaption{\emph{Herschel} Data Log and Photometry\label{tab:hersphot}}
\tablewidth{0pt}
\tablehead{
\colhead{Object} & \colhead{Start} & \colhead{Exposure} & \multicolumn{3}{c}{70~\mum} & \multicolumn{3}{c}{160~\mum} \\
\colhead{} & \colhead{Date (UTC)} & \colhead{Time (s)} & \multicolumn{3}{c}{Flux} & \multicolumn{3}{c}{Flux} \\
\colhead{} & \colhead{} & \colhead{} & \multicolumn{3}{c}{(mJy)} & \multicolumn{3}{c}{(mJy)}}
\startdata
B004 & 2011-07-23 & 2200 & 24.8 & 7.4 & 0  & $-$17 & 24 & 58 \\
B009 & 2011-07-23 & 1080 & 10.1 & 9.1 & 0  & 17 & 34 &  0 \\
B011 & 2011-07-23 & 1528 & 7.5 & 8.3 & 0  & $-$33 & 28 & 33 \\
B014 & 2011-07-23 & 1752 &  7.7 & 8.5 & 0  & 78 & 30 & 84 \\
B024 & 2011-07-23 & 5560 &$-$15.9 & 4.9 & 7.7  & 0\tablenotemark{*} & 16 & 35 \\
B026 & 2011-07-23 & 6680 &  1.3 & 4.3 & 0  & $-$22 & 13 & 91 \\
B029 & 2011-02-25 &  184 &  2.9 & 20.7 & 0 & 4 & 72 &  0 \\
B087 & 2011-07-23 &  856 &  3.7 & 11.3 & 0 & 21 & 37 &  0 \\
B154 & 2011-07-23 & 1006 &  4.3 &  6.2 & 0  & $-$41 & 19 &  0 \\
B182 & 2011-07-23 & 5336 &  2.9 &  4.9 &  0  & 6 & 15 &   0 \\
B188 & 2011-07-23 &  856 &  2.7 & 11.5 & 0 & $-$25 & 37 &  0
\enddata
\tablecomments{The flux measurements are listed with two error estimates. The first 
contains the read noise and shot noise uncertainties from the circular aperture and 
the sky annulus. The second describes the confusion noise, if a significant 
value is measured from the image. The values are added in quadrature to make a total error estimate.}
\tablenotetext{*}{The formal flux measured is a large negative value due to the contamination 
from several bright, nearby sources in the sky annulus. By using larger sky annuli, we do not 
find a significant detection. Therefore, we have set this flux measurement to zero to avoid 
biasing the derived properties.}
\end{deluxetable}

Based on brighter point 
sources in the images, we determine that there is an astrometric offset to the 
images that correlates with the scan angle. The PACS Observer's Manual states that 2\arcsec~
absolute astrometry errors can be expected, but we find offsets of $-$4\farcs1 and $-$0\farcs7 
along the long and short scan axes, respectively. We measure aperture photometry using the 
corrected astrometry. 
 
\subsection{Ancillary Data}
\label{sec:ancil}

Paper I has already presented 
optical through near-IR photometry for these stars. The 
original sources are the OGLE II survey \citep{Udals98}, the Magellanic Clouds 
Photometric Survey \citep{Zarit02}, and 2MASS \citep{Skrut06}. We have also taken the stellar 
temperature subtype as determined from optical spectra in Paper I. One star, B011, was 
not classified in Paper I because of insufficient S/N. Based on the average 
type of the whole sample, we have assigned it a B1 type in the 
following computations. This is compatible with the photometrically derived stellar 
temperature. We have also given B188, a star potentially with weak emission lines, 
an O9 type. We based the stellar properties off the spectral 
types in column 4 of Table \ref{tab:sedprop1}. The masses, luminosities, radii, 
and temperatures are interpolations from the the table for zero-age main sequence stars 
in \cite{Schmi82}. We have corrected the optical and near-IR data for 
the A$_{\rm V}=0.12$ MW extinction near the SMC \citep{Schle98} with the 
reddening curve of \cite{Carde89}. We measured the remaining reddening from 
the B and V bands, and further corrected 
the data with the SMC reddening curve of \cite{Bouch85}. 
The stars were selected to be on the main-sequence by their luminosities, but there may be some 
contamination by giants. There are external 
spectral classifications for three stars. \cite{Evans08} classify B021 as B1--2II and B137 as 
B1--3II, although they flag both as uncertain. \cite{Hunte08} classify B100 as B0.5V. However, 
we give the dust-corrected, absolute V-band magnitudes in column 6 of Table \ref{tab:sedprop1}, and 
they are all consistent with being dwarfs. 

\section{SED Fitting for Different Models}\label{sec:sed}
We now extract physical quantities by fitting the spectral energy distributions (SEDs). The functional forms for the various fits 
are justified below. The most likely candidate models are debris or transition disks, where the 
dust has a circumstellar origin, and cirrus hot spots, where the 
dust has an interstellar origin. Classical Be and Herbig Ae/Be models 
have already been excluded by the optical spectra in Paper I. The cirrus hot spot models 
can be further subdivided into two types. The vast cloud of glowing dust around the Pleiades 
is the prototype for relatively static hot spots. Cases of stars moving at high 
velocities and compressing ISM material into a bow shock are also well known. We follow the calculations 
developed by \cite{Artym97} to model such a situation. The primary difference between the interacting and static 
hot spots is that small dust grains, which would normally be blown out from the system by radiative forces, can 
be retained and emit at higher temperatures in interacting systems. In low velocity cases, the two hot spot models converge. The 
models for debris disks and static hot spots have already been discussed in Paper I. For the 
ease of the reader, we have summarized their justification and discussion in Appendix \ref{ap:eqn}. The properties derived from 
debris disk and static cirrus hot spot models are given in Table \ref{tab:sedprop2}. The
tabulated parameters are not exactly the same as those given in Paper I because we have used the additional data
presented in Section \ref{sec:data} to determine them, but the properties are generally in agreement. 

The forms we use to fit the SEDs are blackbody (BB) and modified blackbody (MBB) functions. 
Both functions fit the IRS data equally well, and the MBB function, appropriate for hot spot models, 
naturally delivers lower dust temperatures. Both models fail for some stars to fit the IRAC data well, which drives the
reduced $\chi^2$ marginally
above its expected value for statistical errors alone ($\bar{\chi^2_{\rm r}}=1$ with a
standard deviation of $\sqrt{(2/N)}$ for N
degrees of freedom). These poor fits are likely caused by the simple assumptions in our models, but
could also be caused by a small underestimate of our uncertainties.
Specifically to the cirrus hot spot models, stochastic heating of small grains may be 
causing some of the near-IR excess we see in the 5.8 and 8.0~\mum~ bands, but we have not included it 
in our modeling. Stochastic heating has been observed in reflection nebulae 
from $\sim$~1--20~\mum\ \citep{Sellg84,Sylve97}. Strong PAH
features may be present in the IRAC bands as well. The one \emph{Herschel} detection at 70~\mum\ is also in excess of our fits. 

\begin{centering}
\begin{deluxetable*}{cr@{$\pm$}lr@{$\pm$}lr@{$\pm$}lr@{$\pm$}lr@{\extracolsep{4pt}}r@{\extracolsep{0pt}}@{$\pm$}lr@{$\pm$}lr@{$\pm$}lr@{$\pm$}lr}
\tabletypesize{\scriptsize}
\tablecaption{Dust properties of the dusty OB stars\label{tab:sedprop2}} \tablewidth{0pt}
\tablehead{ &  \multicolumn{9}{c}{Debris Disk Model} & \multicolumn{9}{c}{Static Cirrus Hot Spot Model} \\
\cline{2-10} \cline{11-19}
                \colhead{ID No.}     &
                \multicolumn{2}{c}{T$_{\rm dust}$}      & 
                \multicolumn{2}{c}{L$_{\rm disk}$/L$_{\rm *}$}      &
                \multicolumn{2}{c}{R$_{\rm BB}$}      &
                \multicolumn{2}{c}{M$_{\rm min}$}      &
                \colhead{$\chi^2_{\rm r}$}      &
                \multicolumn{2}{c}{T$_{\rm dust}$}      &
                \multicolumn{2}{c}{L$_{\rm disk}$/L$_{\rm *}$}      &
                \multicolumn{2}{c}{R$_{\rm MBB}$}      &
                \multicolumn{2}{c}{M$_{\rm dust}$}      &
                \colhead{$\chi^2_{\rm r}$}      \\
                \colhead{}     &
                \multicolumn{2}{c}{(K)}      &
                \multicolumn{2}{c}{(10$^{-4}$)}      &
                \multicolumn{2}{c}{(10$^2$ AU)}      &
                \multicolumn{2}{c}{(M$_{\earth}$)}      &
                \colhead{}      &
                \multicolumn{2}{c}{(K)}      &
                \multicolumn{2}{c}{(10$^{-4}$)}      &
                \multicolumn{2}{c}{(10$^5$ AU)}      &
                \multicolumn{2}{c}{(M$_{\earth}$)}      &
                \colhead{}}
\startdata

B004 & 100.2 &  0.8 &  4.6 &  0.3 & 27.0 &  1.4 &  654 &  121 &  5.1 & 75.7 &  0.5 &  3.5 &  0.2 &  2.0 &  0.1 &  10.3 &   0.6 &  5.1 \\
B009 & 94.8 &  0.9 & 22.7 &  1.6 & 37.7 &  2.9 & 2893 &  641 &  9.0 & 72.2 &  0.5 & 16.9 &  1.2 &  1.3 &  0.1 &  21.6 &   1.4 &  9.3 \\
B011 & 99.1 &  0.8 & 14.2 &  0.9 & 12.4 &  0.8 &  196 &   39 &  6.1 & 75.7 &  0.5 & 10.4 &  0.6 &  1.1 &  0.1 &  10.0 &   0.6 &  6.4 \\
B014 & 113.1 &  0.8 & 22.1 &  1.1 & 16.5 &  1.0 &  541 &  103 & 11.7 & 84.6 &  0.5 & 16.8 &  0.8 &  0.8 &  0.1 &  8.2 &   0.3 & 13.0 \\
B021 & 157.2 &  1.6 &  4.2 &  0.2 &  7.7 &  0.4 &   27 &    5 &  6.7 &105.2 &  0.7 &  3.5 &  0.2 &  0.5 &  0.1 &   0.6 &   0.03 &  6.7 \\
B024 & 180.2 &  3.1 &  7.1 &  0.7 &  3.6 &  0.2 &    8 &    2 &  7.7 &119.8 &  1.5 &  5.8 &  0.6 &  0.3 &  0.1 &   0.3 &   0.04 &  7.4 \\
B026 & 87.5 &  0.9 &  6.5 &  0.5 & 43.8 &  3.4 & 1120 &  248 & 20.3 & 68.1 &  0.5 &  4.7 &  0.3 &  1.5 &  0.1 &  8.5 &   0.6 & 19.9 \\
B029 & 121.7 &  0.6 & 24.7 &  0.8 & 33.7 &  2.6 & 3021 &  646 &  5.4 & 89.6 &  0.3 & 19.9 &  0.7 &  0.8 &  0.1 &  10.2 &   0.3 &  5.0 \\
B034 & 93.9 &  1.2 & 13.0 &  1.4 &  8.1 &  0.3 &   91 &   17 &  7.3 & 74.0 &  0.8 &  8.7 &  0.9 &  1.5 &  0.1 &  14.1 &   1.4 &  7.5 \\
B087 & 124.1 &  0.9 & 14.2 &  0.7 &  9.4 &  0.4 &  134 &   23 &  3.6 & 89.0 &  0.5 & 11.4 &  0.5 &  0.8 &  0.1 &  6.0 &   0.2 &  4.1 \\
B096 & 140.7 &  1.7 &  3.0 &  0.2 &  5.1 &  0.2 &    8 &    1 &  7.0 & 99.5 &  0.9 &  2.5 &  0.2 &  0.6 &  0.1 &   0.6 &   0.08 &  6.4 \\
B100 & 116.3 &  1.0 &  9.4 &  0.6 & 11.5 &  0.7 &  112 &   22 &  5.2 & 84.1 &  0.6 &  7.4 &  0.4 &  0.8 &  0.1 &   3.8 &   0.2 &  6.2 \\
B102 & 118.0 &  1.0 &  3.9 &  0.2 & 25.7 &  1.6 &  513 &  101 &  4.1 & 85.3 &  0.6 &  3.2 &  0.2 &  1.4 &  0.1 &   4.6 &   0.2 &  4.3 \\
B137 & 118.2 &  1.1 &  4.5 &  0.3 & 10.7 &  0.6 &   46 &    8 &  8.2 & 85.2 &  0.6 &  3.7 &  0.2 &  0.8 &  0.1 &   1.7 &   0.08 &  7.9 \\
B148 & 138.3 &  1.5 &  2.8 &  0.2 &  9.4 &  0.5 &   27 &    5 &  5.1 & 97.4 &  0.8 &  2.4 &  0.1 &  0.6 &  0.1 &   0.7 &   0.03 &  4.9 \\
B154 &  93.2 &  1.0 &  6.5 &  0.5 & 26.3 &  2.3 &  406 &   97 & 13.4 & 73.2 &  0.6 &  4.8 &  0.4 &  1.2 &  0.1 &   5.6 &   0.4 & 12.5 \\
B159 & 132.4 &  1.1 &  5.6 &  0.3 & 16.9 &  1.0 &  172 &   33 &  6.5 & 94.3 &  0.6 &  4.6 &  0.2 &  0.7 &  0.1 &   1.7 &   0.08 &  6.8 \\
B182 & 125.2 &  1.2 &  5.2 &  0.3 & 11.4 &  0.5 &   72 &   13 &  4.7 & 90.3 &  0.6 &  4.2 &  0.2 &  0.8 &  0.1 &   2.0 &   0.08 &  4.7 \\
B188 & 115.6 &  0.7 &  6.4 &  0.3 & 33.6 &  2.2 & 1414 &  281 &  3.9 & 85.5 &  0.4 &  5.1 &  0.2 &  1.4 &  0.1 &   7.3 &   0.2 &  4.2 \\
B193 &  89.3 &  0.7 & 11.5 &  0.7 & 67.6 &  5.7 & 5640 & 1283 & 15.3 & 70.7 &  0.4 &  8.0 &  0.5 &  1.7 &  0.1 &  17.2 &   0.9 & 14.4 
\enddata
\tablecomments{Columns 2-6 are inferred properties from 
blackbody function fits to the dust SED. The equilibrium radius of column 4 is from Equation \ref{eq:bbdist}. 
The dust mass estimate of column 5 comes from Equation \ref{eq:mass}. 
Column 6 lists the reduced $\chi^2$ for the fit. Columns 7-11 list the inferred properties from 
modified blackbody function fits to the dust SED. The equilibrium radius of column 9 is from Equation \ref{eq:mbbdist}.
The dust mass estimate of column 11 comes from Equation \ref{eq:emdust}. 
} 
\end{deluxetable*}
\end{centering}

Figure \ref{fig:sed1} shows the data for all the stars overlaid by curves of the 
best-fitting models. We have also plotted 
a zoomed-in inset with the IRS spectra. The absolute flux calibration from the IRS spectra is 
consistent with the 24~\mum\ MIPS photometry in each case. From the mid-IR data alone, the 
spectra look similar to many debris disks 
\citep[e.g.][]{Chen06,Dahm09}. Such comparisons must recognize that the literature debris disks 
are hosted by lower mass stars. It is clear from Equations \ref{eq:bbdist}, \ref{eq:amin}, and \ref{eq:mass} 
that the derived debris disk properties are sensitive to both the mid-IR observables and the 
host star properties. The early B stars we study have approximately 
2.5$\times$ the effective stellar temperature, 35--1800$\times$ the luminosity, and only $\sim$5$\times$ greater 
stellar mass compared to an A0V star. Under the same set of dust signatures, these changes drive 
the thermal equilibrium distance and minimum grain sizes up, along with the minimum dust mass. 

\subsection{Interacting Hot Spot Model}
\label{sec:shockmodel}
\begin{figure}
\centering
\includegraphics [scale=0.48,angle=0]{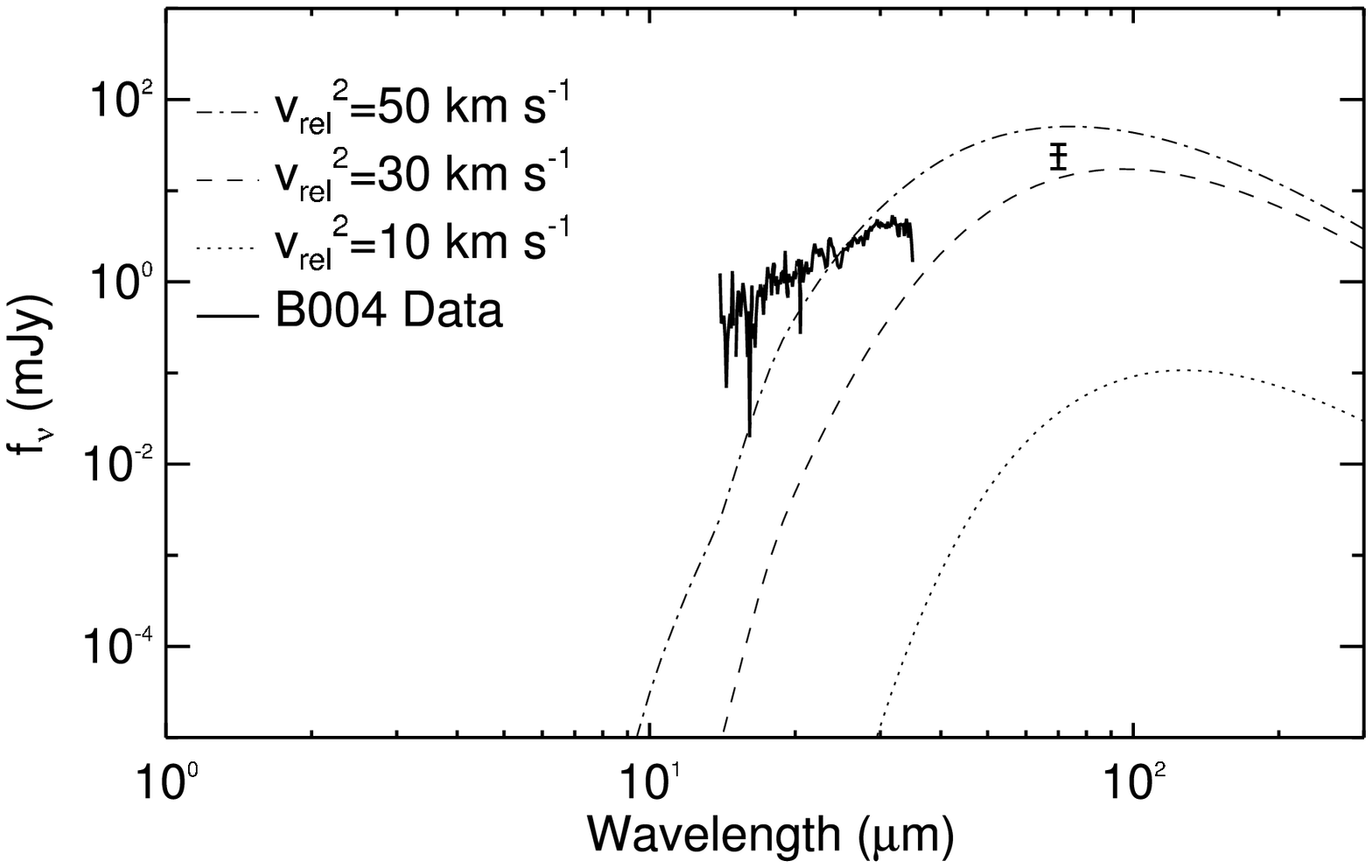}
\caption{Toy models for the thermal dust emission for a B0 star moving through an ISM cloud at several relative 
velocities with other parameters held fixed. The normalizations for the models have been scaled, as described in the key, 
for clarity. The bulk density and the interaction velocity both influence the normalization. The models are described in Section \ref{sec:shockmodel}. The IRS spectrum 
and \emph{Herschel} 70\mum\ data for one star is also shown. While the slopes for the particular parameters shown 
do not exactly match, we consider the agreement plausible given the model simplifications and the full 
range of the available models parameters. The cloud's outer radius, in particular, can be made smaller to create a 
hotter distribution.}
\label{fig:interact}
\end{figure}

We develop some toy models for interacting cirrus hot spots to determine whether plausible values of 
velocity relative to the ISM clouds can match the SEDs. This scenario deviates from the previous two by permitting 
a continuous range of temperatures for differently sized dust grains. We follow the ideas originally developed by 
\cite{Artym97} and applied to data for $\lambda$ Bootis stars in \cite{Kamp02,Gaspa08,Marti09}. 
When a star is moving through a dust cloud, there is a scattering surface with a paraboloidal shape 
for any grain size beyond which the grains reside. The closest point on that surface that a grain 
of size $a$ may approach the star, the radius of avoidance, is given by:
\begin{equation}
r_{\rm av}(a)=\frac{2(\beta(a)-1)MG}{v_{\rm rel}^2}, 
\label{eq:ravoid}
\end{equation}
for a star of mass M, a gravitational constant G, and a relative velocity of v$_{\rm rel}$. $\beta(a)$ is the 
ratio of the radiative force to the gravitational force that we have given in Equation \ref{eq:amin}. However, 
for these small grains the average radiation pressure efficiency, $Q_{\rm pr}(a)$, cannot be approximated as a 
constant, as it can when applied to debris disk models. 
We have numerically estimated its value by using the grain properties for astrophysical silicates in 
\cite{Drain84} and integrating the coefficients over wavelength and weighted 
with a 32,000 K BB function, representing the central 
star, to derive the average radiation pressure efficiency as a function of grain size. We have used a 
mass and luminosity appropriate for a B0 star. The dust is assumed to be at constant density in a 
sphere around the star, and grains of different sizes are evacuated from differently 
sized paraboloidals within the cloud. Each particle is given a temperature to maintain radiative 
equilibrium. The model solely uses silicates. We numerically integrate through the volume and over particle size, and we weight this by a MRN grain 
size distribution \citep{Mathi77}. The free parameters are the outer sphere radius, the relative velocity between the cloud and star, 
and the bulk density. We have set the outer radius to 10$^{6}$ AU, although this is an order of magnitude larger than usually used when 
modeling interacting hot spots around cooler stars. With much smaller clouds and modest velocities, all the dust in the 
MRN grain size range is ejected by the radiation force. Clouds much larger than this would be well-resolved in our 
IRS peak-up images. In Figure \ref{fig:interact} we show toy model spectra 
for several interaction velocities. 

Not surprisingly, larger velocities result in hotter and more luminous emission. The toy models shown are too 
steep when compared to the data for B004, but slightly hotter and better matching distributions can be made by either 
adopting higher velocities or moving in the outer radius so that the contribution of the coolest grains is clipped. Interacting 
hot spots are a viable model for the dusty OB SMC stars.

\section{Local Comparison Samples}\label{sec:MW}
\subsection{Debris Disks}
\label{sec:lit_comp}
\begin{figure}
\centering
\includegraphics [scale=0.53,angle=0]{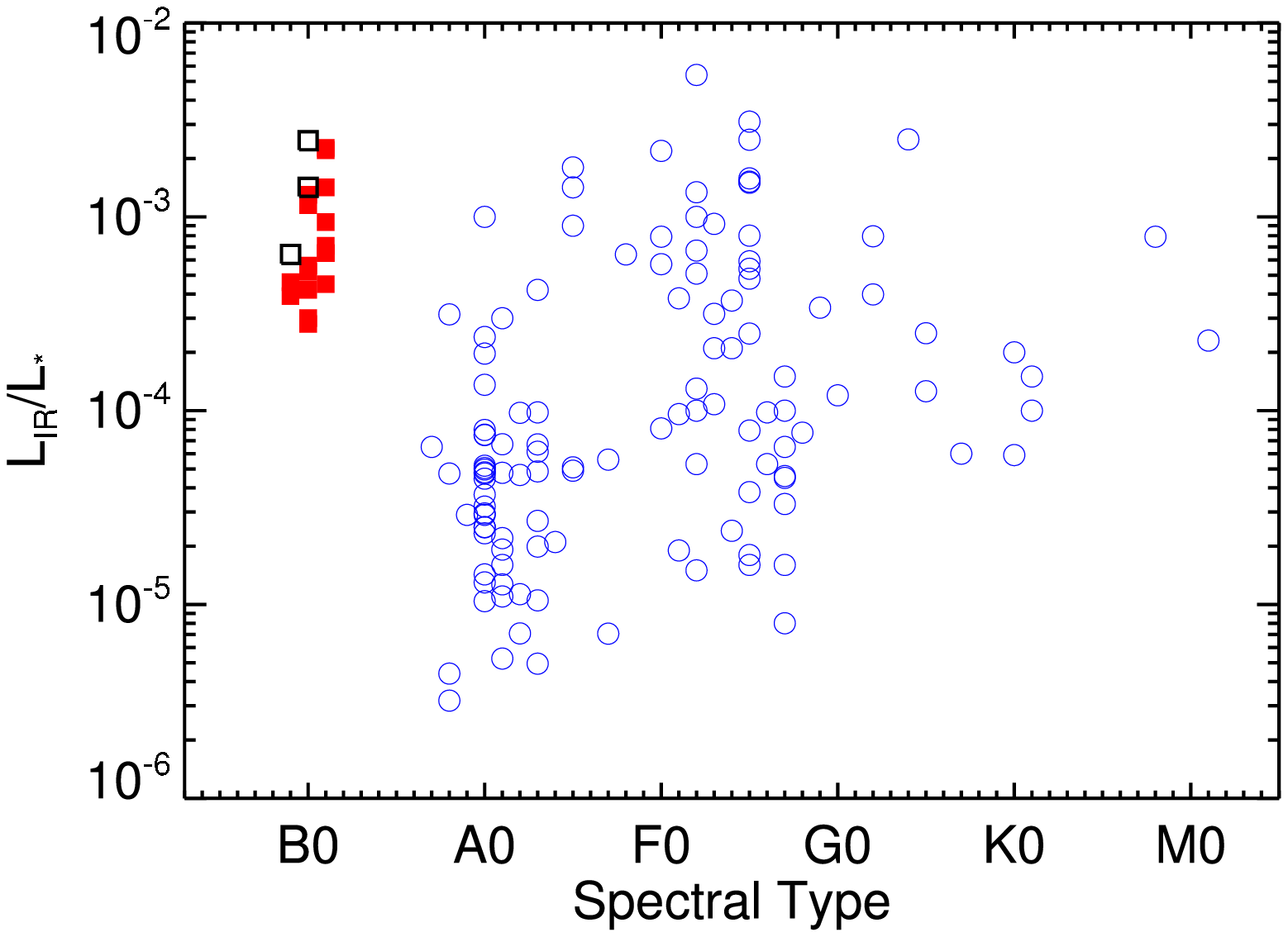}
\caption{Ratio of IR luminosity to stellar bolometric luminosity 
for dusty SMC OB stars (red, closed squares) and local debris disks (blue, open circles). The 
larger values in the dusty SMC OB stars are mostly within the 
upper envelope of confirmed debris disks. Within the local sample, the upper envelope of 
fractional fluxes does appear to bend downward from the G-F to A types. 
The higher average dust 
luminosities for the SMC stars are likely a selection effect since fainter sources 
would not have been detected in the SMC data. 
The three SMC stars resolved in the peak-up images are shown in open, black boxes.
}
\label{fig:debcomp1}
\end{figure}

\begin{figure}
\centering
\includegraphics [scale=0.53,angle=0]{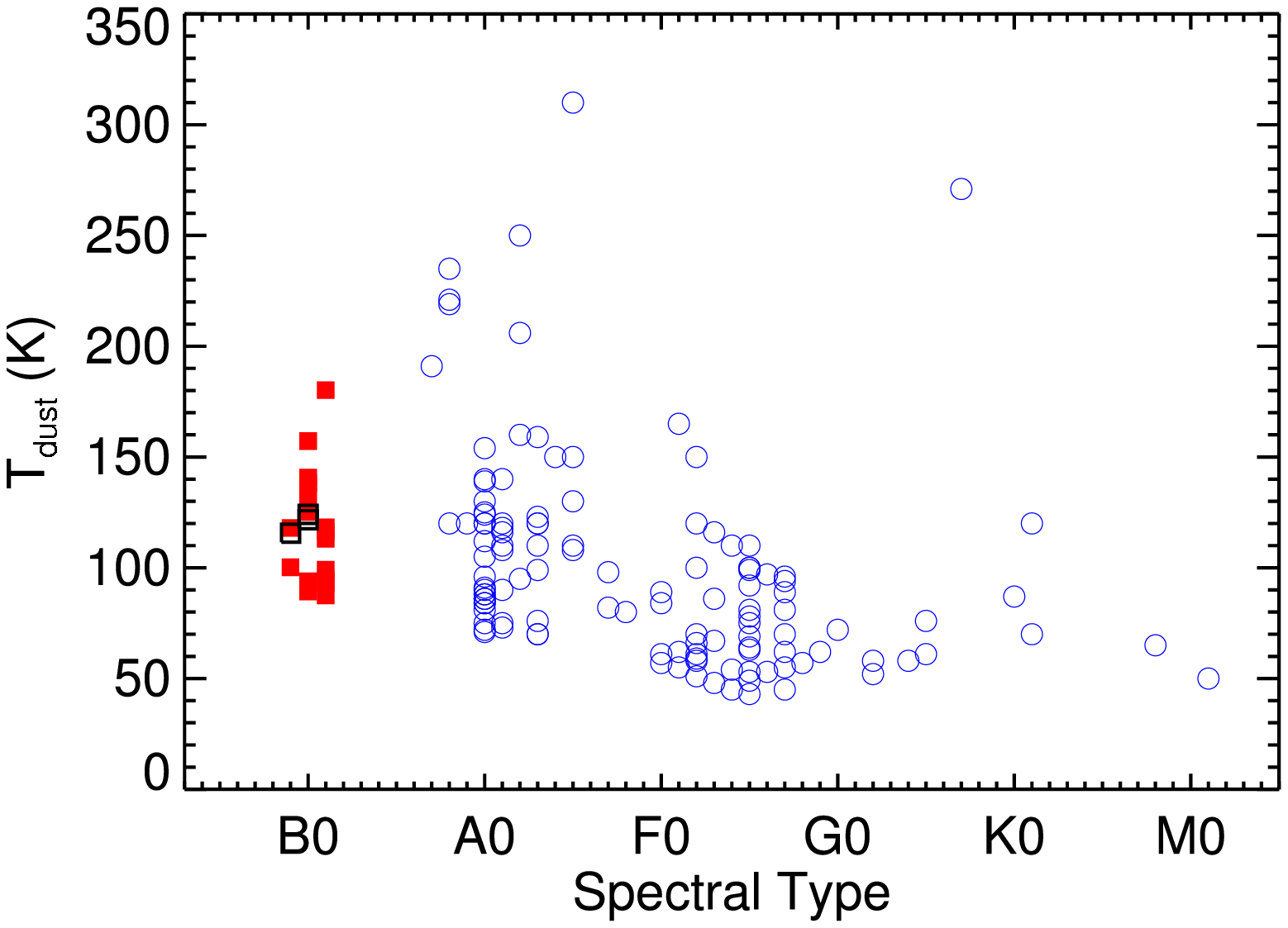}
\caption{Temperature measurements from single valued 
blackbody fits to the dust emission for dusty SMC OB stars (red, closed squares) 
and local debris disks (blue, open circles). No significant difference 
is seen between the two populations.
The three SMC stars resolved in the peak-up images are shown in open, black boxes.
}
\label{fig:debcomp2}
\end{figure}

\begin{figure}
\centering
\includegraphics [scale=0.53,angle=0]{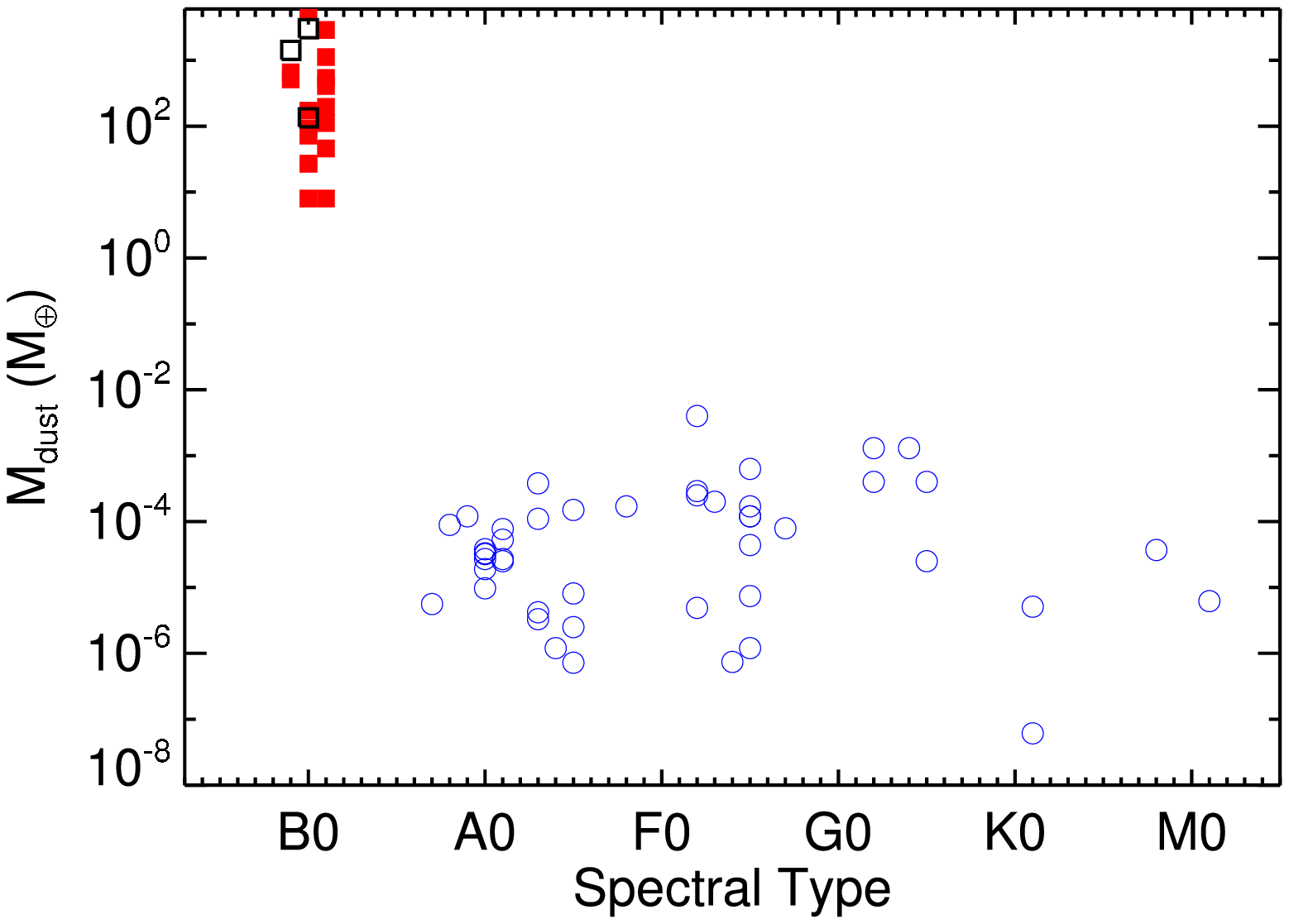}
\caption{Minimum dust mass estimates from blackbody fits to the 
dust emission for dusty SMC OB stars (red, closed squares) and local debris disks (blue, open circles). The 
large jump in inferred dust masses for the early-type OB stars 
appears distinct from an extrapolation of the local sample. 
The plotted masses have all been calculated with the same equations for a minimum mass
composed of single temperature grains with blackbody properties. We infer such high mass
estimates in the SMC OB stars primarily due to the minimum grain size and the
radius for an equilibrium temperature being much higher for the more massive stars.
The elevated fractional luminosities in the SMC OB stars, shown in Figure \ref{fig:debcomp1}, are
a secondary cause of the elevated dust masses. 
We have limited the plot to 
debris disk samples that have \emph{Spitzer} data. Some debris disk samples 
do reach masses of 0.1--1 $M_{\earth}$, such as in Figure 3 of \cite{Wyatt08}, but 
the measurements generally come from submillimeter data and different mass estimators.
The three SMC stars resolved in the peak-up images are shown in open, black boxes.
}
\label{fig:debcomp3}
\end{figure}

\begin{figure}
\centering
\includegraphics [scale=0.53,angle=0]{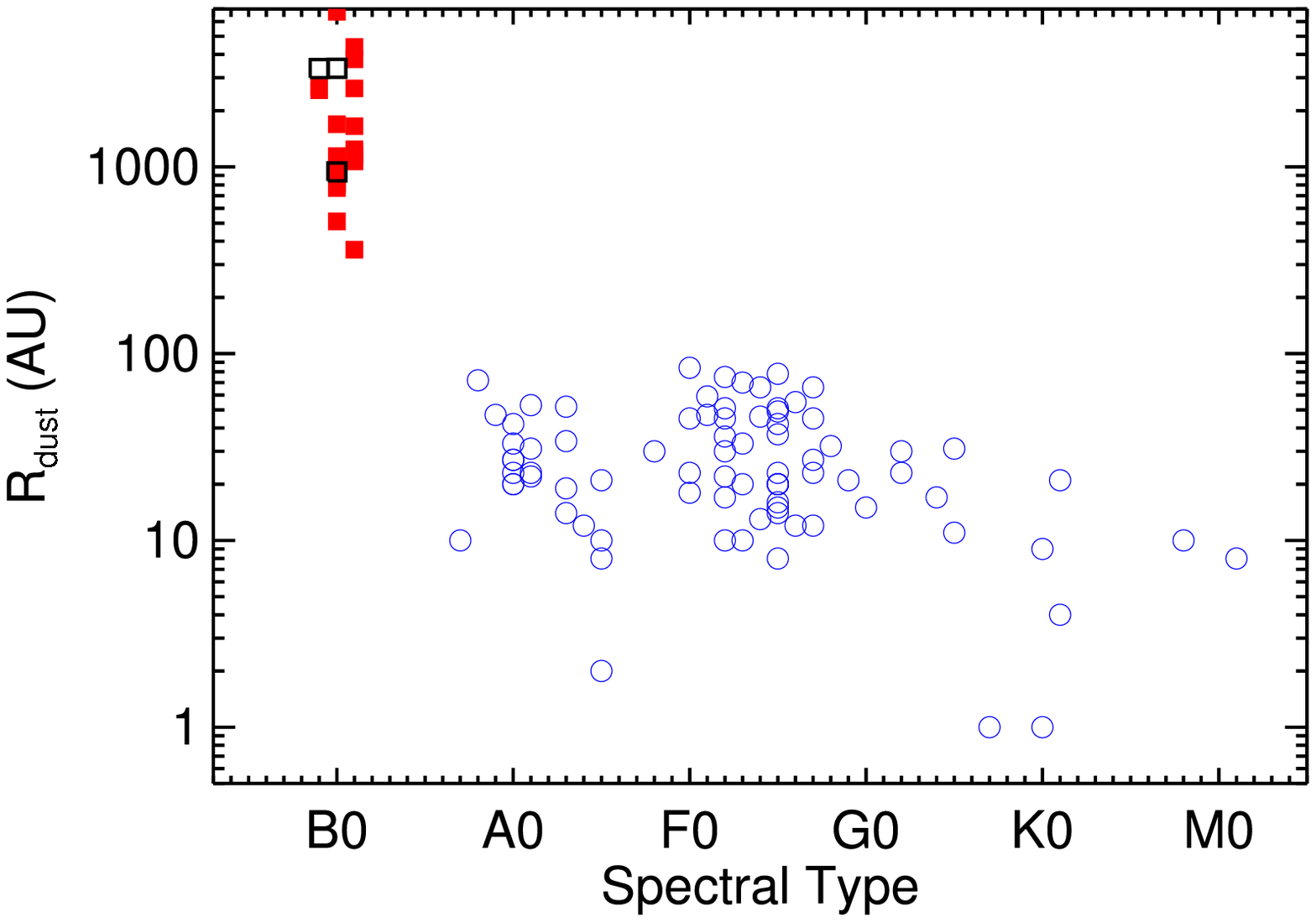}
\caption{Radius at which a single temperature blackbody 
dust grain is in thermal equilibrium 
for dusty SMC OB stars (red, closed squares) and local debris disks (blue, open circles). The 
dusty SMC OB stars have significantly larger radii. One would expect the equilibrium 
radius to be a strong function of stellar type, but the dust temperature 
also determines the radius. The large radii for OB stars are unexpected from the extrapolation 
of cooler, local stars. The three SMC stars resolved in the peak-up images are shown in open, black boxes.
}\label{fig:debcomp4}
\end{figure}

We have compiled the parameters from several 
local debris-disk studies \citep{Beich06,Chen06,Su06,Hille08,Rebul08,Trill08,Moor11} in order 
to compare their properties to the SMC candidates. 
Figures \ref{fig:debcomp1}--\ref{fig:debcomp2} show the ratio of IR to
stellar luminosity and the blackbody
dust temperature as a function of spectral type for the local and SMC samples. The 
properties inferred indirectly from the data, the
minimum dust mass and the dust equilibrium radius in the blackbody model, 
are shown in Figures \ref{fig:debcomp3}--\ref{fig:debcomp4}. 
In the MW, there are no strong trends of debris disk properties over a factor of $\sim$5 in stellar mass and 
a factor of $\sim$700 in luminosity. 


Until recently, no studies have found debris disks around stars more massive than $\sim$3 M$_{\rm \odot}$. \cite{Chen12} 
have searched for debris disks around stars at these higher masses. They analyzed 57 
stars with M $>$ 3 M$_{\rm \odot}$ in the Scorpius-Centaurus association and 
found only four stars with significant mid-IR excesses. Three are Be stars and only one (HIP~81972) is a debris disk 
candidate. HIP~81972 is a known binary, and the infrared emission may be coming from the lower mass
companion. The remaining 53 stars have 24\mum\ flux that is not in excess of the photospheric emission. 
The same study covered slightly less massive stars, and found them to host debris disks more 
frequently. They compile 24\mum\ measurements for A+B stars 
from 2.7 M$_{\rm \odot} >$ M $>$  1.8 M$_{\rm \odot}$ in the same association. 
After removing the objects classified as protoplanetary (based on 16\mum\ excesses or 
emission lines), they find that 72/293 (25\%) host debris disks. Similar work 
covered F+G stars (1.7 M$_{\rm \odot} >$ M $>$  0.8 M$_{\rm \odot}$) 
to find a 47/165 (28\%) debris disk fraction \citep{Chen11}. 
Debris disks around massive stars are evidently much rarer than those around A, F, and G stars. 

HIP~23767 (B5V) is another possible debris disk host \citep{Klopp10}, although the 
dust disk is much hotter (550K) than usual for debris disks, and the star is a binary system with an F type companion 
which may alternatively be hosting the dust.

\subsection{ISM Hot Spots}
A direct comparison to cirrus hot spots is more difficult as 
the modeling in previous papers is more diverse and comparable data, including IRS spectra, have 
not been taken. However, the temperatures have overlap with \cite{Gaust93} and the dust masses are comparable 
(as shown in Paper I). We conclude that the SED properties of present 
samples alone do not allow a clean classification of our SMC stars, but that nearby 
cirrus hot spots suffice as analogs. 

\subsection{Protoplanetary Disks}
\label{sec:proto}
We consider the properties of protoplanetary disks and whether our dusty OB stars may host such disks or their remnants. 
The working demarcation of L$_{\rm disk}$/L$_{*} \leq 10^{-3}$ for debris disks and L$_{\rm disk}$/L$_{*} \geq 10^{-3}$ for
primordial disks \citep{Willi11} suggests that most of our sources are not protoplanetary disks. Four of our twenty 
sources have large enough fractional luminosities to perhaps be protoplanetary disks. The short stellar lifetimes for the 
massive stars in our sample are approximately the same as the lifetimes of protoplanetary
disks. The compilation of protoplanetary disk mass as a function of stellar mass in Figure 5 of \cite{Willi11} shows that
such disks exist with an average mass of M$_{\rm disk}$/M$_{*}=0.01$ up through B star masses, but do not seem to exist around O stars
\citep{Mann09}. These masses are more than sufficient to supply, as remnants, the dust masses we measure. 
Since our source selection has chosen stars without H$\alpha$ emission, this intepretation would 
require a transitional phase where the protoplanetary disks are being disrupted and 
accretion has stopped. Is it plausible that we are 
observing such a phase? A conclusive model would have to consider the sensitivity of our H$\alpha$ observations 
and the rate at which H$\alpha$ emission fades in a disrupting disk. The timescale for disk dissipation 
after primordial accretion stops has been estimated as $\leq$0.5 Myr \citep{Skrut90,Wolk96,Cieza07} and is found to be 
similarly rapid in simulations \citep{Alexa06a,Alexa06b}.

Protoplanetary disks are believed to be destroyed by photoevaporation on an approximate timescale of 1 Myr \citep{Holle94}. 
The literature is rich with numerical simulations that have reinforced the original proposal for 
photoevaporative destruction \citep[e.g.][]{Yorke96,Richl97,Yorke02}. The observations and models have been 
summarized in \cite{Cesar07}, where one particular system, IRAS 20126+4104, has been shown to have properties 
very well matched to the gas-rich phases of protoplanetary disk growth. Notably, the disks may extend from 1,000--10,000 AU, 
such as in the simulations of \cite{Yorke02}. However, OB stars can drive such powerful photoevaporative winds that 
the ultimate fate of the material necessary to build up debris disks is uncertain. \cite{Balog06,Balog07,Balog08} have 
given demonstrations of proplyds where even the dust in the neighboring stellar disks is removed. In some simulations of massive 
stars \citep{Krumh09}, an optically thick disk persists which resists photoevaporation. 

Transition disks have been studied 
with IRS spectra \citep{Espai12}, but no samples have OB stars. \cite{Cieza12} present \emph{Spitzer} photometry 
on forty-one stars including seven B and A stars. None are so hot as our O9--B3 sample. We have 
examined all these stars in the \emph{WISE} catalog, and the only BA stars displaying the extreme colors of our SMC stars are 
the stars numbered in \cite{Cieza12} as 7 and 36 (A5 and B9). \cite{Cieza12} classify these two as a photoevaporating disk and a 
main sequence debris disk, respectively. In \emph{WISE}, source 7 is unresolved. However, source 36 is clearly resolved to a 
large size and is asymmetric. Assuming source 36 to be in the Taurus molecular cloud, the radius of the source in 
the \emph{WISE} filter at 22~\mum\ is 5,000 AU. \cite{Cieza12} discuss that source 36 may be a background 
star and not actually in the Taurus molecular cloud, 
so the size may be larger. 

\begin{figure}
\centering
\includegraphics [scale=0.45,angle=0]{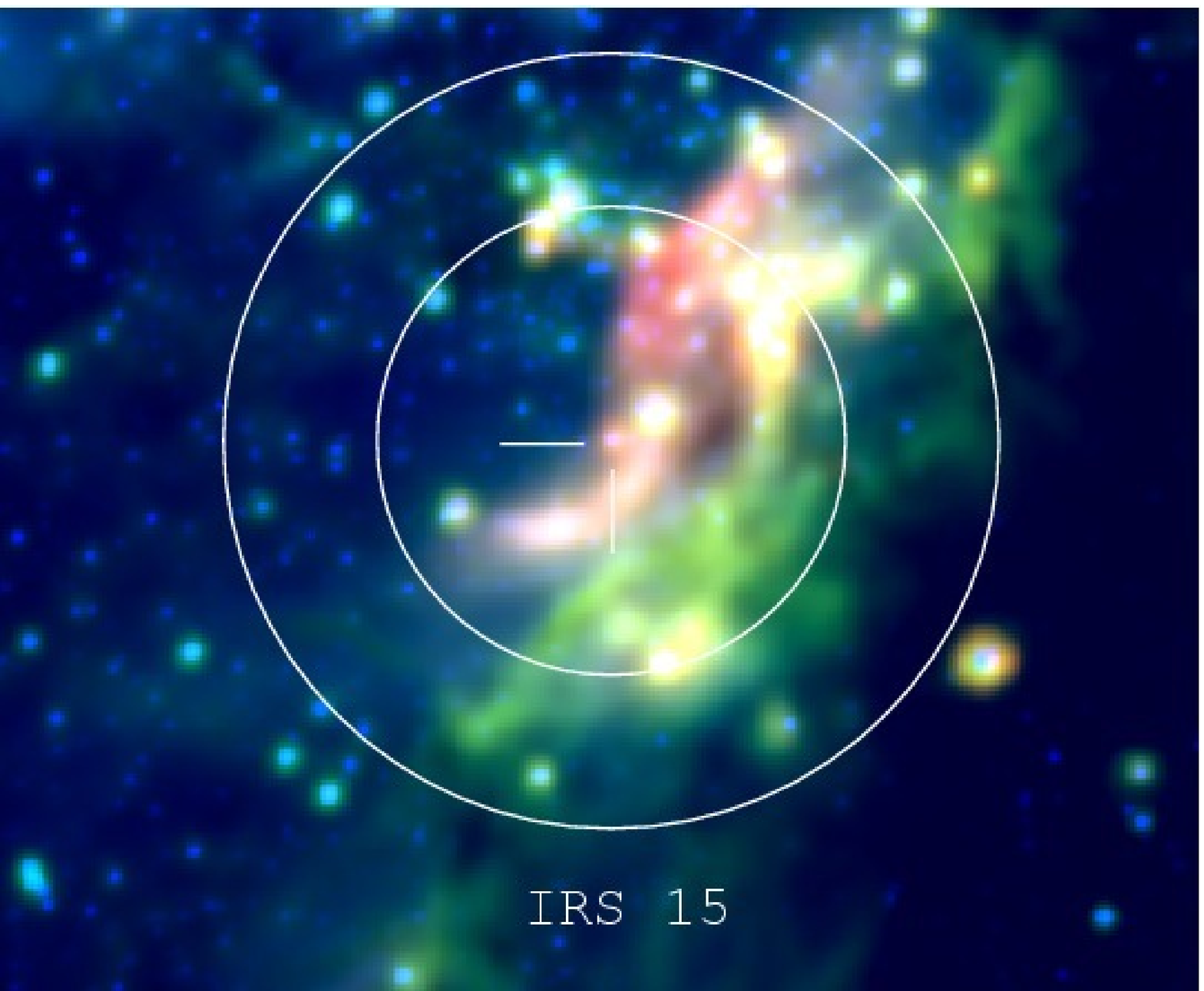}
\caption{A three-color image of IRS~15 with the \emph{2MASS} K$_{\rm s}$  and \emph{WISE} W1 
and W2, bands as blue, green, and red. W3 and W4 are saturated by the 
emission from M17. The crosshair lines are drawn with 30\arcsec\ lengths. 
The white circles represent an aperture and background annulus for the same 
physical scale as available for the SMC 
stars. For IRS~15 at a distance of 2.2 kpc, the inner radius is 1\farcm4. 
IRS~15 has been 
claimed before to be a transition disk with circumstellar dust emission. 
This image shows that interstellar dust, likely being heated by IRS~15, 
contributes significantly to any mid-IR flux measurement with a large aperture. The 
dust emission sweeps out into two tails to the southeast and northwest.
}
\label{fig:IRS15}
\end{figure}

\cite{Chini06} have found a source around a B0.5V star, IRS~15, in the Omega Nebula (M17) at a distance of 2.2 kpc \citep{Chini80} 
which they interpret as a transition disk. 
The star lacks H$\alpha$ emission and has a mid-IR disk extending up to $\approx$10,000 AU. Can this be an analog to our dusty SMC sources? 
If so, does it imply a circumstellar origin for the dust? The answer to both questions is ``no''. 
First, the dust is much hotter than in the dusty SMC sources with a color temperature of 220 K. IRS~15 shows significant excess flux 
in the L band ($\sim$3.5~\mum). Second, while the inner regions analyzed by \cite{Chini06} may have a circumstellar origin, as they claim, the 
star is also interacting with its dusty surroundings. We present a three color image of the source in Figure \ref{fig:IRS15} 
drawn from \emph{2MASS} and \emph{WISE} data. Starting at 16\arcsec\ to the south of the star and extending to the southeast, a 
a curved feature resembling a bow shock can be seen in near-IR wavelengths. The feature is not centered on the star, as one would 
expect for a bow shock in a spherical cloud, but also extends up in a second tail to the northwest. 
The star embedded in IRS~15 is the most plausible candidate, based on 
position and color, for heating this dust. At the physical resolution for the SMC, this 
dusty emission with an interstellar origin will contribute and possibly dominate the flux measurement. 

\subsection{Comparison to All Nearby, Dusty OB Stars}
\label{sec:WISE_comp}
As a complement to comparing with catalogs of specific source types, we 
now build an all sky catalog with a better match in selection 
criteria to the SMC sources. This approach has several advantages over the use of literature samples. 
First, even the classification of these literature samples is somewhat 
contentious. For instance, 31 of the 34 stars in the list of debris-disk candidates by \cite{Backm93} are also listed as 
cirrus hot spots by \cite{Gaust93}. Second, the SMC stars have much larger absolute dust luminosities than the 
published local samples. Third, searches for debris disks tend to target stars later than A0. This is 
partly due to the small number of hot stars showing mid-IR excess in the local volume most 
easily surveyed. However, it 
is possible that the dusty OB stars in the SMC are without analogs or underrepresented in the MW. 
If they do exist, the most 
recent all-sky mid-IR catalog 
of \emph{WISE} \citep{Wrigh10}, with its higher resolution and deeper flux limits than \emph{IRAS}, may measure the extent of the 
larger cirrus hot spots and permit classification. The four \emph{WISE} bands are centered at 3.4~\mum, 4.6~\mum, 
12~\mum, and 22~\mum. The spatial full-width-half-maxima are 6\farcs1, 6\farcs4, 6\farcs5, and 12\farcs0, respectively. 
We will refer to the bands as W1, W2, W3, and W4. 

We have assembled a local comparison sample from the \emph{Hipparcos} catalog \citep{Perry97} with 
parallaxes taken from the latest reduction by \cite{vanLe07}. A simple cut was made on the spectral type as listed 
in the original \emph{Hipparcos} catalog such that only O and B stars were selected. 
Stars with measurements of negative parallaxes were rejected as well, 
leaving 10,676 stars. Next, a 1\arcsec~positional match to the \emph{WISE} catalog was made with the additional requirement 
that the fluxes were detected in the W3 and W4 bands at a 3$\sigma$ level or higher. Our primary interest is in W4 at 22~\mum\ for its 
proximity to the MIPS 24~\mum\ band. The S/N cut left 8171 stars, which predominantly lie in the 
MW disk. In the more restrictive O9--B3 range, there are 1750 stars with mid-IR detections. 
A comparable study by \cite{McDon12} matched catalogs including 
\emph{Hipparcos} and \emph{WISE}, but focused on infrared excesses 
from giant stars rather than massive stars on the main sequence. 

\emph{WISE} has a resolution of 12\arcsec~in the W4 band, and the catalog lists a source as extended when the PSF 
model cannot fit the source in any band with a reduced $\chi^2<$3. We estimate that a source size equal to the 
nominal resolution will cause $\chi^2_{\rm r,W4}>3$ and will trigger the extended flag for that band in the 
\emph{WISE} catalog, although there is some variation with S/N 
and with PSF variation across the 
camera field. 

\begin{figure}
\centering
\includegraphics [scale=0.49,trim=0.6in 0in 0in 0in, clip=true,angle=0]{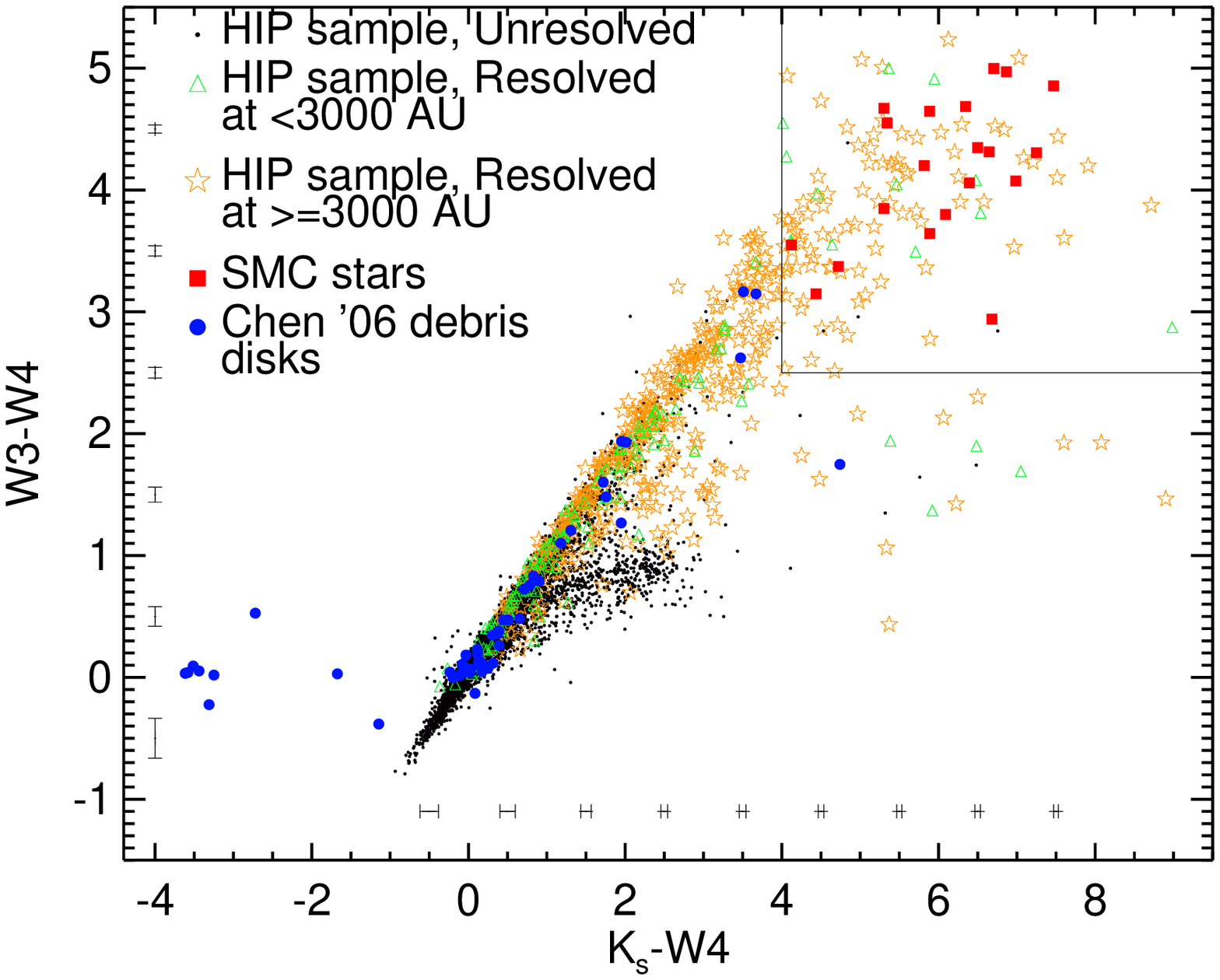}
\caption{Colors for the SMC stars studied here and a local comparison sample from 
\emph{Hipparcos} and \emph{WISE}. We label whether stars are classified as extended in the W4 photometry 
and further break out the \emph{WISE} resolution limit in physical units by considering 
the stars' parallaxes. The box is drawn to encompass the SMC dusty OB stars 
and determines the stars listed in Tables \ref{tab:wise_unres}--\ref{tab:wise_res2}. 
The 3,000 AU cut is made to indicate a conservative size beyond which debris disks are not a 
plausible explanation for the mid-IR excess. Median color errors in steps of 
1 magnitude are shown along each axis. We inspect the objects in the box 
more closely in Section \ref{sec:discuss} and Appendix \ref{ap:list} and find that none with the spectral types of our 
SMC stars are likely to be debris disks. Such extremely red objects are mainly 
comprised of protoplanetary disks and cirrus hot spots. 
}
\label{fig:wisecol}
\end{figure}

The \emph{WISE} catalog has already been cross-matched to 2MASS photometry \citep{Skrut06}, and we use the K$_{\rm s}$ 
fluxes to verify the mid-IR excess in addition to the $W3-W4$ color. 
Figure \ref{fig:wisecol} shows this color-color plot for all 8171 
stars. 
We break the local sample into three subsets: the stars unresolved in W4, the 
stars that are resolved but with a parallax-based resolution limit below 3,000 AU, and resolved stars with 
a resolution limit above 3,000 AU. The cut at 3,000 AU is considered as a conservative size above which debris disks are 
not observed to exist and is justifed in Section \ref{sec:discuss}. 
In fact, debris disks are commonly an order of magnitude smaller \citep[although, in cooler stars;][]{Booth13}, 
and the Solar System's asteroid and Kuiper belts have sizes of $\sim$3 and 40 AU, respectively. The largest 
debris disks known are $\beta$ Pictoris at 550 AU \citep{Backm96}, $\alpha$ Lyrae at 
815 AU \citep{Su05}, and 51 Ophiuchi at 1200 AU \citep{Stark09}. The first two works are based on 
resolved mid-IR images, and the size of 51 Ophiuchi is based on a model to explain the temperature inferred from 
the mid-IR spectra. The largest sizes measure the halos of debris 
disks rather than the site of the parent bodies. 
The sizes of these sources in lower S/N images, such 
as in \emph{WISE}, are substantially smaller. 
We caution that some transition disks, if they exist, may be large enough to violate this distinction. The numerical 
simulations of protoplanetary disks around massive stars \cite[][Figure 9]{Yorke02} create shortlived disks of size 10,000 AU. 
Our choice of the 3,000 AU cut should straddle the sizes of debris disks and the cirrus hot spots. 
\cite{Kalas02} measured stars with coronographic imaging and excluded the central 4\arcsec\ for 
distances of 100--350 pc to find reflection nebulae measured to sizes of 1--2\arcmin. These values translate to $\sim$1000 AU as 
the smallest measureable size in the coronographic sample and cirrus hot spot sizes of $\sim$20,000 AU. 

To identify the closest analogs of the dusty SMC OB stars, we select the 108 stars in Figure \ref{fig:wisecol} with 
$K_s-W4>4$ and $W3-W4>2.5$ and list them by their three subsets in Tables \ref{tab:wise_unres}, 
\ref{tab:wise_res1}, and \ref{tab:wise_res2}. These color cuts were chosen to contain all the dusty SMC 
OB stars included in our study. 
Note that the resolution sizes listed are lower limits to the actual sizes because the surface brightness 
profiles may extend well beyond a resolution element and such direct measurements from the \emph{WISE} 
images are beyond the scope of this study. 

\begin{centering}
\begin{deluxetable*}{crrcrrrrrr}
\tabletypesize{\scriptsize}
\tablecaption{\emph{Hipparcos}/\emph{WISE} OB stars with spatially unresolved mid-IR excess\label{tab:wise_unres}} \tablewidth{0pt}
\tablehead{     \colhead{Hip}     &
                \colhead{$\alpha$}      &
                \colhead{$\delta$}      &
                \colhead{Spectral}      &
                \colhead{$K_s-W4$}      &
                \colhead{$W3-W4$}      &
                \colhead{$\varpi$}      &
                \colhead{$\sigma_{\rm \varpi}$}      &
                \colhead{$\chi^2_{\rm r,W4}$}      &
                \colhead{Resolution}      \\
                \colhead{}          &
                \colhead{(deg,J2000)}          &
                \colhead{(deg,J2000)}          &
                \colhead{Type}          &
                \colhead{(mag)}          &
                \colhead{(mag)}          &
                \colhead{(mas)}          &
                \colhead{(mas)}          &
                \colhead{}          &
                \colhead{(AU)}          }
\startdata
23428 &  75.530807 & -71.336976 & B &  9.98 &  4.35 &  1.31 &  1.35 &      1.5 & $>$9000\\
26062 &  83.378164 &  24.628811 & B8 &  4.54 &  2.85 &  8.78 &  0.61 &      1.0 & 1400\\
77542 & 237.490620 &  -3.921211 & B9 &  4.97 &  2.96 &  8.61 &  0.60 &      1.3 & 1400\\
77716 & 237.999523 &  32.948423 & B2 &  4.84 &  4.38 &  0.10 &  1.64 &      1.1 & $>$7300\\
88496 & 271.062608 & -24.391030 & B2Vne &  6.75 &  2.84 &  1.87 &  1.51 &      1.1 & 6400
\enddata
\tablecomments{Column 4 is the spectral type as taken from \emph{Hipparcos}. Columns 5 and 6 
list the colors taken from \emph{WISE} and \emph{2MASS} with W3 and W4 effective wavelengths 
at 12 and 22~\mum. Column 7 lists the parallax from \emph{Hipparcos} and column 8 lists the error. 
Column 9 lists the reduced $\chi^2$ that the W4 data is compatible with the PSF from the \emph{WISE} 
catalog. Values above 3 are considered to indicate extended emission. Finally, column 10 lists the 
physical size for the 12\arcsec\ W4 resolution. This is the minimum size for extended sources. 
}
\end{deluxetable*}
\end{centering}
\begin{centering}
\begin{deluxetable*}{crrrcrrrrr}
\tabletypesize{\scriptsize}
\tablecaption{\emph{Hipparcos}/\emph{WISE} OB stars with spatially resolved mid-IR excess and size $<$3,000 AU \label{tab:wise_res1}} \tablewidth{0pt}
\tablehead{     \colhead{HIP}     &
                \colhead{$\alpha$}      &
                \colhead{$\delta$}      &
                \colhead{Spectral}      &
                \colhead{$K_s-W4$}      &
                \colhead{$W3-W4$}      &
                \colhead{$\varpi$}      &
                \colhead{$\sigma_{\rm \varpi}$}      &
                \colhead{$\chi^2_{\rm r,W4}$}      &
                \colhead{Resolution}      \\
                \colhead{}          &
                \colhead{(deg,J2000)}          &
                \colhead{(deg,J2000)}          &
                \colhead{Type}          &
                \colhead{(mag)}          &
                \colhead{(mag)}          &
                \colhead{(mas)}          &
                \colhead{(mas)}          &
                \colhead{}          &
                \colhead{(AU)}          }
\startdata
13330 &  42.886541 &  67.815045 & B9V &  6.53 &  3.81 &  4.37 &  0.91 &     80.2 & 2700\\
17465 &  56.142444 &  32.162809 & B5V &  5.94 &  4.91 &  6.58 &  4.09 &    121.9 & 1800\\
19720 &  63.394043 &  10.212458 & B8Vn &  5.47 &  4.05 &  7.56 &  0.81 &     90.4 & 1600\\
24052 &  77.532108 &  53.709788 & B9 &  4.64 &  3.55 &  4.94 &  0.72 &     74.5 & 2400\\
26551 &  84.690112 &  -2.599693 & B0 &  4.01 &  4.55 &  6.38 &  0.90 &    294.1 & 1900\\
26939 &  85.752393 &  -2.312602 & B5V &  5.36 &  5.00 &  5.90 &  1.29 &    219.3 & 2000\\
28711 &  90.958817 &  30.169222 & O9V &  5.71 &  3.50 &  9.18 &  2.85 &     10.0 & 1300\\
31042 &  97.707556 &  -9.654110 & B8 &  6.48 &  4.08 &  4.75 &  1.79 &     48.4 & 2500\\
36369 & 112.294862 &  20.911802 & O6 & 10.56 &  4.21 &  5.60 &  3.75 &     56.7 & 2100\\
47078 & 143.916789 & -50.223235 & B5 &  4.13 &  3.57 &  4.32 &  1.74 &      3.5 & 2800\\
56379 & 173.356003 & -70.194789 & B9Vne &  8.98 &  2.87 & 10.32 &  0.43 &   1191.0 & 1200\\
62913 & 193.406739 & -60.357061 & B3Ib: &  4.05 &  4.27 &  6.03 &  3.09 &    127.7 & 2000\\
83509 & 256.005176 & -51.083654 & B2V &  4.12 &  3.59 &  0.54 &  6.18 &     37.2 & $>$1900\\
110119 & 334.604354 &  63.220020 & B8 &  4.46 &  3.97 & 11.05 &  7.21 &     72.7 & 1100
\enddata
\tablecomments{Column 4 is the spectral type as taken from \emph{Hipparcos}. Columns 5 and 6 
list the colors taken from \emph{WISE} and \emph{2MASS} with W3 and W4 effective wavelengths 
at 12 and 22~\mum. Column 7 lists the parallax from \emph{Hipparcos} and column 8 lists the error. 
Column 9 lists the reduced $\chi^2$ that the W4 data is compatible with the PSF from the \emph{WISE} 
catalog. Values above 3 are considered to indicate extended emission. Finally, column 10 lists the 
physical size for the 12\arcsec\ W4 resolution. This is the minimum size for extended sources. 
}
\end{deluxetable*}
\end{centering}

\section{Discussion}\label{sec:discuss}
With the data and frameworks discussed, we now analyze the evidence for accepting or rejecting 
particular models. First, we examine the SED shapes. Second, we review the 
size distributions of debris disks and static hot spots and present a size cut which reliably 
separates the two populations. Third, we compare the derived properties to those inferred for 
debris disks with similar data. Fourth, we search the literature and utilize the 
size cut to classify nearby sources that serves as analog to the dusty OB SMC stars. Fifth, 
we address the frequency of the near-IR excess pattern in SMC OB stars as compared to the frequencies 
and lifetimes of the models. Finally, we review bow shock models and their gas signature. 

We examine whether the generic properties of the SEDs can select between the 
debris disk and cirrus hot spot models. Cirrus hot-spot models with clouds of constant density that span 
a range of temperatures \citep{Kalas02} 
can produce SEDs that rise in flux density through 100~\mum. 
The dust temperatures in cirrus hot spots are controlled by the 
central star's mass and luminosity, by the distance at which the dust 
particles reside, and the size distribution of grains. 
The radiative force on small grains around OB stars is large enough to 
eventually remove the grains from orbit. We have presented static models 
(Table \ref{tab:sedprop2}) for small ISM grains at scales of $\approx$100,000 AU that have the 
observed dust temperatures. The larger scale environment may be important 
enough at these distances that the steady-state blowout solution need not 
apply. A second type of cirrus hot spot can occur when the star has a 
large velocity relative to the ISM, and the ISM emission is enhanced. 
This permits small grains to stay at a fixed distance from the star and 
reach hotter temperatures than the static model. We have shown how 
such interaction hot spot models can also explain the dusty SMC stars' 
SEDs. 
Additionally, there are debris disks \citep[e.g.][]{Hille08} where 
the SEDs continue to rise
beyond 70~\mum. A sample of A-star debris disks observed 
with \emph{Herschel} for the DEBRIS and DUNES surveys 
finds dust SEDs that generally 
peak around 70~\mum\ \citep{Gaspa13,Booth13,Eiroa13}. 
Most of our data are consistent with a downturn in the 
SED before 70~\mum. However, our \emph{Herschel} error bars are too large to 
strongly rule out a longer wavelength SED peak. 
Both the uncertainties in our data and the large range of permissible dust 
temperatures in the debris disk and cirrus hot spot models nullify 
the SED shape as a classification tool between these source types. 

We have no local examples of
debris disks around O9--B2 stars, so we do not know exactly what their properties 
would be. As discussed in Section \ref{sec:proto}, 
protoplanetary disks around massive stars may be destroyed so violently such 
that 
the planetesimals that 
seed debris disk growth are not formed or retained. We also note that 
the timescale for the sublimation 
of icy grains \citep{Jura98} is much shorter than the collisional 
timescale in massive stars out to several 
thousand AU \citep[as can be derived using the equations and 
constants in][]{Chen06}, which encompasses the radii where debris disk 
parent bodies could possibly reside and which may inhibit the growth 
of debris disks around OB stars. 
One set of models for collisional cascades 
\citep[e.g.][]{Kenyo02,Kenyo04,Kenyo08,Kenyo10} starts with a disk of icy planetesimals from 30--125 AU for 1--3 M$_{\odot}$ 
stars. In these simulations, bright rings manifest where the planetesimals are grinding into dust which move 
outward with time. We know of no collisional cascade simulations for the masses of stars we study here, so 
we extrapolate relations that have been developed for stars and disks in lower mass ranges. We stress that such 
extrapolations are highly uncertain, and we only use them to form a conservative classification rule for the 
maximum extent of debris disks. Equation 6 of \cite{Kenyo10} is:
\footnote{We note that there is a typographical error in their Equation 6 where the sign preceeding 3/2 
in the exponent to $a$ should be positive rather than negative.}
\begin{equation}
\label{eq:tevol1}
t_{\rm gro}\propto x_{\rm m}^{\rm-\gamma-1} a^{\rm n+3/2},
\end{equation}
where $t_{\rm gro}$ is a growth time, $x_{\rm m}$ is a scaling for the mass initially 
in the disk, $\gamma$ is a term for the influence of gas drag on the dynamical friction, $a$ is 
a radius along the disk, and $n$ is the power law exponent for the initial disk mass. This equation is 
valid for a fixed central star mass. Adding the star's mass as an additional variable and assuming that the 
disk surface density is proportional to the stellar mass introduces the dependence that $t_{\rm gro} \propto M_{\rm *}^{-3/2}$. 
\cite{Kenyo08} present a version of this in their Equation 41. 
\cite{Kenyo10} estimate $\gamma\approx$~0.1--0.2 and $n=1$ or 1.5. We use $n=1.5$ here. To provide a 
normalization to the scaling, we use the result from \cite{Kenyo04} that a solar mass star forms a ring at 30 AU 
in $\sim$10--20 Myr. A similar value can be obtained from the time to form the first Pluto-sized object 
\citep[Equation 28 in][]{Kenyo08}. 
At this fixed time, which is approximately the main sequence lifetime for our massive 
stars of M$_{*} \approx 20$M$_{\odot}$, the ring will be at a $\approx4.5\times$ larger radius. Similar 
scaling laws can be found in \cite{Lohne08}. However, a distinction must be made between the 
original site of planetesimal grinding and the sites where small grains primarily emit. When a 
collision fragments a large particle into dust grains, the smaller grains will suddenly 
react to the radiation pressure. This effect pushes particles originally on circular orbits 
into highly eccentric and even unbound orbits. The smaller particles will be, on average, 
ejected onto more eccentric orbits, attain cooler temperatures at larger distances, and 
cause color gradients. These effects have been modeled for Vega by \cite{Mulle10} with a 
ring of source material from 80--120 AU. From their baseline model in Figure 6, 
the FWHM for the surface brightness profiles is broadened to $\approx$150, 200, and 
230 AU for the MIPS 24\mum, 70\mum, and 160\mum\ bands. The halo of emitting grains may 
plausibly extent out from the collisional region by a factor of a few, and we would thus expect the 
maximum extent of debris disks around OB stars to be measureable to $\approx350$ AU. As a conservative cut, 
we consider sources with mid-IR extent beyond 3,000 AU, in Section \ref{sec:WISE_comp}, to exclude a 
debris disk classification. 

We scrutinize the derived dust properties against those in the literature for 
cooler stars in Figures \ref{fig:debcomp1}--\ref{fig:debcomp4}. The dust temperatures of 
the SMC OB stars are consistent with the extrapolation 
from known debris disks, and the relative IR luminosities are within the upper envelope, but 
above an extrapolation, of the debris disk samples. The offset in relative IR luminosity 
is likely a selection effect for the SMC stars. The dust masses and radii, however, are several orders of 
magnitude larger for the dusty OB stars than for any local counterparts. 
We do not see trends in dust mass or radius for the most massive literature stars 
that would anticipate the larger values in our SMC stars. 
The debris disk hypothesis for the SMC stars cannot be strongly rejected from the 
literature samples of disks around cooler stars, but neither are the data 
strongly supportive of such a model. 

Next, we classify the local sources showing mid-IR excesses introduced in Section \ref{sec:WISE_comp}. The catalog match between 
\emph{Hipparcos} and \emph{WISE} produces 8171 OB stars with mid-IR detections and 1750 stars in the
range O9--B3. Based on the previous timescale calculations and the large parallax errors in many stars, we consider 
the nature of the \emph{WISE} sources with sizes less than 3,000 AU to be uncertain 
and a possible mix of cirrus hot spots, debris disks, and protoplanetary disks. When the reported parallaxes are 
below a 1$\sigma$ significance, we quote the resolvable sizes at the 1$\sigma$ limit. 
The extended objects resolved to $>$3,000 AU are considered to be reliably classified cirrus hot spots. 
For consistency with the local sample, we have also computed synthetic photometry to the best-fitting 
stellar and dust models for our SMC stars as described in Section \ref{sec:sed}. The \emph{WISE} W3 and W4 bands are near the 
fourth IRAC and first MIPS bands, so the color correction from our actual \emph{Spitzer} measurements is small. 
It is clear from Figure \ref{fig:wisecol} that the SMC stars have 
extreme colors in both $K_s-W4$ and 
$W3-W4$, and only the tail of the local distribution matches them. The reddest star in the cirrus hot spot sample of 
\cite{Gaust93} has $W3-W4=4.05$, while the vast majority lie at $W3-W4<2$. The debris-disk samples of \cite{Chen06} and 
\cite{Carpe09} also do not match the SMC colors and all lie at $W3-W4<3$. The debris-disk stars fail to match the 
dusty OB star colors both because the OB stars have, on average, higher fractional IR luminosities and 
because the hotter photospheres emit less of their fractional luminosity 
in K$_{\rm s}$ and W3. For example, $\beta$ Pictoris has $W3-W4=-0.05$ and 
${\rm K_{s}-W4}=3.47$, and its SED, in Figure 1 of \cite{Chen07}, can 
be directly compared to our dusty OB stars in Figure \ref{fig:sed1}.  

We classify the 108 reddest local stars in \emph{WISE} from Tables \ref{tab:wise_unres}, \ref{tab:wise_res1}, and \ref{tab:wise_res2} in the 
Appendix \ref{ap:list}. To summarize, none of the 59 stars in the range O9--B3 are likely to host debris disks. We find 
that many of the local sources contain emission lines, and all the others have sizes characteristic of cirrus hot spots. 
Transition disks complicate this conclusion, and until more is known about them in the environment of massive stars, 
we cannot exclude them as being present in our sample of local analogs. The 
known transition disks around hot stars are either too small or two 
hot in their dust emission to match the SMC observations, but 
the small sample sizes for such transition disks preclude a 
firm conclusion. We find only one source with color $W3-W4>2.5$ that may plausibly be a debris disk host by its angular size. 
That source is a B8 star (HIP~26062), and so not a good analog to our SMC sample. Of course, the stars with 
cirrus hot spots may also contain debris disks, but their mid-IR fluxes 
are likely dominated by the hot spots. If O9--B3 stars do host observable debris disks, they may lie at $W3-W4<2.5$.

\begin{figure}
\centering
\includegraphics [scale=0.45,angle=0]{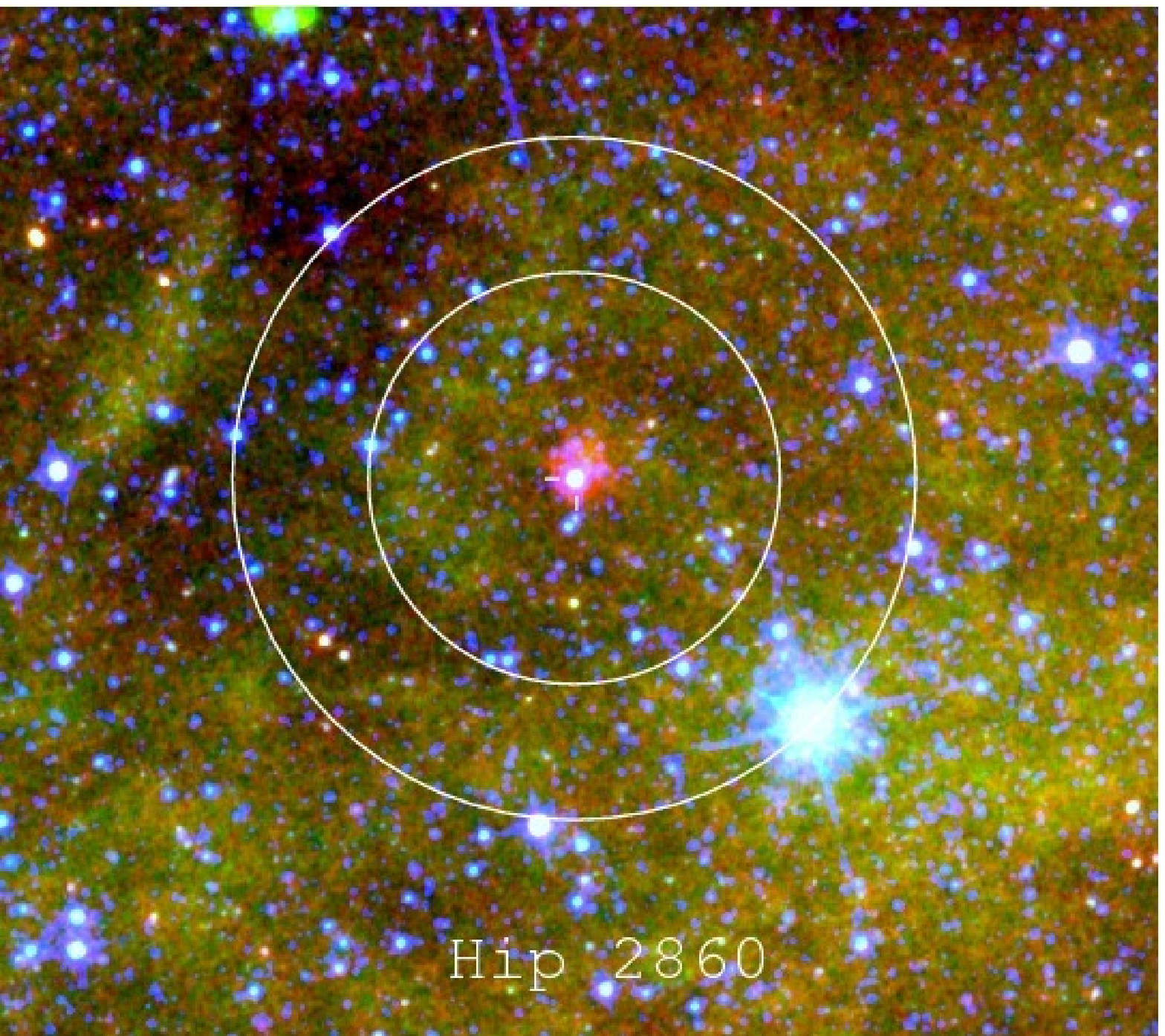}
\caption{A three-color image of HIP~2860 with the \emph{WISE} W2, W3, and W4 
bands as blue, green, and red. This star is not selected as red enough in 
$W3-W4$ to match our dusty SMC stars under the \emph{WISE} PSF-photometry. The 
crosshair lines are drawn with 30\arcsec\ lengths. The white circles represent an 
aperture and background annulus for the same physical scale as available for the SMC 
stars. For HIP~2860 at a distance of 450 pc, the inner radius is 6\farcm7. Even the 
largest aperture fluxes available in the \emph{WISE} catalog do not capture the 
extended W4 flux. The W4 flux extends out to 28,000 AU. This aperture effect 
explains why we find a higher fraction of OB stars with large $W3-W4$ colors in 
the SMC than the MW. HIP~2860 is a likely analog to our SMC sample. 
}
\label{fig:W3col1}
\end{figure}

The fraction of hot stars in our color cut (1\%, 108 of 10,676 stars) is well below the 3\% mid-IR excess 
fraction \citep{Bolat07} that is found in the 
SMC sample. At first glance, this appears as a surprising result. Is the fraction of hot stars with a 
mid-IR excess intrinsically larger in the SMC than the MW? Such a result could indicate that PAHs are 
deficient in the SMC. However, the physical scales probed in the two 
catalogs are vastly different. Consider the 6\arcsec\ FWHM resolution of our 24~\mum\ data, which 
for the SMC measures a physical radius of 183,000 AU. We have measured photometry from 
the \emph{WISE} images for several hot stars with matched physical apertures, and we find that 
the colors are commonly redder in $W3-W4$ for large apertures. One randomly chosen representative 
is HIP~2860 (B2V, $\varpi=2.21$ mas), which has no indication in the literature of hosting a mid-IR excess. 
We show a three-color image for HIP~2860 using 
\emph{WISE} W2, W3, and W4 data in Figure \ref{fig:W3col1} with aperture photometry out to a radius of ${\rm r}=6\farcm85$. The 
images have not been matched in resolution in Figure \ref{fig:W3col1}, but the resolution mismatch is unimportant for the 
large-aperture photometry that we discuss. 
The \emph{WISE} catalog reports that the spatial profile of the star is consistent with the PSF and has a $W3-W4=1.56$ color from the 
PSF-photometry. However, the SMC-matched aperture has a $W3-W4=3.29$ color, and the star is clearly extended in the reddest 
channel in Figure \ref{fig:W3col1}. This star was not considered an analog to our SMC stars by the color cut of Figure \ref{fig:wisecol}, but 
it becomes one with the physically matched aperture photometry. A full analysis of the \emph{WISE} images with matched apertures would 
reveal many more cirrus hot spot analogs. Our intent is only to find debris-disk analogs, and the color cuts of Figure 
\ref{fig:wisecol} will suffice to find them since debris disks will be unresolved or barely resolved in 
\emph{WISE} for the stars under study. To conclude our question, the color corrections for 
extended sources would reveal many more local, dusty OB stars, and the frequency by which OB stars appear 
dusty in the SMC is unlikely to be enhanced over the frequency in the MW. 

Paper I has posited bow shocks around runaway stars as an additional mechanism that 
may be causing the mid-IR emission. Normally, such stars would have strong optical emission lines, and 
Paper I presents 17 stars with clear H$\alpha$ emission and 7 additional candidates for weak emission 
from the 125 stars with spectra taken. The subsample we analyze here is instead drawn from the 88 
main-sequence stars without emission lines. However, Paper I finds two runaway star candidates without 
emission lines from the large radial velocity offsets between the photospheric lines and local HI in the 
literature. 
Even for the classic examples of bow shock stars, the presence of H$\alpha$ emission is mixed. 
$\delta$ Velorum does not show H$\alpha$ in emission \citep{Aleks96}. $\zeta$ Ophiuchi shows 
H$\alpha$ in emission \citep{Touha10} and is classified as a Be star. The Mira-type star R Hydrae does show 
higher order Balmer emission \citep{Merri15} and presumably H$\alpha$ emission, but spectra have not been 
published covering H$\alpha$. \cite{Kobul12} have recently discovered a bow shock around a B1.5V star, and 
the stellar spectrum is without H$\alpha$ emission. Balmer emission lines cannot be reliably 
used to determine the presence or absence of bow shocks around runaway stars. 

Since our search of 
the \emph{WISE} catalog has revealed a fair number 
of analog stars with distorted, extended morphologies, it is plausible that some of our 
SMC stars are manifesting bow shock features in the mid-IR caused by their large velocities 
relative to the local ISM. Higher resolution optical spectroscopy could resolve this interpretation 
with precise radial velocities. Higher resolution imaging might also find conclusive morphological 
features. 

\section{Summary \& Conclusions}
\label{sec:conc}
We have studied the properties of dusty OB stars in the SMC with new mid-IR and far-IR data. Our 
main goal is to determine the nature of the mid-IR excesses. The two most probable 
hypotheses are that they are debris disks or cirrus hot spots. The cirrus 
hot spots may be either due to nearly static ISM overdensities, like 
in the Pleiades, or from regions compressed in bow shocks around runaway stars. 
Classical and Herbig Be stars are excluded by the lack of 
H$\alpha$ emission presented in Paper I. 
Transition disks, a cousin class to debris disks, cannot be rigorously 
rejected but are lacking in compelling analogs. The largest nearby transition disk 
has hotter dust than the sources in the SMC and also is dominated by emission 
from an interaction with the ISM. Without 
information on the angular extent of the dust in the SMC OB stars, the relevant properties 
can only be derived in model-dependent ways. 

Of the twenty stars analyzed with new IRS peak-up images at 16~\mum\ (${\rm FWHM}=3\farcs8$), three are 
significantly extended. This places the dust at scales of $>$10$^5$ AU from the central stars and 
precludes them being debris disks. The angular-resolution data for the remaining seventeen stars are 
inconclusive. We have measured and fit the spectral energy distributions of eleven main-sequence OB stars 
with extant \emph{Spitzer} photometry, 
new \emph{Spitzer} IRS spectroscopy, and limits from \emph{Herschel} photometry. A further 
nine are analyzed with \emph{Spitzer} imaging and spectroscopy but no \emph{Herschel} photometry. 
The mid-IR spectra are featureless. We infer temperatures and dust masses for these cases. 
The dust temperatures do not constrain the mechanism for the mid-IR excess flux, 
but the large dust masses required by the modeling assumptions suggest that 
the sources are not debris disks. The evidence is 
only circumstantial because nearby debris disks around stars as massive as the SMC OB stars 
have not been studied. Such massive disks may exist, but they would require an abrupt jump in 
dust mass compared to debris disks around A stars and later stellar types. The larger 
dust masses are compatible with transition disks.  

The best discriminant, resolving the mid-IR region sizes, remains impossible 
for these distant stars should any host debris or transition disks. 
Instead, we make a search of nearby \emph{Hipparcos} stars 
that can be resolved at meaningful sizes by the all-sky \emph{WISE} survey. We note that the implied colors 
of our SMC stars are unusually red in the \emph{WISE} $W3-W4$ system (12~\mum\ and 22~\mum). We find that only a small 
number of hot main sequence stars in the MW show such strong mid-IR excess (1\% for $W3-W4>2.5$), and that nearly all these 
detections are explained as cirrus hot spots, bow shocks, and protoplanetary disks around young stars 
based on their mid-IR sizes and morphologies. We find only one instance of a plausible debris disk system in this color range, but 
it has a cooler central star (B8) than our SMC sample. From this comparison, the SMC stars under study are most likely to not 
host debris disks. We cannot rule out that transition disks around hot stars may have sizes large enough to 
be present in our MW sample and represent some of our SMC stars, but the current examples are either 
too small or too hot in their dust emission. By concluding that cirrus hot spots make for the best model, 
these dusty OB stars may be useful as a bright target catalog for studying the diffuse interstellar dust of the SMC.  
\acknowledgments
We thank an anonymous referee for a report that significantly improved this work. 
This work is based on observations made with the \emph{Spitzer Space Telescope}, which is 
operated by the Jet Propulsion Laboratory, California Institute of Technology under a contract with NASA. 
Support for this work was provided by NASA through awards issued by JPL/Caltech. 
A.~D.~B.~wishes to acknowledge partial support from a CAREER grant NSF-AST0955836, 
JPL-1433884, and from a Research Corporation for Science Advancement Cottrell Scholar award. K.~S.~ is 
supported by a Marie Curie International Incoming Fellowship. This publication makes use of data products from 
the Wide-field Infrared Survey Explorer, which is a joint project of 
the University of California, Los Angeles, and the Jet Propulsion Laboratory/California Institute of Technology, 
funded by the National Aeronautics and Space Administration. This research has made use of the NASA/IPAC 
Extragalactic Database (NED), which is operated by the 
Jet Propulsion Laboratory, California Institute of Technology, under contract with the National 
Aeronautics and Space Administration, the SIMBAD database, operated at CDS, Strasbourg, France, 
and NASA's Astrophysics Data System Bibliographic Services. 
\emph{Facilities:} \facility{Spitzer Space Telescope}, \facility{Herschel Space Telescope}.

\appendix
\section{Modeling of debris disks and static cirrus hot spots}
\label{ap:eqn}
\subsection{Debris Disk Model}
The mid-IR SEDs of debris disks can be fit with a blackbody (BB) function at a single temperature. 
The radiative force from the central star, as we will discuss, blows out small dust grains in debris 
disks. When the grain sizes are all larger than the observed wavelengths, the grains are 
efficient at both absorbing and emitting and a BB function is appropriate. We have derived the dust temperatures 
in Column 2 of Table \ref{tab:sedprop2} from such fits. 

The thermal equilibrium distance for a single temperature blackbody is given by: 
\begin{equation}
\label{eq:bbdist}
R_{\rm BB}=\frac{1}{2}\left(\frac{T_{*}}{T_{\rm dust}}\right)^2 R_{*}, 
\end{equation}
where $T_{*}$ is the stellar photosphere temperature, $T_{d}$ is the 
dust temperature, and $R_{*}$ is the stellar radius. 
Debris-disk sizes, when resolved, are at most a factor of 3$\times$ larger than 
the blackbody distance estimate \citep{Smith10}.

A simple balance between the forces of gravity and radiation from the central star create the ratio, 
\begin{equation}
\label{eq:amin}
\beta(a)=0.57Q_{\rm pr}(a)\frac{L_{*}}{a M_{*} \delta}
\end{equation}
which forms a stability criterion for a grain with radius $a$ \citep{Burns79,Artym88}. 
We set the radiation coupling coefficient, $Q_{\rm pr}(a)$, to unity since
the geometric limit (2$\pi$a/$\lambda\gg$1) applies for the hot spectra and large grains. The units of $a$ are \mum,
L$_{*}$ is in $L_{\rm \odot}$, M$_{*}$ is in $M_{\rm \odot}$, and $\delta$ is in g cm$^{-3}$. For sizes below the 
minimum grain size, 
the radiation force will overwhelm the gravitational force and eject the small grains. Nominally, 
this condition is $\beta=1$ for which we evaluate $a_{\rm min}$ as given in Column 7 of Table \ref{tab:sedprop1}, although grains can 
still be ejected with $\beta=0.5$ when on elliptical orbits \citep{Burns79}.

A minimum mass estimator commonly used for debris disks can be 
made by assuming that the mid-IR emission comes from a ring or shell of grains of size $a_{\rm min}$ at the 
thermal equilibrium distance \citep{Artym88,Jura95}.
With $a_{\rm min}$ as the smallest dust grain size in cm, $R$ the 
grain distance from the star in cm, $F_{\rm IR}/F_{\rm bol}$ the flux of the dust relative to the star, 
$\delta$ the dust density in g cm$^{-3}$, and $M_{\rm min}$ the dust mass in g:
\begin{equation}
\label{eq:mass}
M_{\rm min}=\frac{16\pi F_{\rm IR} \delta a_{\rm min} R^2}{3 F_{\rm bol}}.
\end{equation}
We use $\delta=3.3$ g cm$^{-3}$ \citep{Drain84}. 
To estimate the dust's infrared luminosity, we numerically integrate the BB 
function from 0.7-100~\mum. We note that this equation is valid for a thin shell 
of dust with spherical particles and may not strictly apply for younger, extended debris disks. 
A slightly better approximation can be made considering 
the full grain size distribution above $a_{\rm min}$ such as in \cite{Chen05}. 
While the grain size distribution in the ISM follows a power law with exponent 
${\rm p}=-3.5$ \citep{Mathi77}, the value for debris disks is slightly steeper at 
${\rm p}=-3.6$ to $-3.7$ \citep{OBrie03,Wyatt11,Gaspa12b}. The upper limit to the 
grain size may be 
considered as infinity, or more usually set to $\approx$1~mm. 
The minimum grain size is already so large in OB stars ($a_{\rm min}\approx$0.5~mm) 
that this averaging makes little difference. The average grain size (and 
mass estimate) for the two upper limits considered will be 1.59$\times$ or 
1.30$\times$ larger, assuming ${\rm p}=-3.7$. We present mass estimates based solely on 
$a_{\rm min}$, but we note that the actual values may be larger. This 
mass estimate is given in Column 5 of Table \ref{tab:sedprop1}. 

\subsection{Static Hot Spot Model}
The MRN model \citep{Mathi77} for the ISM has a power law form for the dust grain number density as a 
function of size with a $p=-3.5$ exponent and a grain size range of 0.005--0.25~\mum. These small grains, 
when in the presence of a hot star, will efficiently absorb radiation but inefficiently radiate it. This 
leads to a MBB form in the SED. We use an emissivity index of 
$\beta_{\rm em}=2$ and the dust mass absorption coefficient evaluated as:
\begin{equation}
\label{eq:kappa}
\kappa_{\rm \lambda}(\lambda)=10 \left(\frac{250}{\lambda}\right)^2, 
\end{equation}
where $\lambda$ is in \mum\ and $\kappa_{\rm \lambda}$ is in cm$^2$ g$^{-1}$. This value was 
originally proposed by \cite{Hilde83} and is used as a standard value in the \emph{Herschel} Gould Belt Survey 
\citep{Andre10,Arzou11,Sadav12}. Values with higher and lower normalizations by a factor of two 
are used in some other works \citep{Wyatt08,Decin11}, and the value will depend on the grain size, composition, 
and wavelength range.
 
We compute a dust mass as simply estimated from: 
\begin{equation}
\label{eq:emdust}
M_{\rm dust}=\frac{I_{\rm \lambda}D^2}{\kappa_{\rm \lambda}B_{\rm \lambda}},
\end{equation}
where $I_{\rm \lambda}$ is the observed flux density, $\kappa_{\rm \lambda}$ is the 
dust absorption coefficient, $B_{\rm \lambda}$ is the blackbody function, and $D$ is the distance to the SMC. 
$M_{\rm dust}$ will be in units of g if $I_{\rm \lambda}$ and $B_{\rm \lambda}$ are 
given in the same units, D is in cm, and $\kappa_{\rm \lambda}$ is in cm$^2$ g$^{-1}$. 
The equation is from \cite{Hilde83}, and is valid under several assumptions. The dust must be
isothermal, and the dust grains must be small relative to the observed wavelengths and    
in the Rayleigh-Jeans regime. For the small grains present in the ISM, these assumptions are valid. 
This mass estimate is listed in column 10 of Table \ref{tab:sedprop2}, and 
implicitly assumes a modified blackbody function due to our chosen form of the dust opacity. 

Once the temperature is derived, the distance for the grains can be found by \citep{Backm93,Akeso09}:
\begin{equation}
R_{\rm MBB}=636^3\frac{\sqrt{L_{*}}}{\lambda_0T^3_{dust}}, 
\label{eq:mbbdist}
\end{equation}
where L$_*$ is the stellar luminosity in units of L$_{\odot}$, $R_{\rm MBB}$ in AU, $\lambda_0$ is the critical 
wavelength and scales with the size of the grain particles in \mum, and 
T$_{\rm dust}$ is the dust temperature in K. We assume $\lambda_0=1$\mum, 
although this value is very unconstrained, for the 
size estimate in column 9 of Table \ref{tab:sedprop2}. We have evaluated the coefficient 
in Equation \ref{eq:mbbdist} from the Equation 2 in \cite{Backm93}\footnote{We note that 
there is a typographical error where the summation term in the denominator should have 
an exponent of -(4+q) instead of -(4+p) in Equation 2 of \cite{Backm93}.} for the $p=0$ and $q=2$ case of an efficient 
absorber and inefficient emitting material. This equation is valid for small grains such that they will 
efficiently absorb the photons from the central star but will inefficiently radiate at the mid-IR 
wavelengths where they are measured. 

\section{Discussion of particular mid-IR excess stars in WISE}
\label{ap:list}
\subsection{Stars with unresolved mid-IR photometry}
Only five stars meet the color cuts stated above and are unresolved in the W4 band. 
\begin{description}
\item[HIP~23428] is the well studied luminous blue variable R71 in the Large 
Magellanic Cloud \citep{Boyer10}. It has a parallax detection 
below 1$\sigma$ significance and is not truly a
``local'' analog.
\item[HIP~26062] has a well determined parallax and is a debris disk candidate.
\item[HIP~77542] (HD~141569) is an extensively studied Herbig Ae/Be star with a
circumstellar disk \citep[e.g.][]{Sylve96a}.
\item[HIP~77716] (BD+33 2642) is a well-known post-AGB star \citep{vanWi97} and common photometric
standard \citep{Stone96}. Its \emph{Hipparcos} parallax has a large uncertainty.
\item[HIP~88496] is an emission-line star.
\end{description}

None of these are likely to be analogs to our 
SMC sample because they are either types that Paper I excluded through optical 
spectroscopy, or are cooler than the 09--B3 stars in our SMC sample. 

\subsection{Stars with resolved mid-IR photometry and uncertain classification}
\label{sec:WISE_res}
Fourteen MW stars are resolved by \emph{WISE}, but at sizes that do not cleanly distinguish their nature. 
\begin{description}
\item[HIP~13330] is
a known reflection nebula \citep{Magak03} with an infrared excess \citep{Oudma92}. The W3 and W4 images
clearly show an elongated source that is much larger than 3,000 AU.
\item[HIP~17465] has several
classifications in the literature, including as a debris disk, but \cite{Rebul07} have shown that
spherical shells of extended IR emission stem from ISM interaction. All four \emph{WISE} images also show a clear
bow shock.
\item[HIP~19720] has been flagged as a known reflection nebula by several authors as well as being an
infrared excess source \citep[e.g][]{Oudma92}. \cite{Kalas02} have classified the nebula as due to
ISM heating.
\item[HIP~24052] shows asymmetric emission with hints of a bow shock in the \emph{WISE} W3 and W4 images. It
is likely to be a cirrus hot spot.
\item[HIP~26551] ($\sigma$ Ori) is near, but offset from, a large patch of dust \citep{vanLo03}.
The gas shows strong optical emission lines and is likely a protoplanetary disk around a binary M1 star and a brown
dwarf \citep{Hodap09}. The disk is being photoevaporated by the more massive, nearby binary (HIP~26551) but 
is otherwise unrelated. 
\item[HIP~26939] is an emission-line star \citep{Weave04}, and \cite{Ponto10} classify the
mid-IR excess as from a protoplanetary disk.
\item[HIP~28711] has little information in the literature, but
appears to be far too faint for its spectral class to lie at its trigonometric parallax distance \citep{Tsvet08}.
However, its infrared excess does not come from a debris disk.
The \emph{WISE} catalog lists the W4 detection as being extended, and we find it to
be extended to 6400 AU FWHM assuming the trigonometric parallax distance.
It also lies near a large H\,{\sc ii} region and is
another example of a cirrus hot spot.
\item[HIP~31042] is a Herbig Ae/Be star \citep{vande98} hosting a reflection nebula \citep{Magak03}.
\item[HIP~47078] is a double star with the mid-IR emission extending over the optical
image of both stars.
\item[HIP~36369] is a planetary nebula (NGC~2392). 
\item[HIP~56379] is a Herbig Be star with a protoplanetary disk \citep{Goto12}.
\item[HIP~62913] is blended with the strong signal from nearby V* DU Cru (M2Iab) and may not truly have a mid-IR excess.
\item[HIP~83509] has been identified as a runaway star by \cite{Mdzin05}. It has an uncertain parallax, but the extent 
of the W4 emission is $>$ 71,000 AU. Beyond the main emission surrounding the star, there is a possible bow-shock to the 
north-east. 
\item[HIP~110119] is a runaway star \citep{Tetzl11}. It is in a double system with HD~211880. Both are 
classified as \emph{WISE} sources, but they blend in the W4 band with HD~211880 being dominant. 
\end{description}

Of the three main-sequence stars here within 09--B3 and that may be analogs to our sample, 
one is likely to be a cirrus hot spot, the second a protoplanetary disk, and the third a bow-shock. 

\begin{figure}
\centering
\includegraphics [scale=0.45,angle=0]{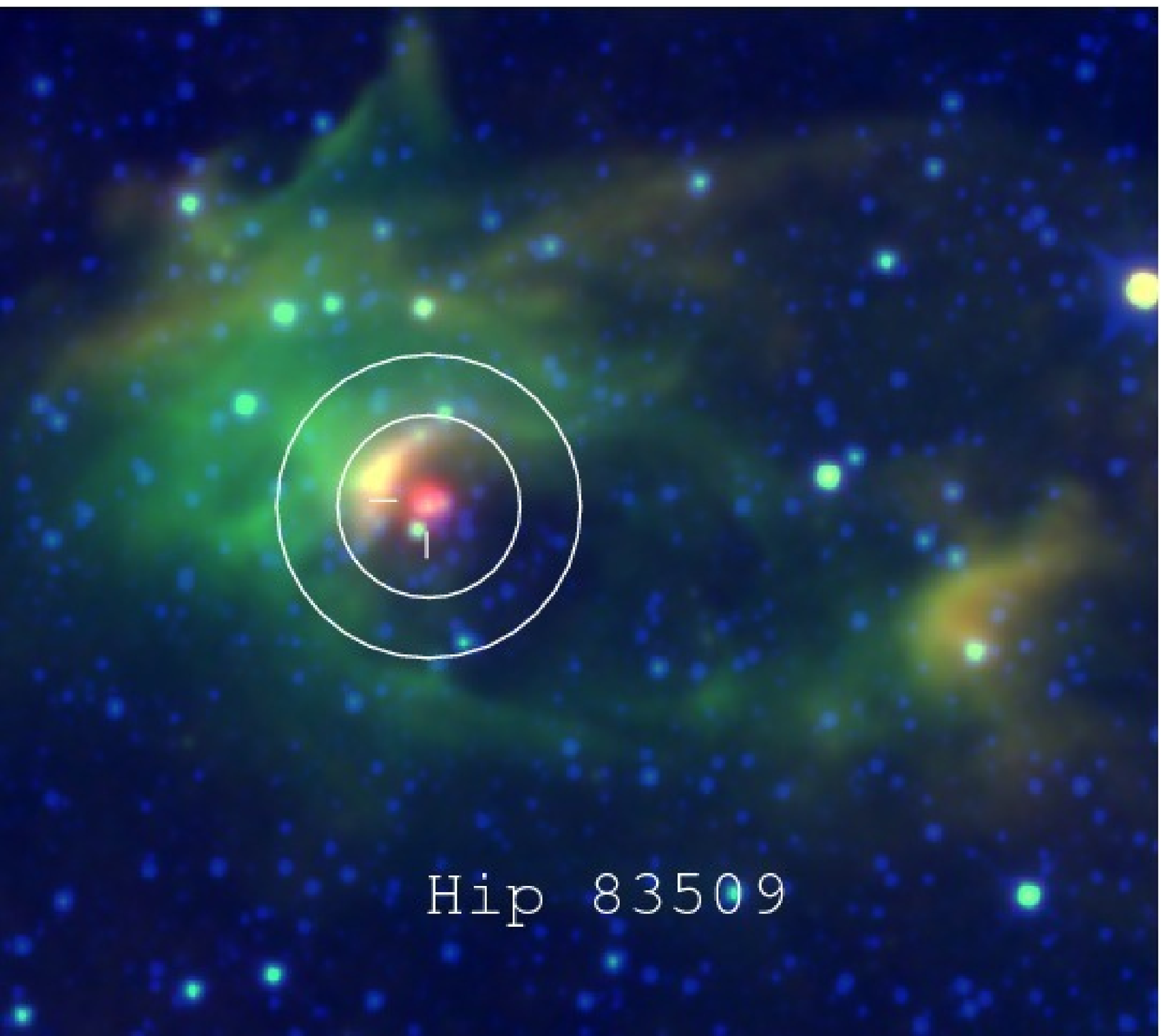}
\caption{A three-color image of HIP~83509 with the \emph{WISE} W2, W3, and W4 
bands as blue, green, and red. The crosshair lines are drawn with 30\arcsec\ lengths. 
The white circles represent an aperture and background annulus for the same 
physical scale as available for the SMC 
stars. For HIP~83509 at a distance of 1.9 kpc, the inner radius is 1\farcm6. 
We note that this star has a highly uncertain parallax (Table \ref{tab:wise_res2}). The star is 
already selected as a cirrus hot spot by its large extent in W4, but examination of the bow shock to the 
northeast and the gap to the southwest of potentially cleared gas reinforces the classification. 
}
\label{fig:W3col3}
\end{figure}
\subsection{Stars with cirrus hot spots}
Eighty-nine MW stars are resolved by \emph{WISE} with sizes larger than 3,000 AU. Fifty-three of these are or may be 
within 09--B3. We have visually assessed their morphology from the \emph{WISE} imaging and searched the literature 
for classifications. Many can be classified as one of the follwoing: obvious bow shocks, 
somewhat asymmetrical shapes around the stars (possible bow shocks), blends with other sources, or regular and 
round. Five (HIP~52628, HIP~56833, HIP~82936, HIP~100193, and HIP~100628) do 
not appear as bright sources in the W4 imaging, despite the 
significant values listed in the \emph{WISE} all-sky survey catalog, 
and may be spurious detections. None are likely to be debris disks based on 
their large W4 sizes, and most can be identified as ISM interactions by their 
morphology. We show one example of a bow shock (HIP~83509) in Figure \ref{fig:W3col3}. The star has already 
been selected by our $W3-W4>2.5$ color cut from the \emph{WISE} PSF-photometry, and including the 
nearby bow shock, as the SMC-matching aperture would, will drive the color redder. HIP~83509 is 
known to illuminate a reflection nebula \citep{Herbs75}. 

\LongTables
\begin{centering}
\begin{deluxetable*}{crrcrrrrrr}
\tabletypesize{\scriptsize}
\tablecaption{\emph{Hipparcos}/\emph{WISE} OB stars with spatially resolved mid-IR excess and size $\geq$3,000 AU \label{tab:wise_res2}} \tablewidth{0pt}
\tablehead{     \colhead{HIP}     &
                \colhead{$\alpha$}      &
                \colhead{$\delta$}      &
                \colhead{Spectral}      &
                \colhead{$K_s-W4$}      &
                \colhead{$W3-W4$}      &
                \colhead{$\varpi$}      &
                \colhead{$\sigma_{\rm \varpi}$}      &
                \colhead{$\chi^2_{\rm r,W4}$}      &
                \colhead{Resolution}      \\
                \colhead{}          &
                \colhead{(deg,J2000)}          &
                \colhead{(deg,J2000)}          &
                \colhead{Type}          &
                \colhead{(mag)}          &
                \colhead{(mag)}          &
                \colhead{(mas)}          &
                \colhead{(mas)}          &
                \colhead{}          &
                \colhead{(AU)}          }
\startdata
876 &   2.693195 &  58.769517 & B7V &  4.22 &  3.58 &  1.69 &  0.89 &     39.8 & 7100\\
2464 &   7.822530 &  71.122005 & B8 &  4.67 &  3.62 &  0.51 &  0.83 &     28.9 & $>$14000\\
7175 &  23.124723 &  67.961186 & B9 &  4.99 &  3.08 &  3.82 &  0.71 &     24.5 & 3100\\
11891 &  38.335785 &  61.521715 & O5e &  5.73 &  4.43 &  2.46 &  0.92 &    178.2 & 4900\\
13487 &  43.418713 &  64.451985 & B8 &  4.37 &  3.38 &  1.77 &  1.13 &     75.9 & 6800\\
15853 &  51.059112 &  61.538570 & B2III-IV &  6.02 &  4.47 &  3.09 &  1.06 &    193.9 & 3900\\
15941 &  51.317762 &  60.483711 & B2III &  5.11 &  4.21 &  1.57 &  1.14 &    120.7 & 7600\\
18701 &  60.097009 &  56.901598 & B0.5V &  5.20 &  3.52 &  0.20 &  1.21 &     62.7 & $>$9900\\
21779 &  70.184969 &  52.616737 & B3 &  5.18 &  3.70 &  0.77 &  1.21 &     24.7 & $>$9900\\
22237 &  71.768536 & -67.114753 & B1.5Ia: &  4.99 &  3.34 &  0.03 &  1.67 &      4.5 & $>$7200\\
25439 &  81.603582 &  39.646548 & B8 &  4.61 &  3.37 &  2.01 &  0.86 &     45.7 & 6000\\
26000 &  83.237848 &  -4.566481 & B2Vvar &  5.53 &  4.46 &  0.66 &  0.73 &     78.9 & $>$16000\\
26237 &  83.846517 &  -4.838358 & B2III\dots &  5.27 &  3.25 &  3.69 &  1.20 &     53.6 & 3300\\
26683 &  85.056416 &  -1.462568 & B3Vn &  5.03 &  4.00 &  3.19 &  0.80 &     92.2 & 3800\\
26742 &  85.234877 &  -1.507181 & B2IV &  4.94 &  3.72 &  3.03 &  0.55 &     28.6 & 4000\\
27040 &  86.021469 &  30.933819 & B3II &  4.58 &  3.96 &  0.55 &  1.18 &     90.2 & $>$10000\\
29115 &  92.098171 &  -6.548652 & B3 &  6.30 &  4.54 &  1.30 &  1.13 &    122.8 & 9200\\
29120 &  92.109496 &  -5.339577 & B1V &  5.46 &  4.20 &  0.70 &  0.81 &     76.1 & $>$15000\\
29127 &  92.133420 &  13.966932 & B1V &  5.13 &  4.34 &  2.31 &  2.96 &     42.1 & $>$4100\\
29147 &  92.232588 &  15.705052 & O7 &  4.05 &  3.34 &  1.03 &  1.04 &     18.7 & $>$12000\\
29216 &  92.414890 &  20.487626 & O6 &  5.01 &  5.07 &  3.28 &  0.71 &    112.5 & 3700\\
29310 &  92.697331 &  13.659469 & B1III &  4.27 &  3.90 &  1.49 &  0.51 &    135.4 & 8100\\
29587 &  93.525801 &  20.169705 & B1:V:nn &  5.38 &  3.89 &  0.78 &  0.91 &     42.8 & $>$13000\\
33432 & 104.283044 & -10.279858 & B7V &  7.52 &  4.44 &  0.79 &  1.06 &     27.5 & $>$11000\\
33457 & 104.338423 & -11.117642 & B3V &  4.06 &  3.14 &  1.39 &  1.60 &      3.9 & $>$7500\\
33735 & 105.119203 &  -8.866022 & B6V &  6.96 &  3.53 &  3.82 &  1.23 &     90.3 & 3100\\
33953 & 105.677561 & -11.453210 & B3n &  4.83 &  3.70 &  2.41 &  1.09 &     27.9 & 5000\\
33987 & 105.782381 & -11.199171 & B5III &  4.40 &  3.44 &  3.52 &  1.10 &     21.7 & 3400\\
34116 & 106.106381 & -10.454372 & B0IV:e &  5.07 &  3.14 &  3.92 &  0.99 &     21.8 & 3100\\
34178 & 106.319780 & -12.326247 & B1II/III &  7.61 &  3.61 &  1.74 &  0.91 &     63.7 & 6900\\
35375 & 109.617448 & -17.189929 & B3II &  4.13 &  3.70 &  0.98 &  1.35 &     49.6 & $>$8900\\
39732 & 121.821833 & -29.078109 & B\dots &  5.44 &  4.05 &  2.38 &  2.02 &      5.8 & 5000\\
40016 & 122.585682 & -49.237347 & B3IV &  4.36 &  3.50 &  1.86 &  0.28 &     71.8 & 6500\\
40024 & 122.613515 & -49.164144 & B6V &  4.07 &  3.37 &  1.71 &  0.57 &     28.0 & 7000\\
42363 & 129.564128 & -39.417647 & B2/B3II &  4.18 &  3.36 &  0.14 &  0.73 &     47.9 & $>$16000\\
43955 & 134.281472 & -43.256186 & B3V &  4.17 &  3.65 &  2.98 &  0.63 &     64.7 & 4000\\
48868 & 149.503763 & -52.893065 & B9III/IV &  4.38 &  2.61 &  1.33 &  0.48 &     32.8 & 9000\\
50272 & 153.969193 & -57.375025 & B1Ia &  5.17 &  4.45 &  0.88 &  0.57 &     91.5 & 14000\\
50843 & 155.724336 & -59.624550 & B2evar &  5.89 &  2.78 &  1.69 &  0.82 &    101.8 & 7100\\
51063 & 156.500793 & -57.826934 & B1.5III &  6.72 &  4.52 &  2.14 &  0.96 &    167.6 & 5600\\
52628 & 161.433835 & -59.407818 & O5e &  4.50 &  4.73 &  1.22 &  0.45 &      8.1 & 9800\\
56021 & 172.225758 & -62.652734 & O6 &  5.28 &  5.00 &  1.81 &  0.76 &    127.5 & 6600\\
56134 & 172.601286 & -63.817226 & O9.5V: &  6.58 &  3.91 &  0.81 &  0.72 &     36.8 & 15000\\
56833 & 174.763656 & -63.429739 & O6 &  4.97 &  4.37 &  0.51 &  0.59 &      9.5 & $>$20000\\
65307 & 200.765817 & -63.061773 & O9.5V: &  5.24 &  3.91 &  0.01 &  1.21 &     18.9 & $>$9900\\
68564 & 210.545313 & -61.957903 & B8/B9II/III &  4.43 &  3.95 &  2.51 &  1.06 &     18.9 & 4800\\
68985 & 211.854676 & -61.354608 & B8/B9V &  8.72 &  3.88 &  3.30 &  1.66 &    107.5 & 3600\\
69582 & 213.640362 & -61.798978 & B5/B6V &  4.83 &  2.80 &  2.11 &  1.82 &     94.6 & 5700\\
75079 & 230.135491 & -59.543772 & B1III &  4.62 &  3.36 &  1.28 &  0.60 &     62.5 & 9400\\
76881 & 235.488359 & -56.612306 & B8/B9III &  5.59 &  4.15 &  2.69 &  1.68 &     35.1 & 4500\\
77452 & 237.199434 & -54.395655 & B2/B3Vnne &  6.26 &  4.11 &  2.62 &  0.86 &     60.5 & 4600\\
78034 & 239.007071 & -66.152566 & Be & 11.73 &  4.42 &  0.94 &  2.30 &     11.4 & $>$5200\\
79936 & 244.730729 & -50.391866 & B9V &  6.21 &  4.31 &  2.88 &  1.31 &     19.0 & 4200\\
81308 & 249.097980 & -45.397421 & B9IV &  4.06 &  4.93 &  1.17 &  1.34 &     46.7 & $>$9000\\
81711 & 250.349725 & -46.385713 & B9IV &  4.24 &  3.02 &  2.37 &  0.95 &     33.2 & 5100\\
82286 & 252.208104 & -39.771304 & B1Ib &  5.72 &  3.83 &  1.74 &  1.86 &     28.9 & $>$6500\\
82936 & 254.227790 & -40.512342 & O7 &  4.47 &  4.12 &  2.53 &  1.08 &     55.9 & 4700\\
85322 & 261.540553 & -34.557386 & B4II &  5.54 &  3.80 &  0.21 &  1.26 &     17.6 & $>$9500\\
85985 & 263.602138 & -32.504438 & B1:V:nn &  5.61 &  4.13 &  0.92 &  1.26 &     22.7 & $>$9500\\
88581 & 271.293962 & -24.398570 & O8 &  5.39 &  4.21 &  0.15 &  0.63 &     42.8 & $>$19000\\
88943 & 272.323748 & -23.988395 & B6III: &  4.24 &  3.61 &  2.18 &  0.67 &     76.0 & 5500\\
89647 & 274.416584 & -19.672176 & B1/B2III &  4.28 &  3.09 &  2.80 &  1.11 &     18.4 & 4300\\
89750 & 274.734115 & -13.808632 & B3Ib &  7.51 &  4.10 &  2.34 &  2.24 &     61.2 & 5100\\
89933 & 275.251235 & -17.151709 & B2/B3Ib/II &  4.83 &  4.52 &  1.53 &  1.10 &     60.0 & 7800\\
90225 & 276.139576 & -13.653548 & B3n+B0 &  7.91 &  4.20 &  0.18 &  1.85 &     49.0 & $>$6500\\
90707 & 277.603649 &   1.223237 & B7V &  4.67 &  2.51 &  3.82 &  1.52 &     65.6 & 3100\\
98418 & 299.954595 &  35.309314 & O7 &  7.03 &  5.09 &  1.89 &  0.98 &    143.8 & 6300\\
100193 & 304.840469 &  40.887906 & B2+\dots &  4.18 &  3.78 &  1.59 &  0.74 &     30.7 & 7500\\
100628 & 306.065481 &  42.300383 & B3n &  4.71 &  2.89 &  0.45 &  1.11 &     11.3 & $>$11000\\
102167 & 310.528596 &  43.184366 & O9p\dots &  6.13 &  5.24 &  1.38 &  0.90 &    241.5 & 8700\\
102219 & 310.671204 &  36.380868 & B0.5Ib &  5.29 &  4.57 &  1.06 &  0.68 &    126.3 & 11000\\
102274 & 310.840022 &  63.209122 & B5 &  6.84 &  4.50 &  0.67 &  0.51 &     97.0 & 18000\\
102410 & 311.294174 &  51.210552 & B0.5IV &  5.77 &  3.74 &  0.74 &  0.82 &     65.7 & $>$15000\\
102449 & 311.397001 &  46.350576 & O9V &  4.75 &  3.32 &  0.77 &  0.86 &     18.5 & $>$14000\\
103061 & 313.221695 &  42.607743 & B0V &  4.52 &  3.63 &  1.67 &  0.70 &     25.9 & 7200\\
103428 & 314.319092 &  48.295773 & B8Ib &  4.51 &  2.87 &  1.63 &  0.64 &     33.9 & 7400\\
105113 & 319.389464 &  60.100448 & B0V &  6.28 &  3.91 &  2.03 &  0.71 &     83.2 & 5900\\
106079 & 322.311852 &  44.338126 & B2V:nne: &  4.03 &  2.53 &  2.12 &  0.48 &     44.6 & 5700\\
106712 & 324.237729 &  68.185363 & B3V &  4.34 &  3.81 &  1.01 &  0.51 &     80.1 & 12000\\
106843 & 324.609513 &  56.973738 & B0V &  4.07 &  3.75 &  1.60 &  0.41 &     91.3 & 7500\\
106956 & 324.935149 &  58.245559 & B5 &  4.53 &  3.87 &  2.13 &  1.21 &     37.8 & 5600\\
107123 & 325.493871 &  58.500058 & B3 &  4.03 &  3.54 &  1.60 &  0.72 &     24.2 & 7500\\
110125 & 334.615847 &  63.222899 & B0.5V &  4.13 &  3.51 &  0.94 &  1.64 &     29.4 & $>$7300\\
110937 & 337.162696 &  58.845777 & B8 &  5.48 &  4.25 &  1.67 &  0.64 &    157.2 & 7200\\
111785 & 339.632630 &  55.834830 & B1:IV:nnpe &  5.84 &  3.37 &  1.76 &  0.84 &     58.1 & 6800\\
112887 & 342.912154 &  51.845116 & B & 11.84 &  4.53 &  0.72 &  3.49 &      3.4 & $>$3400\\
113224 & 343.940806 &  57.601924 & B0.5:IV: &  5.29 &  4.22 &  0.31 &  1.01 &     29.0 & $>$12000\\
113301 & 344.177296 &  62.624876 & B1V &  7.09 &  4.26 &  1.59 &  0.89 &     63.6 & 7500\\
113538 & 344.928631 &  62.777310 & B1.5Vn &  7.21 &  4.23 &  1.58 &  1.53 &     78.9 & 7600
\enddata
\tablecomments{Column 4 is the spectral type as taken from \emph{Hipparcos}. Columns 5 and 6 
list the colors taken from \emph{WISE} and \emph{2MASS} with W3 and W4 effective wavelengths 
at 12 and 22~\mum. Column 7 lists the parallax from \emph{Hipparcos} and column 8 lists the error. 
Column 9 lists the reduced $\chi^2$ that the W4 data is compatible with the PSF from the \emph{WISE} 
catalog. Values above 3 are considered to indicate extended emission. Finally, column 10 lists the 
physical size for the 12\arcsec\ W4 resolution. This is the minimum size for extended sources. 
}
\end{deluxetable*}
\end{centering}
\bibliography{SMC_Bstars}
\bibliographystyle{apj}

\end{document}